\documentclass[a4paper, 11pt]{article}
\usepackage[top=1in, left=1in, right=1in, bottom=1in]{geometry}

\usepackage{amsmath, amssymb, amsthm}
\usepackage{float}
\usepackage{xspace}
\usepackage{paralist}
\usepackage[algo2e,boxed,noline,noend,linesnumbered]{algorithm2e}
\usepackage{multirow}
\usepackage{pgfplots}

\newenvironment{proofof}[1]{\noindent {\bf Proof of #1:  }}{\qed}

\newtheorem{definition}{Definition}[section]
\newtheorem{claim}{Claim}[section]
\newtheorem{corollary}{Corollary}[section]
\newtheorem{theorem}{Theorem}[section]
\newtheorem{lemma}{Lemma}[section]
\newtheorem{fact}{Fact}[section]

\newcommand{\cupdot}{\mathbin{\mathaccent\cdot\cup}}
\newcommand{\widef}{\widehat{f}}
\newcommand{\measpace}{\ensuremath{\mathcal{M}(\R_+^\Omega)}\xspace}
\newcommand{\rand}{\mathsf{R}}
\newcommand{\Ex}{\ensuremath{\text{{\bf E}}}}
\newcommand{\norm}[1]{\ensuremath{\left \| #1 \right \|}}
\newcommand{\cho}{\mathsf{C}}
\newcommand{\ind}{\mathsf{I}}

\newcommand{\eps}{\epsilon}

\newcommand{\ve}{\varepsilon}

\newcommand{\B}{\mathbf{B}}
\newcommand{\R}{\mathbb{R}}
\newcommand{\Z}{\mathbb{Z}}
\newcommand{\I}{\mathcal{I}}

\newcommand{\ceil}[1]{\ensuremath{\lceil #1 \rceil}}
\newcommand{\abs}[1]{\lvert#1\rvert}

\newcommand{\OPT}{\ensuremath{\mathsf{OPT}}\xspace}

\newcommand{\expct}[1]{\ensuremath{\text{{\bf E}$\left[#1\right]$}}}

\newcommand{\floor}[1]{\left \lfloor #1 \right \rfloor}

\newcommand{\Real}{\mathsf{Re}}

\newcommand{\im}{\mathsf{i}}

\newcommand{\txts}{\textstyle}

\newcommand{\U}{\ensuremath{\mathcal{U}}}

\newcounter{note}[section]

\newcommand{\ignore}[1]{}

\newcommand*\samethanks[1][\value{footnote}]{\footnotemark[#1]}

\newcommand{\shortv}[1]{}

\newcommand{\hubert}[1]{\textcolor{red}{(Hubert: #1)}}
\newcommand{\zhiyi}[1]{\textcolor{blue}{(Zhiyi: #1)}}

\title{Online Submodular Maximization with Free Disposal: Randomization Beats $\frac{1}{4}$ for Partition Matroids}

\author{T-H. Hubert Chan\thanks{Department of Computer Science, The University of Hong Kong. {\texttt{\{hubert,zhiyi,sfjiang,nkang,zhtang\}@cs.hku.hk}}} \and Zhiyi Huang\samethanks \and Shaofeng H.-C. Jiang\samethanks \and Ning Kang\samethanks \and Zhihao Gavin Tang\samethanks}
\date{}

\begin{document}

\begin{titlepage}
	
	\maketitle
	
	\begin{abstract}
We study the online submodular maximization problem with free disposal under a matroid constraint. 
Elements from some ground set arrive one by one in rounds, and the algorithm maintains a feasible set that is independent in the underlying matroid.  
In each round when a new element arrives, the algorithm may accept the new element into its feasible set and possibly remove elements from it, provided that the resulting set is still independent.  
The goal is to maximize the value of the final feasible set under some monotone submodular function, to which the algorithm has oracle access.



For $k$-uniform matroids, we give a deterministic algorithm with competitive ratio at least $0.2959$, and the ratio approaches $\frac{1}{\alpha_\infty} \approx 0.3178$ as $k$ approaches infinity, improving the previous best ratio of $0.25$ by Chakrabarti and Kale (IPCO 2014), Buchbinder et al.\ (SODA 2015) and Chekuri et al. (ICALP 2015).
We also show that our algorithm is optimal among a class of deterministic monotone algorithms that accept a new arriving element only if the objective is strictly increased.

Further, we prove that no deterministic monotone algorithm can be strictly better than $0.25$-competitive even for partition matroids, the most modest generalization of $k$-uniform matroids, matching the competitive ratio by Chakrabarti and Kale (IPCO 2014) and Chekuri et al. (ICALP 2015).
Interestingly, we show that randomized algorithms are strictly more powerful by giving a (non-monotone) randomized algorithm for partition matroids with ratio $\frac{1}{\alpha_\infty} \approx 0.3178$.  

Finally, our techniques can be extended to a more general problem that generalizes both the online submodular maximization problem and the online bipartite matching problem with free disposal.  
Using the techniques developed in this paper, we give constant-competitive algorithms for the submodular online bipartite matching problem.
\end{abstract}

	\thispagestyle{empty}
\end{titlepage}

\section{Introduction}


We study online submodular maximization with free disposal under a matroid constraint.
Let $\Omega$ be the ground set of elements, $f : 2^\Omega \rightarrow \R_+$ be a non-negative submodular objective function, and $\I \subseteq 2^\Omega$ be a collection of feasible subsets in $\Omega$ that the algorithm can choose from. The goal is to find $S \in \I$ such that
$f(S)$ is maximized.  
In this paper, we focus on the case when $\I$ forms a matroid, i.e., a set of elements $S$ is feasible if it is independent with respect to the matroid.


The offline version~\cite{citeulike:416655,BF01588971} has been extensively studied due to its many applications,
such as 
the maximum coverage problem with group budget constraints~\cite{Chekuri04,Khuller:1999}, the separable assignment problem~\cite{FleischerGMS06,FleischerGMS11,CalinescuCPV07}, the assignment learning problem \cite{golovin2009online,golovin2014online}, the sequence optimization problem \cite{dey2013contextual}, and the submodular welfare maximization probelm~\cite{Vondrak:2008,DBLP:journals/toc/FeigeV10,DBLP:journals/corr/abs-1202-2792}.

In the online version (without free disposal), the elements in $\Omega$ arrive in rounds in an arbitrary order.
The algorithm maintains a feasible set $S \in \I$, which is initially empty.
In each round, the algorithm must irrevocably decide whether to add the arriving element $u$ into $S$ (provided
that $S + u \in \I$) without knowing the future elements.  
We assume the algorithm has value oracle access to function $f$ on any subset of elements arrived so far.
However, this version of the problem has no non-trivial competitive ratio even
for the simple constraint $|S| \leq 1$.\footnote{Consider a sequence of elements with value $1, \rho, \rho^2, \dots, \rho^n$ for some $\rho \gg 1$ and $n$ unknown to the algorithm.}


Buchbinder et al.~\cite{BuchbinderFS15a} explicitly considered the online version with \emph{free disposal}\footnote{The terms free disposal~\cite{FeldmanKMMP09} and preemption~\cite{BuchbinderFS15a} have both been used in the literature. We will use free disposal throughout this paper.}. 
In this model, in each round,
the algorithm may also remove elements from its current feasible set $S$,
as well as adding the new arriving element into $S$, as long as the resulting $S$ is still in $\I$.
(However, elements that have not been chosen at their arrival, or have been disposed of cannot
be retrieved back.)
They pointed out that a result by
Chakrabarti and Kale~\cite{ChakrabartiK14} 
implies a $0.25$-competitive algorithm for maximizing monotone submodular functions online under a $k$-uniform matroid constraint, i.e.,
for some positive integer $k$, $\I$ consists of all subsets with cardinalities at most $k$.
Buchbinder et al.~\cite{BuchbinderFS15a} also proposed a different $0.25$-competitive algorithm 
which leads to a $0.0893$-competitive randomized algorithm for non-monotone submodular functions under a $k$-uniform matroid constraint. They also showed several hardness results for various settings.

\noindent \emph{Streaming Model.}
Chakrabarti and Kale~\cite{ChakrabartiK14} and Chekuri et al.~\cite{DBLP:conf/icalp/ChekuriGQ15} 
considered streaming version of this problem in which the algorithm has limited memory.
They consider even more general independent systems than matroids, and their algorithms for the case of matroids can be interpreted as an online algorithm with free disposal that is $0.25$-competitive.
We summarize the previous results in Table \ref{tab:previous}.

\begin{table}[H]
	\centering
	\begin{tabular}{|c|c|c|c|}
		\hline
		Matroid & Objective $f$ & Algorithm & Hardness \\
		\hline
		\multirow{2}{*}{$k$-Uniform} & Monotone & $0.25$~\cite{BuchbinderFS15a, ChakrabartiK14, DBLP:conf/icalp/ChekuriGQ15} & $0.5$~\cite{BuchbinderFS15a} \\
		\cline{2-4}
		& General & $0.0893$~\cite{BuchbinderFS15a} & $0.438$~\cite{BuchbinderFS15a} \\
		\hline
		\multirow{2}{*}{General} & Monotone & 0.25~\cite{ChakrabartiK14,DBLP:conf/icalp/ChekuriGQ15} & $0.5$~\cite{BuchbinderFS15a} \\
		\cline{2-4}
		& General &  - 
		& $0.438$~\cite{BuchbinderFS15a} \\
		\hline
	\end{tabular}
	\vspace{-5pt}
	\caption{A summary of previous results}
	\label{tab:previous}
	\vspace{10pt}
\end{table}

\paragraph{Our Contributions.}
We make contribution to the problem by improving both the upper and lower bounds on the competitive ratios in various settings.
A summary is given in Table~\ref{tab:results}.

\ignore{
	\noindent \textbf{General Matroid.}
	We give the first non-trivial competitive algorithm for online submodular maximization under general matroid constraint in the free disposal model.
	For monotone submodular functions, we show that our algorithm is $0.25$-competitive in Section~\ref{sec:matroid}.
	Our algorithm also follows a greedy framework
	as in~\cite{BuchbinderFS15a} with the following twist.
	Loosely speaking, a main novel idea in our algorithm is that decisions are made based on
	an auxiliary function $w(\cdot)$ that
	measures the marginal value of an item $u$
	with respect to \emph{all the items} the algorithm
	has accepted at some point before $u$ arrives.
	To ensure the objective does not decrease after each round,
	the actual auxiliary function we use are more complicated
	and value an accepted element $u$ in subsequent rounds at least as much as when it arrived. 
	At a high level,
	our algorithm remembers how important each accepted item $u$ is and does not devalue it no matter what new items are accepted after $u$.
	In contrast, existing algorithms~\cite{BuchbinderFS15a} decide whether to accept a new element $u$ and which element $u'$ in the current
	feasible set $S$ to remove based 
	on how much the new value $f(S-u'+u)$ improves with respect to the existing value $f(S)$,
	independent of the arrival order of elements.
}


\begin{table}
	\centering
	\begin{tabular}{|c|c|c|c|c|c|}
		\hline
		\multirow{2}{*}{Matroid} & \multicolumn{3}{|c|}{Algorithm} & \multicolumn{2}{|c|}{Hardness for Det. Alg.} \\
		\cline{2-6}
		& \multicolumn{2}{|c|}{Det. Alg.} & Rand. Alg. & General Alg. & Monotone Alg. \\
		\hline
		\multirow{2}{*}{$k$-Uniform} & Worst $k$ & $k \to \infty$ & \multirow{2}{*}{$\frac{1}{\alpha_\infty} \approx 0.318$} & \multirow{2}{*}{$0.5$~\cite{BuchbinderFS15a}} & {\footnotesize $\frac{1}{\alpha_\infty}$} $\approx 0.318$ \\ 
		\cline{2-3}
		& $0.296$ (Thm.~\ref{th:uni_ratio}) & {\footnotesize $\frac{1}{\alpha_\infty}$} $\approx 0.318$ & & & (Thm.~\ref{theorem:submodular_hardness}) \\
		\cline{1-3} \cline{5-6}
		\multirow{2}{*}{Partition} & \multicolumn{2}{|c|}{\multirow{2}{*}{0.25~\cite{BuchbinderFS15a, ChakrabartiK14, DBLP:conf/icalp/ChekuriGQ15}}} & \multirow{2}{*}{(Thm.~\ref{th:rand_alg})} & $0.382$ & $0.25$ \\ 
		& \multicolumn{2}{|c|}{} & &  (Thm.~\ref{thm:lowerbound_general}) & (Thm.~\ref{thm:lowerbound_monotone}) \\
		\hline
	\end{tabular}
	\vspace{-5pt}
	\caption{A summary of the main results in this paper.
		The objective is monotone.
		The deterministic algorithms are monotone. The hardness results are with respect to deterministic algorithms.}
	\label{tab:results}
	\vspace{10pt}
\end{table}

\noindent \textbf{Monotone Algorithm.}~ 
A deterministic algorithm is  \emph{monotone} if,
after each round, it either keeps the same the set of chosen elements, or makes changes that strictly increase the objective (see the 
precise Definition~\ref{defn:mono_alg}).

\emph{Why monotone algorithms?}
First of all, monotonicity of algorithms is a natural requirement for some applications.
Consider the example of managing a soccer team proposed by Buchbinder et al.~\cite{BuchbinderFS15a}.
It would be difficult to talk the board and the fan base into a transfer of players without immediate benefits to the team.
Further, to our knowledge, all known algorithms in the literature for monotone submodular objectives are deterministic and monotone.
Hence, it would be interesting to fully understand the potential of this family of algorithms.

Our first contribution is an improved algorithm for the case of $k$-uniform matroids.  
We propose a deterministic monotone algorithm (Section~\ref{sec:cardinality}) that is at least $0.2959$-competitive for monotone submodular functions, improving the previous ratio of $0.25$~\cite{BuchbinderFS15a, ChakrabartiK14, DBLP:conf/icalp/ChekuriGQ15}.
As $k$ tends to infinity, our competitive ratio approaches $\frac{1}{\alpha_\infty} \approx 0.3178$ (from below),
where $\alpha_\infty$ is the unique root of
$\alpha = e^{\alpha-2}$ that is greater than 1.  
Further, we obtain a matching hardness result (Section~\ref{sec:hardness_uni}) in the sense that for any $\epsilon > 0$, there is some sufficiently large $k > 0$ such that no deterministic monotone algorithm has competitive ratio at least $\frac{1}{\alpha_\infty} + \epsilon$ under a $k$-uniform matroid constraint.

For general matroids, we show that 
no deterministic monotone algorithm is strictly better than $0.25$-competitive even for partition matroids, the most modest generalization of $k$-uniform matroids (Section~\ref{sec:hardness_partition}).
Our hardness result matches the competitive ratio by \cite{ChakrabartiK14, DBLP:conf/icalp/ChekuriGQ15}.

\ignore{
	An interesting insight from our results is that for online submodular maximization with monotone objectives using deterministic monotone algorithms, it suffices to consider a linear objective to compare the value of the feasible set with that of the union of an optimal set together with \emph{all the elements that have been accepted by the algorithm at some point}.
	The insight for monotone algorithms in retrospect help us better understand the limitations of general algorithms,
	and we show a hardness of $0.382$ for (not necessarily monotone) deterministic algorithms (Section~\ref{sec: hardness_general}).
}

\noindent 
\textbf{Randomized Algorithms on Partition Matroids.}~
Given the hardness for deterministic monotone algorithms, it is natural to ask whether we could get a better competitive ratio using randomized (and non-monotone) algorithms.
We consider the setting that the adversary first fixes the arrival order before the algorithm samples its randomness.
We give affirmative answer to this question for the case of partition matroids.
While a partition matroid can be viewed as a union of disjoint uniform matroids, our $k$-uniform algorithm fails to generalize directly due to the case of a union of $1$-uniform matroids.
We handle a single $1$-uniform matroid using the trivial algorithm that keeps the most valuable element, but this trick no longer works when there is a union of many $1$-uniform matroids.

Our high-level idea is to use randomized algorithms to effectively allow picking only a fraction of each element and, thus, treating each partition as effectively having large size (w.r.t.\ tiny fractions of the elements).
There are some technical obstacles.
First of all, any rounding scheme that does not incur an intrinsic loss in the objective, e.g., pipage rounding~\cite{DBLP:journals/jco/AgeevS04, DBLP:journals/jacm/GandhiKPS06}, fails to work in the online setting.
As a result, we settle for an online rounding scheme that loses a $1 - \frac{1}{e}$ factor in the objective in the worst case.
However, due to the intrinsic loss from rounding, a na\"ive competitive analysis gives only the product of $\frac{1}{\alpha_\infty}$ and $1 - \frac{1}{e}$ which is smaller than $\frac{1}{4}$.
We avoid losing an extra $1 - \frac{1}{e}$ factor observing that the scenario that gives rise to a $\frac{1}{\alpha_\infty}$ ratio for the fractional algorithm and the scenario that incurs a rounding loss of $1 - \frac{1}{e}$ do not occur simultaneously. 
To instantiate this observation, we introduce a novel inequality (Lemma \ref{lemma:f_and_hatf}) that allows us to directly compare the optimal objective and the expected value of $f$ for the fractional solution after the rounding.


\noindent \textbf{Dichotomy between Deterministic and Randomized Algorithms.}~ 
Our improved competitive ratio for partition matroids shows that (non-monotone) randomized algorithms are strictly more powerful, as our randomized algorithm on partition matroids has ratio $\frac{1}{\alpha_\infty} \approx 0.3178$, which is achieved by our ``continuous'' algorithm. Conventional discrete algorithms can approach this ratio arbitrarily closely.

\noindent 
\textbf{Extensions.}~
Using the new insights we get for monotone objectives, we further introduce a randomized algorithm that is $0.1145$-competitive for non-monotone objectives under uniform matroid constraints (Section~\ref{sec:non-mono}), improving the previous $0.0893$ ratio~\cite{BuchbinderFS15a}.

\noindent \textbf{Generalized Online Bipartite Matching.}
Our techniques in fact solve a more general problem that generalizes both the online submodular maximization problem and the online bipartite matching problem with free disposal that was first proposed in~\cite{FeldmanKMMP09}.
In this submodular online bipartite matching problem, each offline node
corresponds to some agent $\lambda \in \Lambda$,
and each online node corresponds to an item $u \in \Omega$.
Each agent~$\lambda$ has an evaluation
function $f^\lambda : 2^\Omega \rightarrow \R_+$,
and is also associated with a matroid $(\Omega, \I^\lambda)$.


The online submodular maximization problem is a special case with only one agent, and the online bipartite matching problem is a special case when each agent is under a $1$-uniform matroid constraint.  
We show that each of our $\frac{1}{\alpha}$-competitive deterministic online algorithms for
a single offline node that is defined in Section~\ref{sec:cardinality} and Section~\ref{sec:matroid}
induces $\frac{1}{\alpha+1}$-competitive algorithms for submodular online bipartite matching (Section~\ref{sec:multi-offline}) respectively.

\noindent \textbf{Streaming Model.} 
In contrast to previous approaches \cite{BuchbinderFS15a, ChakrabartiK14, DBLP:conf/icalp/ChekuriGQ15}, our improved algorithm (Algorithm~\ref{alg:k_uni}) for uniform matroid cannot be fitted into the streaming model. As we shall see, it is crucial for the algorithm to remember all the items that have been selected, where the limited space is insufficient. This might represent a separation between the streaming and the online version of the model.

\noindent \textbf{Paper Organization.} 
The monotone algorithm for 
a $k$-uniform matroid is given in Section~\ref{sec:cardinality}.  
The randomized algorithm for
a partition matroid is given in Section~\ref{sec:part_cont}.
The hardness results for uniform matroids and general matroids are given in Section~\ref{sec:hardness_uni} and Section~\ref{sec:hardness_partition}, respectively.
For completeness, we also reprove the $\frac{1}{4}$ competitive ratio under a general matroid in Section~\ref{sec:matroid}, which is useful for our submodular online bipartite matching problem in Section~\ref{sec:multi-offline}.
Section~\ref{sec:non-mono} gives randomized algorithms for non-monotone objective functions.


\paragraph{Other Related Work.}
We have already discussed the related work on online submodular maximization.
There is a vast literature on submodular maximization in different settings. 
We will review some of the results that are most related.


In the offline setting, Buchbinder et al. introduced 
a $0.5$-approximate randomized algorithm for maximizing a non-monotone submodular function 
with no constraint~\cite{Buchbinder:2012:TLT:2417500.2417835}. Feige et al. had previously proved that $0.5$ is the best possible for this setting~\cite{Feige07maximizingnon-monotone}.
Recently, Buchbinder and Feldman~\cite{BuchbinderF15} obtained a deterministic algorithm 
which also achieves the optimal $0.5$ ratio.
For a uniform matroid constraint, Nemhauser et al.~\cite{citeulike:416655} showed a $(1-\frac{1}{e})$-approximate algorithm for monotone objectives, which is optimal~\cite{citeulike:416654}.
Feige~\cite{Feige:1998} further proved that even when the objective is a coverage function, no algorithm can achieve better than $(1-\frac{1}{e})$, assuming $\textsc{P} \neq \textsc{NP}$.
For maximizing a monotone submodular function under a general matroid constraint, the simple greedy algorithm is $\frac{1}{2}$-approximate \cite{BF01588971}.
C\u{a}linescu et al.~\cite{CalinescuCPV07} found an algorithm that is $(1-\frac{1}{e})$-approximate. 
Recently, Filmus and Ward~\cite{abs-1204-4526} introduced a simpler algorithm with the same ratio.
Finally, for non-monotone objectives under a matroid constraint, the best known approximation ratio is $\frac{1}{e}$~\cite{FeldmanNS11}, and the best hardness result is $0.478$~\cite{GharanV11}. 
For maximizing a non-monotone submodular function with multiple matroid constraints, Lee et al.~\cite{LeeMNS09} presented a $\frac{1}{l+2+\frac{1}{l}+\varepsilon}$-approximate algorithm under $l$ matroid constraints. 



Our work is also closely related to the literature of submodular matroid secretary problem, which can be formulated as online submodular maximization without free disposal but assuming the elements arrive in random order.
The submodular secretary problem has been widely studied recently \cite{BarmanUCM12, FeldmanNS11, GuptaRST10, Ma0W13}, and constant-competitive algorithms have been found on some special cases, such as on a uniform matroid constraint \cite{BateniHZ13}, or when the objective function is to maximize the largest weighted element in the set \cite{10.2307/1402748, 10.2307/2283044}. 
However, there is no constant ratio for the general submodular matroid secretary problem till now.
Feldman and Zenklusen~\cite{FeldmanZ15} reduced the problem to the matroid secretary problem with linear objective functions, which implies an $O(\log\log(\text{rank}))$-competitive algorithm for the submodular matroid secretary problem, matching the current best result for the matroid secretary problem~\cite{Lachish14, Feldman:2015:SOC:2722129.2722208}.




There is a long line of research on the online bipartite cardinality matching problem~\cite{KarpVV90,GoelM08,Birnbaum2008,KarandeMT11,MahdianY11}, and the vertex-weighted version~\cite{AggarwalGKM11,DevanurJK13}. 
$\big(1-\frac{1}{e}\big)$-competitive algorithms are known for both cases.
For the most general edge-weighted version, random arrival order or free-disposal is necessary for any non-trivial competitive ratio.
When online nodes arrive in random order, Wang et al.~\cite{DBLP:journals/corr/WangW16d} discovered an algorithm that is $\big(1-\frac{1}{e}\big)$-competitive; 
while in the free-disposal model, the same competitive ratio can only achieved by assuming that the offline nodes have large capacity ~\cite{FeldmanKMMP09,devanur2013whole}, or under the small bid assumption~\cite{DBLP:journals/corr/WangW16d}.
It remains an important open question whether there is an online algorithm with a competitive ratio unconditionally strictly better than $\frac{1}{2}$ for the free-disposal model of the edge-weighted problem.


Finally, the buyback problem is similar to our model, except that costs are associated with disposals. The most related work is by Babaioff et al. \cite{babaioff2008selling,babaioff2009selling}, in which a matroid constraint is considered. For other buyback works, see e.g. \cite{constantin2009online,han2014online,iwama2002removable,varadaraja2011buyback,ashwinkumar2009randomized}.


\section{Preliminaries}
\label{sec:prelim}

We consider elements coming from some ground set $\Omega$,
and a non-negative submodular function $f: 2^\Omega \rightarrow \R_+$. 
We further assume that $f$ is monotone, i.e., $S \subseteq T$ implies that $f(S) \leq f(T)$.
Moreover, we assume that there is a matroid $(\Omega, \I)$,
and, without loss of generality, every singleton in $\Omega$ is independent.
Given $a \in \Omega$ and $S \subseteq \Omega$,
we denote $S - a := S \setminus \{a\}$, $S+a := S \cup \{a\}$ and $f(a|S) := f(S+a) - f(S)$.
We assume value oracle access to the function $f$ and independence oracle access to the matroid, i.e., given a
subset $S \subseteq \Omega$, an oracle returns the value $f(S)$
and answers whether $S \in \I$.  For a positive integer $n$,
we denote $[n] := \{1,2, \ldots, n\}$.

\noindent \textbf{Online maximization problem with free disposal.}
The algorithm maintains an independent set $S$, which is
initially empty.  Elements from $\Omega$ arrive in a finite sequence,
whose length is not known by the algorithm.  In each round when
an element $u$ arrives, the algorithm may remove some elements from $S$,
and may also include the current element $u$ into $S$, as long as $S$
remains independent in $\I$.  The objective is to maximize $f(S)$
at the end of the sequence.  We denote by $\OPT$ an independent subset
of elements in the sequence that maximizes the function $f$.
An algorithm has a competitive ratio $r \leq 1$, if at the end of every finite
sequence, the set $S$ satisfies $\expct{f(S)} \geq r \cdot f(\OPT)$.




Our deterministic algorithm in Section~\ref{sec:cardinality} is strictly monotone in the sense that it accepts an arriving item only if there is absolute advantage in doing so.  
This is formalized in Definition~\ref{defn:mono_alg}.  
Our hardness results in Section~\ref{sec:hardness_uni} and \ref{sec:hardness_partition} apply to any strictly monotone algorithms.

\begin{definition}{\sc (Strict Monotonicity)}
\label{defn:mono_alg}
An algorithm is strictly monotone if,
in each round, the algorithm includes the new arriving element
into the feasible set $S$ (and possibly removing some elements
from $S$) only if the objective value $f(S)$ strictly increases
compared to its value at the beginning of the round.
\end{definition}



\noindent \textbf{Auxiliary Set and Weight Function.~}  
Suppose we consider some algorithm.
Recall that the algorithm maintains some independent set $S$.  To facilitate
the analysis, at the end of each round, we consider an auxiliary set $A$ that keeps track of all the elements that have ever been added into $S$, but might have been removed since then.
For an element $u$ in the sequence,
at the beginning of the round in which $u$ arrives,
let $S(u)$ be the independent set maintained by the algorithm at this moment,
and let $A(u)$ be the set of elements that have been added
into $S$ (but might have been removed since then) at this moment.

We remark that the sets $S(u)$ and $A(u)$ are dependent on the algorithm,
and so are the following quantities.  We denote $w(u) := f(u|A(u))$ as the marginal value of $u$
when it arrives with respect to all the elements that have
ever been picked by the algorithm at this moment.
For some element $u$ that was added in some previous round,
we measure its value with respect to the current set $S$
by $w_S(u) := f(u | A(u) \cap S)$.
In general, given a weight function $\omega : \Omega \rightarrow \R$,
for $T \subseteq \Omega$, we denote $\omega(T) := \sum_{u \in T} \omega(u)$.

\noindent \textbf{Element Naming Convention.}
For $i \geq 1$, let $u_i$ denote the $i$-th element added to $S$ by the
algorithm (not the element arriving in the $i$-th round).

Let $S_i$ and $A_i$ denote the sets of elements contained in $S$ and $A$ respectively, where $S$ and $A$ are those at the end of the round in which $u_i$ arrives.
We denote the value function $w_i := w_{S_i}$.

\begin{lemma}{\sc (Relating $w$ and $f$)}
	\label{lemma:w_and_f}
	The functions $f$ and $w$ satisfy the following for each $n \geq 1$.
	\begin{compactitem}
		\item[(a)] $w(S_n) \leq w_n(S_n) = f(S_n) - f(\varnothing)$.
		\item[(b)] $w(A_n) = f(A_n) - f(\varnothing)$.
	\end{compactitem}
\end{lemma}

\begin{proof}
	For statement (a), the inequality follows because
	for each $v \in S_n$, $w(v) = f(v|A(v)) \leq f(v | A(v) \cap S_n) = w_n(v)$,
	where the inequality holds because $f$ is submodular.
	If we write the elements of $S_n = \{v_1, v_2, v_3, \ldots, \}$
	in the order they arrive,
	then for each $v_i \in S_n$,
	we have $w_n(v_i) = f(v_i | A(v_i) \cap S_n) = f(v_i |\{v_1, \ldots, v_{i-1}\})$.
	Hence, a telescoping sum gives $w_n(S_n) = f(S_n) - f(\varnothing)$.
	
	For statement (b), we write the elements of $A_n = \{u_1, u_2, \ldots, u_n\}$
	in the order they arrive, and observe that
	$w(u_i) = f(u_i | A(u_i)) = f(u_i|\{u_1, \ldots, u_{i-1}\})$.
	Hence, a similar telescoping sum gives $w(A_n) = f(A_n) - f(\varnothing)$.
\end{proof}

\begin{lemma}{\sc (Monotone $w_S(u)$)}
\label{lemma:monotone_ws}
Suppose an element $u$ arrives in some round.  Then,
in subsequent rounds, the value $w_S(u)$ does not decrease, when $S$ is modified by the algorithm.
\end{lemma}

\begin{proof}
Observe that in each round, the algorithm may remove elements from $S$,
and may add the new element to $S$. 
Since the elements in the sequence are distinct,
it follows that in the subsequent rounds after $u$ arrives,
the set $A(u) \cap S$ can only shrink.  
Since $f$ is submodular, it follows that $w_S(u) := f(u | A(u) \cap S)$ does not decrease, as the algorithm updates~$S$.
\end{proof}

\section{Improved Algorithm for k-Uniform Matroid}
\label{sec:cardinality}

In this section, we consider the special case of a $k$-uniform matroid, i.e.,
a set is independent \emph{iff} its cardinality is at most $k$.  
Observe that the trivial algorithm that keeps 
the singleton with the largest value achieves ratio $\frac{1}{k}$.
Since we wish to obtain a ratio better than $\frac{1}{4}$,
we consider $k \geq 4$ in this section.

\noindent \textbf{Defining $\alpha_k$.}
We define $\alpha_k$ to be the unique root
in the interval $(3,4)$ of the
equation $a = (1+\frac{a-2}{k+1})^{k+1}$.
We shall show that the competitive ratio is $\frac{1}{\alpha_k}$.
It can be shown that $\alpha_k$ is decreasing (see Lemma~\ref{lemma:alpha_k_dec}).
Moreover, as $k$ tends to infinity, the equation defining $\alpha_k$
becomes $a = e^{a - 2}$, which has root
$\alpha_\infty \approx 3.14619$.  
For simplicity, we write $\alpha := \alpha_k$ in the rest
of this section.

\begin{algorithm2e}
	\caption{Modified Algorithm for $k$-Uniform Matroids}
	\label{alg:k_uni}
	Initialize $S$ and $A$ to empty sets. \\
	\For{each round when $u$ arrives}{
		{

		\If {$w(u) > \frac{1}{k} \left( \alpha \cdot w_S(S) - w(A) \right)$} { 
		
		\eIf{$|S| = k$}{
			$u' := \arg\min_{v \in S}w_S(v)$;   (element in $S$ to be replaced)
		}{
		
			$u' := \bot$;  (no element in $S$ is replaced)
		}

		$S \leftarrow S - u' + u$; 
		$A \leftarrow A + u$;
		}
	}
	}
\end{algorithm2e}

\noindent \textbf{Replacement Condition.}~
\ignore{
For technical reasons, we assume that before ``real'' elements arrive,
the sets $S$ and $A$ are initialized to contain
$k$ dummy elements $\{u_1, u_2, \ldots, u_k\}$. For notation simplicity, we assume that they are the first $k$ elements accepted by the algorithm. However, since they do not really exist, this do not break the strict monotonicity of our algorithm.
The function $f$
is extended naturally such that any dummy elements are ignored.
In each round when the new element $u$ cannot be added to the
current feasible $S$, for $k$-uniform matroid, every element in $S$
could be potentially replaced by $u$.
}
The replacement condition
is $w(u) > \frac{1}{k} \left( \alpha \cdot w_S(S) - w(A) \right)$.
This means that even when $|S| < k$,
if the arriving element $u$
does not have enough value $w(u)$, then it will not be accepted.
When the algorithm decides to accept $u$,
if $|S| = k$, then 
the element $u'$ in $S$ with minimum value under $w_S$ will be replaced;
if $|S| < k$, for notational convenience, we set $u'$ to a dummy element $\bot$.
The function $f$
can be extended naturally such that any dummy elements are ignored,
and so $w_S(\bot) = 0$.

An important technical result is the following Lemma~\ref{lemma:large_replace},
which we use to argue about certain monotone properties of our algorithm.
Intuitively, it says that
we only accept an element if it is significantly better than the replaced one.
We defer its proof in Section~\ref{sec:proof_large_replace}.

\begin{lemma}{\sc (Monotone Replacement)}
\label{lemma:large_replace}
Suppose at the beginning of the round when $u$ arrives,
the feasible set is $S$, and
Algorithm~\ref{alg:k_uni} 
includes $u$ and discards $u'$ (which could be $\bot$) from $S$. 
Then, 
$w(u) > \frac{\alpha}{\alpha-1} \cdot w_S(u')$.
Observe this implies that $w(u) > w_S(u')$.
\end{lemma}



Using Lemma~\ref{lemma:large_replace}, by showing the following Lemma~\ref{lemma:alg_monotone},
we can conclude that
Algorithm~\ref{alg:k_uni} is strictly monotone.

\begin{lemma}{\sc (Monotonicity)}
	\label{lemma:alg_monotone}
	Suppose in each round
	the algorithm only replaces $u'$ (which could be $\bot$) in $S$ with the new $u$
	such that $w(u) > w_S(u')$. Then,
	for any $n \geq 0$, $f(S_{n+1}) > f(S_n)$.
\end{lemma}



\begin{proof}
	%
	We write $u = u_{n+1}$ and denote
	$S_{n+1} = S_n - u' + u$, for some $u' \in S_n$. 
	(We use the convention that a dummy element $\bot \in S_n$
	and $A(\bot) = \emptyset$.)
	
	Then, we have $f(S_{n+1}) = f(S_n - u') + f(u|S_n - u')
	\geq f(S_n - u') + f(u | A(u))$, where the inequality follows
	because $S_n - u' \subseteq A(u)$ and $f$ is submodular.
	
	We next observe that $w(u) = f(u | A(u))$.
	Moreover, by the hypothesis of the lemma,
	$w(u) > w_{n}(u') = f(u' | A(u') \cap S_n)$, because 
	the algorithm replaces $u'$ by $u$.
	Moreover, since $A(u') \cap S_n \subseteq S_n - u'$ 
	and $f$ is submodular, we have $f(u' | A(u') \cap S_n) \geq f(u'|S_n - u')$.
	
	Combining all the inequalities,
	we have $f(S_{n+1}) \geq f(S_n - u') + f(u | A(u)) >
	f(S_n - u') + f(u' | A(u') \cap S_n) \geq f(S_n - u') + f(u'|S_n - u') = f(S_n)$,
	as required.
\end{proof}

Using Lemma~\ref{lemma:large_replace},
we also show the following monotone property, which
is useful in proving the competitive ratio.

\begin{lemma}{\sc (Monotone Threshold)}
\label{lemma:new_A_S}
The sequence $\{\alpha \cdot w_n(S_n) - w(A_n)\}_{n \geq 0}$
is monotonically increasing.
\end{lemma}

\begin{proof}
Fix $n \geq 0$.
Observe that 
$A_{n+1} = A_{n} + u_{n+1}$, and $w(A_{n+1}) - w(A_n) = w(u_{n+1})$.
Hence, to prove
$\alpha \cdot w_{n+1}(S_{n+1}) - w(A_{n+1}) \geq \alpha \cdot w_n(S_n) - w(A_n)$,
it suffices to show that
$\alpha \cdot (w_{n+1}(S_{n+1}) - w_{n}(S_{n})) \geq w(u_{n+1})$.

\ignore{
The easy case is when
$S_{n+1} = S_{n} + u_{n+1}$.
This happens when $S_{n+1}$ and $A_{n+1}$ contain
$n+1$ dummy elements, and hence both sides of the inequality is zero.
}


We write $S_{n+1} = S_{n} - u' + u_{n+1}$,
where $u' \in S_n$.  (Again, by convention,
if $u' = \bot$ is dummy, we assume $u' \in S_n$.)

Observe that
$\alpha \cdot (w_{n+1}(S_{n+1}) - w_{n}(S_{n}))
\geq \alpha \cdot (w_{n+1}(u_{n+1}) - w_n(u'))
\geq \alpha \cdot (w(u_{n+1}) - w_n(u'))$,
where the first inequality
follows from $w_{n+1} \geq w_n$
and the second follows from $w_{n+1}(u_{n+1}) \geq w(u_{n+1})$ (both
of which follows from the submodularity of $f$).
Finally, Lemma~\ref{lemma:large_replace}
implies that $\alpha \cdot (w(u_{n+1}) - w_n(u')) \geq w(u_{n+1})$,
which completes the proof.
\end{proof}

\begin{theorem}
\label{th:uni_ratio}
Algorithm~\ref{alg:k_uni} is $\frac{1}{\alpha}$-competitive.
\end{theorem}

\begin{proof}
We suppose that the algorithm has included $n$ elements into $A$.
Then, the feasible solution at the end is $S_n$, and we have
$f(\OPT) \leq f(A_n \cup \OPT) \leq f(A_n) + \sum_{u \in \widehat{\OPT}} f(u | A_n)$,
where $\widehat{\OPT} := \OPT \setminus A_n$
are the elements in an optimal solution that
are discarded immediately in the rounds that they arrive.

For $u \in \widehat{\OPT}$,
by the submodularity of $f$,
$f(u|A_n) \leq w(u)$,
which, since $u$ is discarded in the round it arrives, is at most 
$\frac{1}{k}(\alpha \cdot w_{S(u)}(S(u)) - w(A(u)))$.
This quantity is at most 
$\frac{1}{k}(\alpha \cdot w_{n}(S_n) - w(A_n))$,
by Lemma~\ref{lemma:new_A_S}.

Since $|\widehat{\OPT}| \leq k$,
we have 
$f(\OPT) \leq f(A_n) + \alpha \cdot w_n(S_n) - w(A_n)
\leq \alpha \cdot f(S_n) - (\alpha-1) \cdot f(\varnothing)
\leq \alpha \cdot f(S_n)$,
where the second inequality follows
from Lemma~\ref{lemma:w_and_f}.
\end{proof}

\begin{corollary}
For monotone $f$ with uniform matroid,
there exists a deterministic algorithm with
competitive ratio at least $\min_k \max\{\frac{1}{k}, \frac{1}{\alpha_k}\} = \frac{1}{\alpha_4} > 0.2959 $.
\end{corollary}

\subsection{Proof of Lemma~\ref{lemma:large_replace}}
\label{sec:proof_large_replace}

Define $\beta := \alpha^{\frac{1}{k+1}} = 1 + \frac{\alpha-2}{k+1}$.
One can check that $k \beta^{k+1} - (\alpha + k - 1) \beta^k + \alpha = 0$.

For ease of notation, we assume that there are
$k$ dummy elements $\{u_1, u_2, \ldots, u_k\}$.
The function $f$
is extended naturally such that any dummy elements are ignored.
For $1 \leq i \leq k$, we use the convention that
$S_i = A_i = \{u_1, u_2, \ldots, u_i\}$.
Therefore, the real algorithm starts at $n = k+1$.

\ignore{
Observe that for $0 \leq n < k$,
we have $S_n = A_n$.
Hence, the 
acceptance rule
reduces to 
$w(u_{n+1}) > \frac{1}{k} \left( \alpha \cdot w_n(S_n) - w(S_n) \right) \geq 0$,
where the last inequality follows from $w_n(S_n) \geq w(S_n)$, which
is implied by the submodularity of $f$.  Hence,
Lemma~\ref{lemma:large_replace} holds trivially
when the element $u'$ to be replaced is dummy.
}

We prove a stronger statement that for all $n \geq k$, we have the following.

\begin{compactitem}
	\item[(A)] If $u_{n+1}$ replaces $u' \in S_n$ (which could be dummy),
	then $w(u_{n+1}) > \frac{\alpha}{\alpha - 1} \cdot {w_{n}(u')}$.
	
	\item[(B)] $w(A_{n+1}) \geq \beta \cdot w(A_n)$.
\end{compactitem}


Observe that the first $k$ dummy elements ensure that
$w(A_i)=0$ for $0 \leq i \leq k$, and hence
statement~(B) actually holds for $0 \leq n < k$.

For contradiction's sake, we consider the smallest integer $n$ (at least $k$)
for which at least one of the above statements does not hold.


We next prove the following claim.

\noindent \emph{Claim.} For all
$I \subseteq A_n$
such that $|I| \leq k$,
$w_{n}(S_n) \geq w(I)$. 

\begin{proof}
By our assumption, for all $k \leq i < n$,
if $u_{i+1}$ replaces $u_{i+1}'$, then
$w(u_{i+1}) > \frac{\alpha}{\alpha - 1} \cdot w_i(u_{i+1}')$,
where $u_{i+1}'$ is an element attaining $\min_{u \in S_i} w_i(u)$.

Observe that if an element $u$ stays in the set $S_j$ (for $j \geq i+1$),
then $w_j(u)$ does not decrease as $j$ increases (Lemma~\ref{lemma:monotone_ws}).
Moreover, observe that $\min_{v \in S_j} w_j(v)$ is non-decreasing as $j$ increases.
Hence,
for any element $u \in S_j$, we must have $w_j(u) \geq w_i(u_{i+1}') \geq w(u_{i+1}')$.

Hence, it follows that if $u \in S_n$ and $u'$ is an element
that is replaced at some point, then $w_n(u) \geq w(u')$.

Hence, if we set $P := S_n \cap I$,
we have $w_n(S_n) = w_n(P) + w_n(S_n \setminus P) 
\geq w(P) + w(I \setminus P) = w(I)$, as required.
\end{proof}


Hence, we can pick
$I = A_n \setminus A_{n-k}$,
and have
$w_{n}(S_n) \geq w(A_n) - w(A_{n-k}) \geq (1 - \frac{1}{\beta^k}) w(A_n)$,
where the last inequality holds because $w(A_{i+1}) \geq \beta \cdot w(A_i)$ holds for $i < n$.
Since the algorithm
replaces an element from $S_n$ with $u_{n+1}$,
we have $w(u_{n+1}) > \frac{1}{k}(\alpha \cdot w_n(S_n) - w(A_n))$.
Combining this with the above
lower bound for $w_n(S_n)$,
we have:
\begin{align}
	w(u_{n+1})
	&> \frac{1}{k} \cdot \{\alpha (1 - \frac{1}{\beta^k}) -1 \} \cdot w(A_n) \nonumber\\
	&= (\beta - 1) \cdot w(A_n),
	\label{eq:wun}
\end{align}

where the last equality follows from the choice of $\beta$.

We first show that statement~(B) must hold for $n$.
From
$w(A_{n+1}) - w(A_{n}) = w(u_{n+1})$ and inequality~(\ref{eq:wun}),
we have $w(A_{n+1}) \geq \beta \cdot w(A_n)$, as required.
It remains to show that statement (A) must also hold for $n$.
We prove the following lemma.

\begin{lemma}
	\label{lemma:min_fraction}
	For all $0\leq i\leq k-1$,
	$\min_{u\in S_n}{w_n(u)} \leq \frac{w_{n-i}(S_{n-i})}{k-i}$.
\end{lemma}

\begin{proof}
	The claim holds trivially for $i=0$. We next fix $i>0$.
	
	We next show that for $j < n$,
	$w_{j+1}(S_{j+1}) - w_{j}(S_{j}) \leq w_{j+1}(u_{j+1})$.
	When $S_{j+1} = S_{j} + u_{j+1}$,
	we have $w_j(S_j) = w_{j+1}(S_j)$, and so equality holds.
	Suppose $S_{j+1} = S_{j} - u' + u_{n+1}$ for some $u' \in S_j$.
	Then, from Lemma~\ref{lemma:w_and_f},
	$w_{j+1}(S_{j+1}) - w_{j}(S_{j}) = f(S_{j+1}) - f(S_{j})
	= f(S_{j} - u' + u_{j+1}) - f(S_{j})
	\leq f(S_{j} + u_{j+1}) - f(S_{j})
	= f(u_{j+1} | S_j)
	\leq f(u_{j+1} | S_j - u')
	= w_{j+1}(u_{j+1})
	$,
	where the first inequality follows
	from the monotonicity of $f$.
	
	Hence, summing the above inequality over appropriate indices,
	we have
	\begin{align*}
	\sum_{j=0}^{i-1}{(w_{n-j}(S_{n-j}) - w_{n-j-1}(S_{n-j-1}))} \leq \sum_{j=0}^{i-1}{w_{n-j}(u_{n-j})}.
	\end{align*}
	
	After rearranging, we have
	$
	w_{n-i}(S_{n-i}) \geq w_n(S_{n}) -  \sum_{j=0}^{i-1}{w_{n-j}(u_{n-j})}.
	$
	
	Define $P = \{0\leq j \leq i-1 : u_{n-j} \in S_n\}$ and
	$Q = \{0, 1,\ldots, i-1\} \setminus P$. 
	
	By Lemma~\ref{lemma:monotone_ws},
	for $j \in P$, $w_n(u_{n-j}) \geq w_{n-j}(u_{n-j})$.
	Hence, we have
	$w_{n}(S_n) - \sum_{j=0}^{i-1}{w_{n-j}(u_{n-j})}
	\geq w_n(S_n - \{u_{n-j}\}_{j\in P}) - \sum_{j\in Q}{w_{n-j}(u_{n-j})}
	$.

	Fix $j\in Q$. Observe that in some round $l$,
	$u_{n-j}$ is replaced by some element $u$,
	where $w_{n-j}(u_{n-j}) \leq w_{l-1}(u_{n-j}) \leq \min_{v\in S_l}{w_{l}(v)}$.
	The first inequality comes from Lemma~\ref{lemma:monotone_ws}.
	Moreover, the minimum weight is only increasing during the execution of our algorithm,
	because when the algorithm needs to replace an element in $S$, it will choose
	$\arg\min_{v \in S}w_S(v)$.
	Hence, it follows that 
	$w_{n-j}(u_{n-j}) \leq \min_{v\in S_n}{w_{n}(v)}$.

	
	Therefore, $w_{n-i}(S_{n-i}) 
	\geq w_{n}(S_n - \{u_{n-j}\}_{j\in P}) - \sum_{j\in Q}w_{n-j}(u_{n-j})$
	
	$ \geq (|S_n| - |P| - |Q|) \cdot \min_{u\in S_n}{w_n(u)} = (k-i) \cdot \min_{v\in S_n}{w_n(v)}$, as required.
\end{proof}

\noindent \textbf{Proving Statement (A).}
Define $\gamma := \frac{(\alpha - 2)(\alpha - 1)}{\alpha} \cdot \frac{k}{k+1}$.
Observe that $\alpha \gamma > 2 $ (see Lemma~\ref{lemma:technical}(a)). 

The easy case is when $w_n(S_n) \leq \gamma \cdot w(A_n)$.
Then, from~(\ref{eq:wun}),
we have $w(u_{n+1}) > (\beta - 1) \cdot w(A_n) \geq \frac{\beta-1}{\gamma} \cdot w(S_n) = \frac{\alpha}{\alpha - 1} \cdot \frac{w(S_n)}{k}
\geq \frac{\alpha}{\alpha - 1} \cdot \min_{v\in S_n}{w_n(v)}$,
where the last inequality comes from Lemma~\ref{lemma:min_fraction}.
Hence, we can assume $w_n(S_n) > \gamma \cdot w(A_n)$ from now on.
Recall that since $u_{n+1}$ is selected by the algorithm,
we have $w(u_{n+1}) > \frac{1}{k} \cdot (\alpha w_n(S_n) - w(A_n))
\geq \frac{1}{k} \cdot (\alpha \gamma - 1) \cdot w(A_n)$.  Hence,
we next give a lower bound on $w(A_n)$
with respect to $\mathsf{m} := \min_{u\in S_n}{w_n(u)}$.

Suppose $0< i \leq n$ is the smallest integer such that $w_{n-i}(S_{n-i}) \leq \gamma \cdot w(A_{n-i})$.  Such an integer must exist because the first dummy element implies that $w_1(S_1) = w(A_1)= 0$.
For $0 \leq j < i$,
we have $w_{n-j}(S_{n-j}) > \gamma \cdot w(A_{n-j})$.
Since the algorithm replaces an element from $S_{n-j}$ with $u_{n-j+1}$, 
it follows that
$w(A_{n-j+1}) - w(A_{n-j})
= w(u_{n-j+1})
\geq \frac{1}{k}(\alpha \cdot w_{n-j}(S_{n-j}) - w(A_{n-j}))
\geq \frac{1}{k}(\alpha \gamma - 1) \cdot w(A_{n-j})
$. 

Define $\delta := 1+\frac{1}{k}(\alpha \gamma- 1)$.
Hence, for $0\leq j < i$,
$w(A_{n-j+1}) \geq \delta \cdot w(A_{n-j})$.

Define the function $\vartheta(x) := (1-x) \delta^{kx}$ for $x \in [0,1]$,
and $\lambda := 1 - \frac{1}{k \ln \delta}$.  Observe
that $\vartheta$ is increasing on $(0, \lambda)$ and decreasing on $(\lambda, 1)$.
Hence, $\vartheta$ attains its maximum at $\lambda$.  We consider two cases.

\paragraph{Case 1.} $i \leq  \lambda k$. 
In this case,
we have 

$w(A_n) \geq \delta^{i-1} \beta \cdot w(A_{n-i}) \geq
\frac{\delta^{i-1} \beta}{\gamma} \cdot w_{n-i}(S_{n-i})
\geq \frac{\beta k}{\delta \gamma} \cdot \delta^i (1 - \frac{i}{k}) \cdot
\mathsf{m} =
\frac{\beta k}{\delta \gamma} \cdot \vartheta(\frac{i}{k}) \cdot
\mathsf{m}$,
where the last inequality follows from
Lemma~\ref{lemma:min_fraction}.

To finish with this case,
we have
\begin{align*}
w(u_{n+1})
&> \frac{1}{k} \cdot (\alpha \gamma - 1) \cdot w(A_n)
\geq \frac{1}{k} \cdot (\alpha \gamma - 1) \cdot \frac{\beta k}{\delta \gamma} \cdot \vartheta(\frac{i}{k}) \cdot
\mathsf{m} \\
&\geq \frac{\beta}{\delta \gamma} 
\cdot (\alpha \gamma -1) \cdot \vartheta(0) \cdot \mathsf{m}
\geq \frac{\alpha}{\alpha-1} \cdot \mathsf{m},
\end{align*}
where the last inequality follows from Lemma~\ref{lemma:technical}(b).

\paragraph{Case 2.} $i > \lambda k$.
In this case, set $\ell := \floor{\lambda k}$.  Then,
we have

$w(A_n) \geq \delta^{\ell} \cdot w(A_{n-\ell}) \geq
\delta^{\ell} \cdot w_{n-\ell}(S_{n-\ell})
\geq k \cdot \vartheta(\frac{1}{k} \cdot \floor{\lambda k}) \cdot 
\mathsf{m}$, where
the last inequality follows from Lemma~\ref{lemma:min_fraction},
and the penultimate inequality follows
from Lemma~\ref{lemma:w_and_f} and the monotonicity of~$f$.
Hence,
to finish with this case,
we have
\begin{align*}
w(u_{n+1})
> (\alpha \gamma -1 ) \cdot \vartheta(\frac{1}{k} \cdot \floor{\lambda k}) \cdot \mathsf{m}
\geq \frac{\alpha}{\alpha-1} \cdot \mathsf{m},
\end{align*}
where the last inequality follows from
Lemma~\ref{lemma:technical}(c).

This finishes the proof of statement (A).

\begin{lemma}{\sc (Technical Inequalities)}
	\label{lemma:technical}
	We have the following technical inequalities.
	\begin{compactitem}
		\item[(a)] $\alpha \gamma > 2$.
		\item[(b)] $\frac{\beta}{\delta \gamma} 
		\cdot (\alpha \gamma -1) \geq \frac{\alpha}{\alpha - 1}$.
		\item[(c)] $(\alpha \gamma -1 ) \cdot \vartheta(\frac{1}{k} \cdot \floor{\lambda k}) 
		\geq \frac{\alpha}{\alpha-1}$.
	\end{compactitem}
\end{lemma}

\begin{proof}
	For (a), observe that $\alpha_{\infty}>3.14$. For $k\geq 5$, $\alpha \gamma > (3.14-2)\cdot(3.14-1)\cdot \frac{5}{6}>2$. For $k=4$, $\alpha_4>3.37$, and we also have $(\alpha_4 - 2)(\alpha_4 - 1)\cdot 0.8 >2$.

	\noindent For (b), we prove the equivalent inequality $\gamma (\alpha - \frac{\delta \alpha}{\beta (\alpha-1)}) \geq 1$.
	For $k\geq 8$, we have 
	
	$\gamma(\alpha - \frac{\delta}{\beta}\frac{\alpha}{\alpha-1})
	> \gamma(3.14 - \frac{3.14}{2.14}\cdot \frac{\delta}{\beta})
	\geq \gamma (3.14 - \frac{3.14}{2.14}\cdot \frac{k+\alpha\gamma - 1}{k+\alpha-1}\cdot \frac{k+1}{k})
	\geq \gamma (3.14 - \frac{3.14}{2.14}\cdot \frac{k+1}{k})
	\geq \frac{(2.14)(1.14)}{3.14}\cdot \frac{k}{k+1}\cdot (3.14 - \frac{3.14}{2.14}\cdot \frac{k+1}{k})
	= (2.4396)\cdot\frac{k}{k+1} - 1.14
	\geq (2.4396)\cdot \frac{8}{9} - 1.14
	>1.02>1$.
	
	For $4\leq k \leq 7$, the values are at least $1.40$, $1.39$, $1.38$, $1.37$, respectively.
	
	\noindent For~(c), when $4\leq k < 1000$, we verify the inequality by plotting the function $G(k) := (\alpha\gamma - 1) \cdot \vartheta(\frac{1}{k} \cdot \floor{\lambda k})
	- \frac{\alpha}{\alpha-1}$ in Figure~\ref{fig:tec_ine_c}.
	
	Now we can assume $k\geq 1000$. We will prove that $\vartheta(\frac{\floor{k\lambda}}{k}) > \frac{\alpha}{(\alpha-1)(\alpha\gamma-1)}$.
	
	We observe that $3.14 <\alpha < 3.15$. Hence $\alpha\gamma> 1.14\cdot 2.14 \cdot \frac{1000}{1001} > 2.437$, and $\alpha \gamma < 1.15\cdot 2.15 = 2.4725$.
	Then $\delta > 1 + \frac{1.437}{k}$, and $\delta < 1 + \frac{1.4725}{k}$. Furthermore, $\lambda = 1 - \frac{1}{k\ln{\delta}} \geq 1 - \frac{1}{k\ln(1 + \frac{1.437}{k})} \geq 1 - \frac{1}{1000 \ln(1+\frac{1.437}{1000})} > 0.3036$, and $\lambda < 1 - \frac{1}{k \ln(1+\frac{1.4725}{k})} \leq 1 - \frac{1}{1.4725} < 0.321$.
	
	Therefore,
	$\vartheta(\frac{1}{k}\cdot \floor{\lambda k})
	> \vartheta(\lambda - \frac{1}{k})
	= (1+\frac{1}{k} - \lambda)\cdot \delta^{k\lambda - 1}
	\geq \frac{1-\lambda}{\delta} \cdot \delta^{k\lambda}
	\geq \frac{1-0.321}{1+\frac{1.4725}{k}}\cdot (1+\frac{1.437}{k})^{0.3036 k}
	\geq \frac{1-0.321}{1+\frac{1.4725}{1000}}\cdot (1+\frac{1.437}{1000})^{0.3036\cdot 1000}
	> 1.04
	$.
	On the other hand, $\frac{\alpha}{(\alpha-1)(\alpha\gamma-1)} < \frac{3.14}{2.14\cdot 1.437} < 1.03$. This gives the inequality.
	

	
	\begin{figure}[H]

\centering
\begin{tikzpicture}
\begin{axis}[
    xlabel={$k$},
    ylabel={$G(k)$},
    xmin=4, xmax=1000,
    ymin=0.1, ymax=0.3,
    legend pos=north west,
    ymajorgrids=true,
    grid style=dashed,
]
 
\addplot[
    color=blue,
    ]
    coordinates {
(4,0.290350)(5,0.250904)(6,0.218275)(7,0.222466)(8,0.207593)(9,0.192695)(10,0.178841)(11,0.185534)(12,0.177229)(13,0.168837)(14,0.172155)(15,0.166973)(16,0.161438)(17,0.163172)(18,0.159700)(19,0.155829)(20,0.156725)(21,0.154280)(22,0.151458)(23,0.151873)(24,0.150089)(25,0.147965)(26,0.148088)(27,0.146752)(28,0.145115)(29,0.143258)(30,0.144033)(31,0.142747)(32,0.141259)(33,0.141775)(34,0.140750)(35,0.139540)(36,0.139871)(37,0.139043)(38,0.138048)(39,0.138243)(40,0.137569)(41,0.136742)(42,0.136836)(43,0.136282)(44,0.135590)(45,0.135607)(46,0.135150)(47,0.134566)(48,0.133877)(49,0.134145)(50,0.133651)(51,0.133059)(52,0.133249)(53,0.132828)(54,0.132317)(55,0.132444)(56,0.132085)(57,0.131641)(58,0.131717)(59,0.131409)(60,0.131023)(61,0.131057)(62,0.130793)(63,0.130456)(64,0.130055)(65,0.130229)(66,0.129934)(67,0.129579)(68,0.129710)(69,0.129452)(70,0.129138)(71,0.129232)(72,0.129006)(73,0.128726)(74,0.128790)(75,0.128591)(76,0.128343)(77,0.128379)(78,0.128205)(79,0.127984)(80,0.127997)(81,0.127845)(82,0.127647)(83,0.127409)(84,0.127508)(85,0.127332)(86,0.127117)(87,0.127192)(88,0.127035)(89,0.126841)(90,0.126895)(91,0.126755)(92,0.126580)(93,0.126616)(94,0.126491)(95,0.126333)(96,0.126352)(97,0.126241)(98,0.126098)(99,0.126104)(100,0.126005)(101,0.125876)(102,0.125719)(103,0.125781)(104,0.125665)(105,0.125521)(106,0.125568)(107,0.125463)(108,0.125333)(109,0.125366)(110,0.125272)(111,0.125152)(112,0.125174)(113,0.125089)(114,0.124980)(115,0.124991)(116,0.124914)(117,0.124815)(118,0.124816)(119,0.124747)(120,0.124657)(121,0.124546)(122,0.124588)(123,0.124505)(124,0.124403)(125,0.124435)(126,0.124360)(127,0.124266)(128,0.124289)(129,0.124220)(130,0.124134)(131,0.124148)(132,0.124086)(133,0.124007)(134,0.124013)(135,0.123957)(136,0.123885)(137,0.123796)(138,0.123833)(139,0.123767)(140,0.123684)(141,0.123714)(142,0.123653)(143,0.123576)(144,0.123599)(145,0.123543)(146,0.123472)(147,0.123488)(148,0.123437)(149,0.123371)(150,0.123381)(151,0.123334)(152,0.123274)(153,0.123277)(154,0.123235)(155,0.123179)(156,0.123111)(157,0.123139)(158,0.123088)(159,0.123024)(160,0.123046)(161,0.122999)(162,0.122940)(163,0.122956)(164,0.122913)(165,0.122858)(166,0.122869)(167,0.122830)(168,0.122779)(169,0.122785)(170,0.122749)(171,0.122702)(172,0.122703)(173,0.122670)(174,0.122626)(175,0.122573)(176,0.122594)(177,0.122553)(178,0.122503)(179,0.122519)(180,0.122482)(181,0.122435)(182,0.122447)(183,0.122413)(184,0.122369)(185,0.122377)(186,0.122346)(187,0.122305)(188,0.122309)(189,0.122280)(190,0.122242)(191,0.122243)(192,0.122216)(193,0.122181)(194,0.122138)(195,0.122154)(196,0.122121)(197,0.122081)(198,0.122093)(199,0.122063)(200,0.122025)(201,0.122034)(202,0.122006)(203,0.121971)(204,0.121976)(205,0.121951)(206,0.121918)(207,0.121920)(208,0.121897)(209,0.121866)(210,0.121828)(211,0.121844)(212,0.121815)(213,0.121779)(214,0.121792)(215,0.121765)(216,0.121732)(217,0.121741)(218,0.121717)(219,0.121685)(220,0.121692)(221,0.121669)(222,0.121640)(223,0.121644)(224,0.121623)(225,0.121595)(226,0.121596)(227,0.121577)(228,0.121551)(229,0.121520)(230,0.121532)(231,0.121508)(232,0.121479)(233,0.121489)(234,0.121466)(235,0.121438)(236,0.121446)(237,0.121425)(238,0.121399)(239,0.121404)(240,0.121385)(241,0.121360)(242,0.121363)(243,0.121345)(244,0.121322)(245,0.121323)(246,0.121306)(247,0.121285)(248,0.121258)(249,0.121268)(250,0.121248)(251,0.121223)(252,0.121231)(253,0.121212)(254,0.121188)(255,0.121194)(256,0.121177)(257,0.121154)(258,0.121158)(259,0.121142)(260,0.121121)(261,0.121123)(262,0.121108)(263,0.121088)(264,0.121088)(265,0.121075)(266,0.121056)(267,0.121033)(268,0.121042)(269,0.121024)(270,0.121003)(271,0.121009)(272,0.120993)(273,0.120973)(274,0.120978)(275,0.120963)(276,0.120943)(277,0.120946)(278,0.120932)(279,0.120914)(280,0.120916)(281,0.120903)(282,0.120886)(283,0.120865)(284,0.120873)(285,0.120858)(286,0.120838)(287,0.120845)(288,0.120830)(289,0.120811)(290,0.120817)(291,0.120803)(292,0.120785)(293,0.120789)(294,0.120776)(295,0.120759)(296,0.120761)(297,0.120749)(298,0.120734)(299,0.120734)(300,0.120723)(301,0.120709)(302,0.120690)(303,0.120698)(304,0.120684)(305,0.120667)(306,0.120672)(307,0.120659)(308,0.120643)(309,0.120647)(310,0.120635)(311,0.120620)(312,0.120623)(313,0.120612)(314,0.120597)(315,0.120599)(316,0.120588)(317,0.120574)(318,0.120575)(319,0.120565)(320,0.120552)(321,0.120536)(322,0.120542)(323,0.120530)(324,0.120515)(325,0.120520)(326,0.120509)(327,0.120494)(328,0.120498)(329,0.120487)(330,0.120474)(331,0.120476)(332,0.120466)(333,0.120453)(334,0.120454)(335,0.120445)(336,0.120433)(337,0.120433)(338,0.120425)(339,0.120413)(340,0.120399)(341,0.120404)(342,0.120394)(343,0.120380)(344,0.120384)(345,0.120374)(346,0.120362)(347,0.120364)(348,0.120355)(349,0.120343)(350,0.120345)(351,0.120336)(352,0.120325)(353,0.120326)(354,0.120318)(355,0.120307)(356,0.120294)(357,0.120299)(358,0.120289)(359,0.120276)(360,0.120281)(361,0.120271)(362,0.120259)(363,0.120263)(364,0.120254)(365,0.120243)(366,0.120245)(367,0.120237)(368,0.120226)(369,0.120227)(370,0.120220)(371,0.120210)(372,0.120210)(373,0.120203)(374,0.120193)(375,0.120181)(376,0.120186)(377,0.120177)(378,0.120166)(379,0.120170)(380,0.120161)(381,0.120151)(382,0.120153)(383,0.120145)(384,0.120135)(385,0.120137)(386,0.120130)(387,0.120120)(388,0.120121)(389,0.120114)(390,0.120105)(391,0.120106)(392,0.120099)(393,0.120091)(394,0.120080)(395,0.120084)(396,0.120076)(397,0.120066)(398,0.120069)(399,0.120062)(400,0.120052)(401,0.120054)(402,0.120047)(403,0.120038)(404,0.120040)(405,0.120033)(406,0.120024)(407,0.120025)(408,0.120019)(409,0.120011)(410,0.120011)(411,0.120005)(412,0.119997)(413,0.119988)(414,0.119991)(415,0.119984)(416,0.119975)(417,0.119978)(418,0.119971)(419,0.119962)(420,0.119964)(421,0.119958)(422,0.119950)(423,0.119951)(424,0.119945)(425,0.119937)(426,0.119938)(427,0.119932)(428,0.119925)(429,0.119915)(430,0.119919)(431,0.119912)(432,0.119904)(433,0.119907)(434,0.119900)(435,0.119892)(436,0.119894)(437,0.119888)(438,0.119880)(439,0.119882)(440,0.119876)(441,0.119869)(442,0.119870)(443,0.119864)(444,0.119857)(445,0.119857)(446,0.119852)(447,0.119846)(448,0.119837)(449,0.119841)(450,0.119834)(451,0.119826)(452,0.119829)(453,0.119823)(454,0.119816)(455,0.119818)(456,0.119812)(457,0.119805)(458,0.119806)(459,0.119801)(460,0.119794)(461,0.119795)(462,0.119790)(463,0.119784)(464,0.119784)(465,0.119779)(466,0.119773)(467,0.119766)(468,0.119768)(469,0.119763)(470,0.119756)(471,0.119758)(472,0.119752)(473,0.119746)(474,0.119747)(475,0.119742)(476,0.119736)(477,0.119737)(478,0.119732)(479,0.119726)(480,0.119726)(481,0.119722)(482,0.119716)(483,0.119716)(484,0.119712)(485,0.119706)(486,0.119699)(487,0.119702)(488,0.119697)(489,0.119690)(490,0.119692)(491,0.119687)(492,0.119681)(493,0.119682)(494,0.119678)(495,0.119672)(496,0.119672)(497,0.119668)(498,0.119662)(499,0.119663)(500,0.119659)(501,0.119653)(502,0.119647)(503,0.119649)(504,0.119644)(505,0.119638)(506,0.119640)(507,0.119635)(508,0.119629)(509,0.119631)(510,0.119627)(511,0.119621)(512,0.119622)(513,0.119618)(514,0.119612)(515,0.119613)(516,0.119609)(517,0.119604)(518,0.119604)(519,0.119600)(520,0.119595)(521,0.119589)(522,0.119592)(523,0.119587)(524,0.119581)(525,0.119583)(526,0.119579)(527,0.119573)(528,0.119575)(529,0.119570)(530,0.119565)(531,0.119566)(532,0.119562)(533,0.119557)(534,0.119558)(535,0.119554)(536,0.119549)(537,0.119549)(538,0.119546)(539,0.119541)(540,0.119536)(541,0.119538)(542,0.119534)(543,0.119528)(544,0.119530)(545,0.119526)(546,0.119521)(547,0.119522)(548,0.119518)(549,0.119513)(550,0.119514)(551,0.119510)(552,0.119506)(553,0.119506)(554,0.119503)(555,0.119498)(556,0.119493)(557,0.119495)(558,0.119491)(559,0.119486)(560,0.119488)(561,0.119484)(562,0.119479)(563,0.119480)(564,0.119477)(565,0.119472)(566,0.119473)(567,0.119469)(568,0.119465)(569,0.119465)(570,0.119462)(571,0.119458)(572,0.119458)(573,0.119455)(574,0.119451)(575,0.119446)(576,0.119448)(577,0.119444)(578,0.119439)(579,0.119441)(580,0.119437)(581,0.119433)(582,0.119434)(583,0.119430)(584,0.119426)(585,0.119427)(586,0.119424)(587,0.119419)(588,0.119420)(589,0.119417)(590,0.119413)(591,0.119413)(592,0.119410)(593,0.119407)(594,0.119402)(595,0.119404)(596,0.119400)(597,0.119396)(598,0.119397)(599,0.119394)(600,0.119389)(601,0.119390)(602,0.119387)(603,0.119383)(604,0.119384)(605,0.119381)(606,0.119377)(607,0.119377)(608,0.119375)(609,0.119371)(610,0.119371)(611,0.119368)(612,0.119365)(613,0.119360)(614,0.119362)(615,0.119359)(616,0.119355)(617,0.119356)(618,0.119353)(619,0.119349)(620,0.119350)(621,0.119347)(622,0.119343)(623,0.119344)(624,0.119341)(625,0.119337)(626,0.119337)(627,0.119335)(628,0.119331)(629,0.119327)(630,0.119329)(631,0.119326)(632,0.119321)(633,0.119323)(634,0.119320)(635,0.119316)(636,0.119317)(637,0.119314)(638,0.119310)(639,0.119311)(640,0.119309)(641,0.119305)(642,0.119305)(643,0.119303)(644,0.119299)(645,0.119300)(646,0.119297)(647,0.119294)(648,0.119290)(649,0.119292)(650,0.119289)(651,0.119285)(652,0.119286)(653,0.119283)(654,0.119280)(655,0.119281)(656,0.119278)(657,0.119274)(658,0.119275)(659,0.119273)(660,0.119269)(661,0.119270)(662,0.119267)(663,0.119264)(664,0.119264)(665,0.119262)(666,0.119259)(667,0.119255)(668,0.119257)(669,0.119254)(670,0.119250)(671,0.119251)(672,0.119249)(673,0.119245)(674,0.119246)(675,0.119244)(676,0.119240)(677,0.119241)(678,0.119239)(679,0.119235)(680,0.119236)(681,0.119234)(682,0.119231)(683,0.119231)(684,0.119229)(685,0.119226)(686,0.119222)(687,0.119224)(688,0.119221)(689,0.119218)(690,0.119219)(691,0.119216)(692,0.119213)(693,0.119214)(694,0.119211)(695,0.119208)(696,0.119209)(697,0.119206)(698,0.119204)(699,0.119204)(700,0.119202)(701,0.119199)(702,0.119196)(703,0.119197)(704,0.119194)(705,0.119191)(706,0.119192)(707,0.119190)(708,0.119187)(709,0.119188)(710,0.119185)(711,0.119182)(712,0.119183)(713,0.119181)(714,0.119178)(715,0.119178)(716,0.119176)(717,0.119173)(718,0.119173)(719,0.119172)(720,0.119169)(721,0.119166)(722,0.119167)(723,0.119165)(724,0.119161)(725,0.119163)(726,0.119160)(727,0.119157)(728,0.119158)(729,0.119156)(730,0.119153)(731,0.119154)(732,0.119152)(733,0.119149)(734,0.119149)(735,0.119147)(736,0.119145)(737,0.119145)(738,0.119143)(739,0.119140)(740,0.119137)(741,0.119139)(742,0.119136)(743,0.119133)(744,0.119134)(745,0.119132)(746,0.119129)(747,0.119130)(748,0.119128)(749,0.119125)(750,0.119126)(751,0.119124)(752,0.119121)(753,0.119122)(754,0.119120)(755,0.119117)(756,0.119117)(757,0.119116)(758,0.119113)(759,0.119111)(760,0.119112)(761,0.119109)(762,0.119107)(763,0.119108)(764,0.119105)(765,0.119103)(766,0.119104)(767,0.119102)(768,0.119099)(769,0.119099)(770,0.119098)(771,0.119095)(772,0.119095)(773,0.119094)(774,0.119091)(775,0.119089)(776,0.119090)(777,0.119088)(778,0.119085)(779,0.119086)(780,0.119084)(781,0.119081)(782,0.119082)(783,0.119080)(784,0.119078)(785,0.119078)(786,0.119076)(787,0.119074)(788,0.119074)(789,0.119073)(790,0.119070)(791,0.119071)(792,0.119069)(793,0.119067)(794,0.119064)(795,0.119065)(796,0.119063)(797,0.119061)(798,0.119061)(799,0.119060)(800,0.119057)(801,0.119058)(802,0.119056)(803,0.119054)(804,0.119054)(805,0.119052)(806,0.119050)(807,0.119050)(808,0.119049)(809,0.119047)(810,0.119047)(811,0.119045)(812,0.119043)(813,0.119041)(814,0.119042)(815,0.119040)(816,0.119037)(817,0.119038)(818,0.119036)(819,0.119034)(820,0.119035)(821,0.119033)(822,0.119031)(823,0.119031)(824,0.119029)(825,0.119027)(826,0.119028)(827,0.119026)(828,0.119024)(829,0.119024)(830,0.119023)(831,0.119021)(832,0.119018)(833,0.119019)(834,0.119017)(835,0.119015)(836,0.119016)(837,0.119014)(838,0.119012)(839,0.119013)(840,0.119011)(841,0.119009)(842,0.119009)(843,0.119008)(844,0.119006)(845,0.119006)(846,0.119004)(847,0.119002)(848,0.119000)(849,0.119001)(850,0.118999)(851,0.118997)(852,0.118998)(853,0.118996)(854,0.118994)(855,0.118995)(856,0.118993)(857,0.118991)(858,0.118991)(859,0.118990)(860,0.118988)(861,0.118988)(862,0.118987)(863,0.118985)(864,0.118985)(865,0.118984)(866,0.118982)(867,0.118980)(868,0.118980)(869,0.118979)(870,0.118977)(871,0.118977)(872,0.118976)(873,0.118974)(874,0.118974)(875,0.118973)(876,0.118971)(877,0.118971)(878,0.118970)(879,0.118968)(880,0.118968)(881,0.118967)(882,0.118965)(883,0.118965)(884,0.118964)(885,0.118962)(886,0.118960)(887,0.118961)(888,0.118959)(889,0.118957)(890,0.118958)(891,0.118956)(892,0.118954)(893,0.118955)(894,0.118953)(895,0.118951)(896,0.118952)(897,0.118950)(898,0.118949)(899,0.118949)(900,0.118948)(901,0.118946)(902,0.118946)(903,0.118945)(904,0.118943)(905,0.118941)(906,0.118942)(907,0.118940)(908,0.118938)(909,0.118939)(910,0.118938)(911,0.118936)(912,0.118936)(913,0.118935)(914,0.118933)(915,0.118933)(916,0.118932)(917,0.118930)(918,0.118930)(919,0.118929)(920,0.118928)(921,0.118926)(922,0.118926)(923,0.118925)(924,0.118923)(925,0.118924)(926,0.118922)(927,0.118920)(928,0.118921)(929,0.118920)(930,0.118918)(931,0.118918)(932,0.118917)(933,0.118915)(934,0.118915)(935,0.118914)(936,0.118913)(937,0.118913)(938,0.118912)(939,0.118910)(940,0.118908)(941,0.118909)(942,0.118907)(943,0.118906)(944,0.118906)(945,0.118905)(946,0.118903)(947,0.118904)(948,0.118902)(949,0.118901)(950,0.118901)(951,0.118900)(952,0.118898)(953,0.118898)(954,0.118897)(955,0.118896)(956,0.118896)(957,0.118895)(958,0.118893)(959,0.118891)(960,0.118892)(961,0.118891)(962,0.118889)(963,0.118890)(964,0.118888)(965,0.118887)(966,0.118887)(967,0.118886)(968,0.118884)(969,0.118884)(970,0.118883)(971,0.118882)(972,0.118882)(973,0.118881)(974,0.118879)(975,0.118879)(976,0.118878)(977,0.118877)(978,0.118875)(979,0.118876)(980,0.118875)(981,0.118873)(982,0.118873)(983,0.118872)(984,0.118871)(985,0.118871)(986,0.118870)(987,0.118868)(988,0.118869)(989,0.118867)(990,0.118866)(991,0.118866)(992,0.118865)(993,0.118864)(994,0.118862)(995,0.118863)(996,0.118861)(997,0.118860)(998,0.118860)(999,0.118859)(1000,0.118858)

    };
 
\end{axis}

\end{tikzpicture}
\caption{Plot of $G(k)$}
\label{fig:tec_ine_c}
\end{figure}

\end{proof}
\section{Randomized Algorithm for Partition Matroid}
\label{sec:part_cont}

We consider $\Omega := \cupdot_{l \in [L]} \Omega_l$,
which is a disjoint union of $L$ sets.  Suppose for $l \in [L]$,
capacity $k_l$ is associated with the set $\Omega_l$.
Then, the \emph{partition matroid} $(\Omega, \I)$ is defined
such that a set $S \subset \Omega$ is independent in $\I$
\emph{iff} for all $l \in [L]$, $|S \cap \Omega_l| \leq k_l$.
For an element $u \in \Omega$,
we denote $l(u) \in [L]$ such that $u \in \Omega_{l(u)}$.
We consider a monotone submodular $f: 2^\Omega \rightarrow \R_+$
objective function.

In this section, we consider randomized algorithms for the online problem.
We first define a continuous variant of the problem and describe
a corresponding online algorithm.  We observe in
Section~\ref{sec:cardinality} that the competitive ratio
for $k$-uniform matroid is $\frac{1}{\alpha_k}$,
where $\alpha_k$ approaches  the root $\alpha_\infty \approx 3.14619$ of $a = e^{a - 2}$,
as $k$ tends to infinity.  By considering the continuous variant
of the problem, we are essentially considering arbitrarily large $k$
in order to achieve ratio $\frac{1}{\alpha_\infty}$.
For simplicity, in this section, we write $\alpha := \alpha_\infty$.
Moreover, we shall
describe a rounding procedure that gives us an online randomized algorithm
for the original problem.

\noindent \textbf{Continuous Variant.}
The algorithm maintains a vector $S \in \R_+^\Omega$
such that initially $S = \vec{0}$.
A vector $S$ is \emph{feasible} (with respect to $\I$)
if for all $l \in [L]$, $\sum_{u \in \Omega_l} S_u \leq k_l$.
The interpretation is that we can take a fractional number of copies
(even larger than 1) of an item.
When an item $u$ arrives, the algorithm may increase
the coordinate $S_u$ corresponding to the item $u$
and possibly decrease the coordinates $S_v$
for other items $v$ to maintain feasibility.

The objective function $\widef: \R_+^\Omega \rightarrow \R_+$
is induced by the original function as follows.
Given $S \in \R_+^\Omega$,
denote $\rand(S) \subset \Omega$
as the random subset sampled by including each element
$u \in \Omega$ independently with probability
$1 - \exp(-S_u)$.  Then,
$\widef(S) := \expct{f(\mathsf{R}(S))}$.

\noindent \textbf{Measure Interpretation.}  We also interpret
$S$ as a subset of the product measure space $\Omega \times \R_+$
(where $\Omega$ has the cardinality measure and $\R_+$ has
the standard Lebesgue measure).  Specifically,
we identify a vector $S \in \R_+^\Omega$
with the following subset:
$\{(u,t): u \in \Omega \wedge  t \in (0,S_u]\}$. (We use half-open intervals
to make the rounding description more convenient later.)
Observe that there is a natural 1-1 correspondence
between vectors in $\R_+^\Omega$ and \emph{valid} subsets in \measpace
defined as follows.

\begin{definition}{\sc (Valid Subset)}
\label{defn:valid_subset}
A subset $B \subset \Omega \times \R_+$ is valid if
for all $u \in \Omega$, there exists $t_u \geq 0$
such that $\{u\} \times (0, t_u] \subseteq B$
and for all $t > t_u$, $(u,t) \notin B$.

We use \measpace to denote the collection of valid subsets of $\Omega \times \R_+$.
\end{definition}

Observe that valid subsets in \measpace are closed under union and intersection.
Hence,  it makes sense to consider the submodularity of the function $\widef$ 
interpreted as having domain \measpace.

\begin{lemma}{\sc (Monotonicity and Submodularity of $\widef$)}
\label{lemma:sub_widef}
Suppose $f: 2^\Omega \rightarrow \R_+$ is monotone and submodular.
Then, $\widef : \measpace \rightarrow \R_+$ is also monotone and submodular.
\end{lemma}

\begin{proof}  This can be proved by a coupling argument.
Suppose $\omega$ is sampled from $[0,1]^\Omega$ uniformly at random.
Given $P \in \measpace$, denote $\rand_\omega(P) := \{u \in \Omega: \omega_u \leq 1 - \exp(-P_u)\}$.
Hence, it follows that $\widef(P) = \Ex_\omega[f(\rand_\omega(P))]$.

Then, the results follows because of the following facts that can be verified easily
for any $P,Q \in \measpace$ and $\omega \in \Omega$.

\begin{compactitem}
\item[1.] If $P \subseteq Q$, then $\rand_\omega(P) \subseteq \rand_\omega(Q)$.
\item[2.] $\rand_\omega(P \cup Q) = \rand_\omega(P) \cup \rand_\omega(Q)$.
\item[3.] $\rand_\omega(P \cap Q) = \rand_\omega(P) \cap \rand_\omega(Q)$.
\end{compactitem}

Hence, the monotonicity and submodularity of $\widef$ follow from those of $f$ immediately.
\end{proof}

It will be clear from the context whether we use the vector or the measure
interpretation for $S$.  For instance, $S_u$ is the $u$-th coordinate
of the vector, and $(u,t) \in S$ means that $S_u \geq t$.
We use $\norm{S}$ to denote the measure of $S \in \measpace$.
For $l \in [L]$, we denote $S|_{\Omega_l} := \{(u, t) \in S : u \in \Omega_l\}$.
Then, the feasibility of $S$ can be
expressed as $\norm{S|_{\Omega_l}} \leq k_l$ for all $l \in [L]$.

\noindent \textbf{Increment.}
Given valid $B \in \measpace$,
an element $u \in \Omega$ and $t \geq 0$,
we use $B \oplus (u,[t]) := B \cup (\{u\} \times (B(u), B(u) + t])$
to denote adding extra $t$ units of element $u$ to $B$.

\noindent \textbf{Marginal Value.}
Given valid $B', B \in \measpace$,
we denote the marginal value $\widef(B'|B) := \widef(B' \cup B) - \widef(B)$.
The marginal value of an element $u \in \Omega$ with respect to $B$
is $\widef(u|B) := \lim_{t \rightarrow 0^+} \frac{\widef(B \oplus (u,[t])) - \widef(B)}{t}$.

\noindent \textbf{Auxiliary Set $A$.}
Observe that as the algorithm increases $S_u$ from 0 to some value $t$,
we can interpret this as adding $(u,\tau)$ to $S$ continuously for $\tau$ from $0$ to $t$.
Similarly, as the algorithm decreases $S_v$ from $t_2$ to $t_1$,
we can interpret this as removing $(v, \tau)$ from $S$ continuously for $\tau$
from $t_2$ to $t_1$.  While the algorithm modifies $S$, we use an auxiliary set $A$
to keep track of all pairs $(u,t)$ that have ever been added to $S$, but could have
already been removed at some point.

\noindent \textbf{Value Function $w$.}
Suppose in the round that $u$ arrives,
the algorithm has so far increased $S_u$ to some value $t \geq 0$.
In order to decide whether to further increase $S_u$,
we denote $z = (u,t)$ and use a value function $w(z) := \widef(u|A(z))$,
where $A(z)$ is the set of pairs that have ever been added to $S$ by the algorithm up to this moment.
Observe that $w$ is dependent on the behavior of the algorithm,
and can be interpreted as a function $w: A \rightarrow \R_+$,
where $A$ is the auxiliary set.
Hence, for any subset $B \subseteq A$,
we denote $w(B) := \int_{B} w(z) dz$ as the Lebesgue integral.

\begin{lemma}{\sc (Relating $w$ and $\widef$)}
	\label{lemma:cont_w_and_f}
	Suppose at some instant, $S$ is the feasible set maintained
	by the (continuous) algorithm, and $A$ is the auxiliary set
	defined above in the same instant.
	Then, the following holds.
	\begin{compactitem}
		\item[(a)] $w(S) \leq \widehat{f}(S) - \widehat{f}(\varnothing)$.
		\item[(b)] $w(A) = \widehat{f}(A) - \widehat{f}(\varnothing)$.
	\end{compactitem}
\end{lemma}

\begin{proof}
We treat the measure $\norm{A}$ as a way
to keep track of time $\tau$.

For statement (a),
for $z =(u,t) \in S$, $w(z) = \widef(u | A(z)) \leq \widef(u |A(z) \cap S)$,
where the last inequality follows from the
submodularity of $\widef$.  Hence, integrating over $z \in S$,
we have $w(S) \leq \widef(S) - \widef(\emptyset)$.

For statement (b),
for $z = (u,t) \in A$, $w(z) = \widef(u|A(z))$.
Hence, integrating over $z \in A$ gives
$w(A) = \widef(A) - \widef(\emptyset)$.
\end{proof}
 
\begin{definition}{\sc (Knapsack for Rounding)}
\label{defn:knapsack}
To facilitate the description of the rounding procedure,
we can view the algorithm as storing the pairs in $S$ in
a knapsack $\B := \cup_{l \in [L]} \{l\} \times (0, k_l]$,
where each interval $(0, k_l]$ is also equipped with the Lebesgue measure.
Specifically, when pairs $(u,t)$ are added to $S$ continuously (and other pairs
possibly removed), we associate $(u,t)$ with a point $\varphi(u,t) \in \B$ such that
the following conditions hold.

\begin{compactitem}
\item[1.] Element $u \in \Omega_{l(u)}$ is put in the correct part, i.e., $\varphi(u,t) = (l(u), s)$, for some $s \in (0, k_{l(u)}]$.
\item[2.] At any moment, $\varphi|_S: S \rightarrow \B$ is injective. (Half-intervals are used
to satisfy this property.)
\item[3.] For any (measurable) subset $B \subseteq S$, $\norm{\varphi(B)} = \norm{B}$.
\end{compactitem}
\end{definition}

We remark that there is a natural way to replace pairs in $S$ and assign values to $\varphi$
such that the above conditions hold.  Hence, in the description of the algorithm,
we do not explicitly mention~$\varphi$.

\begin{definition}{\sc (Online Rounding Procedure)}
\label{defn:rounding}
Before any item arrives, a random subset $\mathcal{Z} \in \B$ is sampled in the
knapsack as follows.
For each $l \in [L]$, $k_l$~points are sampled uniformly at random independently
from $\{l\} \times (0, k_l]$ and included in $\mathcal{Z}$.

At any moment when the (continuous) algorithm is maintaining $S \in \measpace$,
we can imagine that the randomized algorithm (which must maintain feasibility
in the original partition matroid $(\Omega, \I)$) is keeping 
$\widetilde{S} := \{u \in \Omega: \exists t, (u,t) \in S \wedge \varphi(u,t) \in \mathcal{Z}\}$.
\end{definition}

\begin{theorem}{\sc (Randomized Online Algorithm for Partition Matroid)}
\label{th:rand_alg}
The rounding procedure in Definition~\ref{defn:rounding}
can be applied to Algorithm~\ref{alg:cont} to produce
a randomized algorithm with competitive ratio $\frac{1}{\alpha_\infty}$.
\end{theorem}

\subsection{Continuous Online Algorithm for Partition Matroid}

\noindent \textbf{Algorithm Model.}  
Without loss of generality, we assume that the algorithm knows
the capacity $k_l$ for each part $\Omega_l$,
and when an element $u$ arrives, it also knows to which part $l(u)$
the element belongs.  This is because the algorithm can keep on accepting elements
until a conflict is detected, at which point it can tell which elements are in the full part and its capacity.
We assume that oracle accesses to
the objective function $\widef(\cdot)$ and its marginals
$\widef(\cdot | \cdot)$ (which involves first derivatives).
For ease of exposition, we do not discuss how these quantities
can be approximated by sampling the original function $f$.  Moreover,
we assume that the algorithm can monitor and change variables continuously.

\begin{algorithm2e}
	\caption{Continuous Online Algorithm for Partition Matroids}
	\label{alg:cont} 
	Initialize $S$ and $A$ to empty. \\
	\For{each round when $u$ in $\Omega_l$ arrives}{
		{
		
		\tcc{We try to increase $S_u$ by including pairs $(u,t)$ in $S$ and possibly decrease
		other $S_v$ by removing some pairs from $S$.
		We use the parameter $t = S_u$ to keep track of how much we increase
		$S_u$.}
		
		At the beginning of the round, $t = 0$.

		\While {$w(u,S_u) > \frac{1}{k_l} \left( \alpha \cdot w(S|_{\Omega_l}) - w(A|_{\Omega_l}) \right)$} { 
		
		Increase $S_u=t$ and $A_u=t$ (by including $(u,t)$) continuously.
		
		\If{$\norm{S|_{\Omega_l}} = k_l$}{
		
			\tcc{We need to remove some pairs from $S$ to maintain feasibility.}
			
			Define $T := \{ v \in \Omega_l: (v, S_v) \in \arg \min_{z \in S|_{\Omega_l}} w(z)\}$. \tcc{$T$
			can change as $S$ changes;
			observe that $v \in T$ implies that $S_v > 0$.}
			
			For $v \in T$, we simultaneously decrease $S_v$(i.e., removing $(v, S_v)$ from $S$) at an appropriate rate such that
			as $S$ changes, no element leaves $T$ unless some $v \in T$ has $S_v$ dropping to 0;
			on the other hand, it is possible that some new element can join $T$ as $S$ changes.
						
			More precisely, define $w_v(\tau) := w(v, \tau)$,
			whose value was defined earlier at the moment when $(v,\tau)$ was included in $S$ and $A$.
			Observe that $w_v(\tau)$
			is a decreasing function of $\tau$. 
			
			Let $\eta_v := |w_v'(S_v)|$.
			Then, for each $v \in T$, $S_v$ is decreased at rate (with respect to $t$)
			given by $\frac{d S_v}{d t} = - \frac{\eta_v^{-1}}{\sum_{s \in T} \eta_s^{-1}}$.

		}

		}
	}
	}
\end{algorithm2e}

\begin{lemma}{\sc (Continuous Replacement)}
\label{lemma:cont_replace}
Suppose during the round that $u \in \Omega_l$ arrives,
$S_u$ is currently being increased,
i.e., $w(u,S_u) > \frac{1}{k_l} \left( \alpha \cdot w(S|_{\Omega_l}) - w(A|_{\Omega_l}) \right)$.
Moreover, suppose at this moment $\norm{S|_{\Omega_l}} = k_l$.  Then, we have
\begin{equation*}
w(u,S_u) > \frac{\alpha}{\alpha-1} \min_{z \in S|_{\Omega_l}} w(z).
\end{equation*}
\end{lemma}

\begin{lemma}{\sc (Monotone Threshold)}
	\label{lemma:cont_A_S}
	For each $l \in [L]$, the quantity
	$(\alpha \cdot w(S|_{\Omega_l}) - w(A|_{\Omega_l}))$
	is monotonically increasing during the execution of algorithm.
\end{lemma}

\begin{proof}
Fix $l \in [L]$.
We use the parameter $\tau := \norm{A|_{\Omega_l}}$ to keep track of time.
(Observe that $\tau$ does not change if elements from other  $\Omega_{l'}$'s
are considered.)
Define $G(\tau)$ as the quantity $\alpha \cdot w(S|_{\Omega_l}) - w(A|_{\Omega_l})$
at the instant when $\norm{A|_{\Omega_l}} = \tau$.
Suppose at the instant $\tau$,
for some $u \in \Omega_l$,
a pair $z=(u,t)$ is being included into $S$, i.e.,
$S_u$ is increasing and $\tau$ is moving forward.
Hence, $w(A|_{\Omega_l})$ is increasing at rate $w(z)$.

If at this moment $\norm{S|_{\Omega_l}} < k_l$,
then no pair is being removed from $S$,
and we have $G'(\tau) = (\alpha - 1) \cdot w(z) \geq 0$.

Otherwise, pairs with value $\mathsf{m} := \min_{z' \in S|_{\Omega_l}} w(z')$
are being removed from $S|_{\Omega_l}$.
Hence,
$G'(\tau) = \alpha( w(z) - \mathsf{m}) - w(z)
\geq \alpha (w(z) - \frac{\alpha -1}{\alpha} \cdot w(z)) - w(z)
= 0$, where the inequality follows from Lemma~\ref{lemma:cont_replace}.
\end{proof}

The next lemma compares the objective value of
a subset $O \subset \Omega$ with that of a valid set $A \in \measpace$.

\begin{lemma}
\label{lemma:f_and_hatf}
Suppose $f$ is monotone and submodular. Then, for any finite $O \subset \Omega$  and  valid $A \in \measpace$,
we have
\[
f(O) \leq \widehat{f}(A) + \sum_{v \in O}\widehat{f}(v | A).
\]
\end{lemma}

\begin{proof}
We prove by induction on the cardinality of $O$. The statement holds trivially when $|O| = 0$, because $f$ is monotone.

Fix $u \in O$, and let $O' = O - u$.
We assume the statement holds for $O'$. 
Define $g: 2^\Omega \to \mathbb{R}_+$ as $g(X) := f(X+u)$,
which is also monotone and submodular.
For $v \in O$,
define $A^{v} := \{(x,t) \in A: x = v\}$
and $A^{-v} := \{(x,t) \in A: x \neq v\}$.
Moreover, we have
\begin{align}
\widehat{f}(A)
&= \widehat{f}(A^{v} \cup A^{-v}) \nonumber\\
& = (1 - e^{-A_v}) \expct{f(\mathsf{R}(A^{-v}) + v)}+ e^{-A_v} \expct{f(\mathsf{R}(A^{-v}))} \nonumber \\
& = (1 - e^{-A_v}) \expct{f(v|\mathsf{R}(A^{-v}))} + \expct{f(\mathsf{R}(A^{-v}))},
\label{eq:A_func}
\end{align}
Hence, we can interpret $\widef(A)$
as a function of $A_v$:
\begin{align*}
\tau \mapsto (1 - e^{-\tau}) \expct{f(v|\mathsf{R}(A^{-v}))} + \expct{f(\mathsf{R}(A^{-v}))}.
\end{align*}

Differentiating this function with respect to $\tau$,
we have the following claim.

\noindent \emph{Claim.}
For any $v \in \Omega$,
$\widehat{f}(v|A) = e^{-A_v} \expct{f(v|\rand(A^{-v}))}$.

In particular, for $v \neq u$,
we have
\begin{align}
\label{eq:widehatg}
\widehat{f}(v|A)
&= e^{-A_v} \expct{f(v|\rand(A^{-v}))} \nonumber\\
&\geq e^{-A_v} \expct{f(v|\rand(A^{-v})+ u)}
= \widehat{g}(v|A^{-u}),
\end{align}
where the inequality follows from the submodularity of $f$.
%
Hence, using (\ref{eq:A_func}) for the first equality below, we have
\begin{align*}
\widehat{f}(A) + &\sum_{v \in O}\widehat{f}(v|A)
 =
(1 - e^{-A_u}) \expct{f(u|\mathsf{R}(A^{-u}))} \\
&\quad\quad + \expct{f(\mathsf{R}(A^{-u}))} 
+ \widehat{f}(u|A) + \sum_{v \in O'}\widehat{f}(v|A) \\
& = \expct{f(u|\mathsf{R}(A^{-u}))} + \expct{f(\mathsf{R}(A^{-u}))} \\
&\quad \quad + \sum_{v \in O'}\widehat{f}(v|A)\\
& = \expct{f(\mathsf{R}(A^{-u}) + u)} + \sum_{v \in O'}\widehat{f}(v|A) \\
& = \widehat{g}(A^{-u})  + \sum_{v \in O'}\widehat{f}(v|A) \\
& \geq \widehat{g}(A^{-u}) + \sum_{v \in O'}\widehat{g}(v|A^{-u}) \\
& \geq g(O') = f(O),
\end{align*}
where the first inequality
follows from (\ref{eq:widehatg}) and
the last inequality follows from induction hypothesis.
\end{proof}
\begin{lemma}{\sc (Competitive Ratio of Continuous Algorithm)}
	\label{lemma:cont_ratio}
Suppose $\OPT \subset \Omega$
is an independent subset of items
that have ever arrived,
and $S$ is the feasible set maintained
by Algorithm~\ref{alg:cont} at the end.
Then, $f(\OPT) \leq \alpha \cdot \widef(S)$.
\end{lemma}
\begin{proof}
We use $\tau := \norm{A}$ to keep track of time.
For instance, we denote $A(\tau)$ as the auxiliary set $A$
at the instant when $\norm{A} = \tau$,
and denote $S(\tau)$ as the $S$ at the same instant.
We use $\widehat{\tau}$ to denote the instant at the end.
For $u \in \OPT$, we use $\tau_u$ to denote
the instant when the algorithm stops including pairs $(u,t)$ involving $u$.

For $u \in \OPT \cap \Omega_l$,
by the submodularity of $\widehat{f}$, 
$\widehat{f}(u|A(\widehat{\tau})) \leq \widef(u|A(\tau_u))$, which is at most
\begin{align*}
\frac{1}{k_l}(\alpha \cdot w(S(\tau_u)|_{\Omega_l}) - w(A(\tau_u)|_{\Omega_l})),
\end{align*}
because the algorithm does not accept pairs involving $u$ after time $\tau_u$.
This last quantity is at most $\frac{1}{k_l}(\alpha \cdot w(S(\widehat{\tau})|_{\Omega_l}) - w(A(\widehat{\tau})|_{\Omega_l}))$,  by Lemma~\ref{lemma:cont_A_S}.
Using Lemma~\ref{lemma:f_and_hatf}, we have
\begin{align*}
f(\OPT) \leq \widehat{f}(A(\widehat{\tau})) + \sum_{u \in \OPT}\widehat{f}(u|A(\widehat{\tau})).
\end{align*}
%

Since $|\OPT \cap \Omega_l| \leq k_l$, we have 
\begin{align*}
f(\OPT)
&\leq \widehat{f}(A(\widehat{\tau})) + \sum_{l \in [L]} \sum_{u \in \OPT \cap \Omega_l} \frac{1}{k_l}(\alpha \cdot w(S(\widehat{\tau})|_{\Omega_l}) - w(A(\widehat{\tau})|_{\Omega_l})) \\
& \leq \widehat{f}(A(\widehat{\tau})) + \sum_{l \in [L]}
(\alpha \cdot w(S(\widehat{\tau})|_{\Omega_l}) - w(A(\widehat{\tau})|_{\Omega_l}))  \\
& =  \widehat{f}(A(\widehat{\tau})) + \alpha \cdot w(S(\widehat{\tau}))- w(A(\widehat{\tau})) \\
& \leq \alpha \cdot \widef(S(\widehat{\tau})),
\end{align*}

where the last inequality comes from Lemma~\ref{lemma:cont_w_and_f}.
\end{proof}

\begin{lemma}{\sc (Rounding Preserves Ratio)}
\label{lemma:rounding}
Suppose the rounding procedure described in Definition~\ref{defn:rounding}
takes valid $S \in \measpace$ and produces $\widetilde{S} \subset \Omega$.
Then, $\expct{f(\widetilde{S})} \geq \widef(S)$.
\end{lemma}

\begin{proofof}{Theorem~\ref{th:rand_alg}}
Lemma~\ref{lemma:cont_ratio} shows that the competitive ratio
of the continuous algorithm is $\frac{1}{\alpha_\infty}$.  
Lemma~\ref{lemma:rounding} shows that the rounding procedure
can produce a randomized algorithm for the original discrete problem
with the same guarantee on the competitive ratio.
\end{proofof}

\subsection{Large Replacement: Proof of Lemma~\ref{lemma:cont_replace}}

Suppose we fix $l \in [L]$.
For ease of notation, we write $\widehat{S} := S|_{\Omega_l}$,
$\widehat{A} := A|_{\Omega_l}$ and $k := k_l$.
We use $\widehat{A}(\tau)$ to denote
the $\widehat{A}$ at the instant when it has measure $\tau$
and we use $\widehat{S}(\tau)$ to denote the corresponding $\widehat{S}$
at the same instant.
We can imagine that $\tau$ increases as
pairs pertaining to $\Omega_l$ are added to $S$ and $A$.
To simplify the argument,
we imagine that when $\tau$ is increased from 0 to $k$,
$\widehat{A}$ is filled with dummy pairs such that
$\widehat{A}(k) = \{\bot\} \times (0, k]$
for some dummy element $\bot$ that has 0 value.

We prove a stronger statement that 
for $\tau \geq k$, suppose currently
there is some $u \in \Omega_l$ such that 
$S_u$ is 
being increased. Then, the following holds.

\begin{compactitem}
\item[(A)] $w(u,S_u) > \frac{\alpha}{\alpha-1} \min_{z \in S} w(z)$.

\item[(B)] $w(u,S_u) = \frac{d w(\widehat{A}(\tau))}{d \tau} > \theta \cdot w(\widehat{A}(\tau))$,
where $\theta := \frac{\alpha - 2}{k}$.
\end{compactitem}

Observe that because of the dummy pairs,
we have $w(\widehat{A}(\tau)) = 0$ for $\tau \in [0,k]$.
Hence, statement~(B) holds with equality for $\tau \in [0,k]$.

For contradiction's sake,
we consider the infimum $\tau_0$ over $\tau \geq k$
for which at least one of the above statements does not hold.
Since all involved quantities are continuous in $\tau$,
one of the above statements does not hold for $\tau_0$.

\begin{claim}
For all $B \subseteq \widehat{A}(\tau_0)$ such that 
$\norm{B} \leq k$, $w(\widehat{S}(\tau_0)) \geq w(B)$.
\end{claim}

\begin{proof}
For $\tau < \tau_0$, statement~(A) must hold.
Hence, when $S_u$ is increased while $S_v$ is decreased,
it must be the case that $w(u,S_u) \geq \frac{\alpha}{\alpha - 1} \cdot w(v,S_v)$.
This means the pair entering $S$ has larger $w(\cdot)$ value than the pair leaving $S$.
Therefore, it must be case that $w(\widehat{S}(\tau_0))$ has the maximum $w(\cdot)$ value among all $B \subseteq \widehat{A}(\tau_0)$ having measure $\norm{B}=k$.
\end{proof}

Hence, from the claim,
we have $w(\widehat{S}(\tau_0)) \geq w(\widehat{A}(\tau_0)) - w(\widehat{A}(\tau_0 - k))$.

Since statement~(B) holds (maybe with equality) for $\tau < \tau_0$,
by integrating from $\tau = \tau_0 -k$ to $\tau_0$,
we have $w(\widehat{A}(\tau_0)) \geq e^{\theta k} \cdot w(\widehat{A}(\tau_0 - k))$.
Therefore, $w(\widehat{S}(\tau_0)) \geq (1 - e^{- \theta k}) \cdot w(\widehat{A}(\tau_0))$.

Next, when $w(\widehat{A})$ is about to increase,
we must have some $u \in \Omega_l$ being considered
such that $w(u, S_u) > \frac{1}{k} \cdot \{\alpha \cdot w(\widehat{S}(\tau_0)) - w(\widehat{A}(\tau_0))\} \geq 
\frac{1}{k} \cdot \{\alpha (1 - e^{- \theta k}) - 1\} \cdot w(\widehat{A}(\tau_0))
= \theta \cdot w(\widehat{A}(\tau_0))$,

where the last equality follows because $\theta = \frac{\alpha - 2}{k}$
and $e^{\alpha - 2} = \alpha$.  Hence, statement~(B) must hold.

Denote $\mathsf{m} := \min_{z \in \widehat{S}(\tau_0)} w(z)$.

\begin{lemma}
	\label{lemma:cont_fraction}
	For $0 \leq t \leq k \leq \tau_0$,
	$\mathsf{m} \leq \frac{w(\widehat{S}(\tau_0-t))}{k-t}$.
\end{lemma}

\begin{proof}
We denote $z_\tau$ as the pair that is being added
to $\widehat{S}(\tau)$.  Since at the same time,
some pair may possibly be removed from $\widehat{S}(\tau)$,
we have $\frac{d}{dt} w(\widehat{S}(\tau)) \leq w(z_\tau)$.
Integrating this from $\tau = \tau_0 - t$ to $\tau_0$,
we have
\begin{align}
\label{eq:wA}
w(\widehat{S}(\tau_0)) - w(\widehat{S}(\tau_0-t)) \leq 
w(\widehat{A}(\tau_0)) - w(\widehat{A}(\tau_0-t)).
\end{align}

Denote $P := (\widehat{A}(\tau_0) \setminus \widehat{A}(\tau_0-t)) \cap \widehat{S}(\tau_0)$ and $Q := (\widehat{A}(\tau_0) \setminus \widehat{A}(\tau_0-t)) \setminus P$.  In other words,
$P$ is the set of pairs that arrive between $\tau_0 - t$ and $\tau_0$
and still stay in $\widehat{S}(\tau_0)$,
and $Q$ is the set of pairs arriving within the same time frame, but have been removed
from $\widehat{S}$ before $\tau_0$.
Observe that $\norm{P \cup Q} = t$.

Since pairs with minimum $w(\cdot)$ value are removed from $\widehat{S}$,
we have for all $z \in Q$, $w(z) \leq \mathsf{m}$.
Hence, rearranging (\ref{eq:wA}),
we have

$w(\widehat{S}(\tau_0-t)) \geq w(\widehat{S}(\tau_0)) - w(P) - w(Q)
\geq (k - t) \cdot \mathsf{m}$, as required.
\end{proof}

\noindent \textbf{Proving Statement (A)}.
Define $\gamma := \frac{(\alpha -2)(\alpha - 1)}{\alpha}$.

The easy case is when $w(\widehat{S}(\tau_0)) \leq \gamma \cdot w(\widehat{A}(\tau_0))$.
Statement~(B) implies that

$w(u,S_u) > \theta \cdot w(\widehat{A}(\tau_0)) = \frac{\alpha}{\alpha - 1} \cdot \frac{w(\widehat{S}(\tau_0))}{k} \geq \frac{\alpha}{\alpha - 1} \cdot \mathsf{m}$,
where the last inequality follows from Lemma~\ref{lemma:cont_replace} (with $t=0$).

From now on, we consider $w(\widehat{S}(\tau_0)) > \gamma \cdot w(\widehat{A}(\tau_0))$.
We have

$w(u, S_u) > \frac{1}{k} \cdot \{\alpha \cdot w(\widehat{S}(\tau_0)) - w(\widehat{A}(\tau_0))\} > \frac{1}{k} \cdot (\alpha \gamma - 1) \cdot  w(\widehat{A}(\tau_0))$.

Let $0 < t \leq \tau_0$ be the smallest $t$
such that $w(\widehat{S}(\tau_0 -t)) \leq \gamma \cdot w(\widehat{A}(\tau_0 - t))$.
We know such a $t$ exists because 
$w(\widehat{S}(0)) = w(\widehat{A}(0)) = 0$.

Denoting $z_\tau$ as the pair that is being added
to $\widehat{S}(\tau)$,
we have for $\tau \in (\tau_0 - t, \tau_0]$,
\begin{align}
\label{eq:widehatA_grow}
\frac{d}{d \tau} w(\widehat{A}(\tau)) = w(z_\tau) > \frac{1}{k} \cdot (\alpha \gamma - 1)
\cdot w(\widehat{A}(\tau)).
\end{align}

Define the function $\vartheta(x) := e^{(\alpha \gamma-1)x} (1 - x)$ for $x \in [0,1]$,
and $\lambda := 1 - \frac{1}{\alpha \gamma - 1}$,
where $\lambda \in [0,1]$ because $\alpha \gamma > 2$ (Fact~\ref{fact:cont_technical}~(a)).
Observe
that $\vartheta$ is increasing on $(0, \lambda)$ and decreasing on $(\lambda, 1)$.
Hence, $\vartheta$ attains its maximum at $\lambda$
and $\vartheta(\lambda) = \frac{1}{\alpha \gamma - 1} \cdot e^{\alpha \gamma - 2}$.  We consider two cases.

\noindent \textbf{Case 1.} $t \leq \lambda k$.
After integrating (\ref{eq:widehatA_grow}) on $\tau \in (\tau_0 - t, \tau_0]$,
we have 

$w(\widehat{A}(\tau_0)) \geq \exp\{(\alpha \gamma - 1) \cdot \frac{t}{k}\} \cdot
w(\widehat{A}(\tau_0-t))$.

Applying the definition of $t$,
we have $w(\widehat{A}(\tau_0-t)) \geq \frac{1}{\gamma} \cdot w(\widehat{S}(\tau_0-t))
\geq \frac{k-t}{\gamma} \cdot \mathsf{m}$,
where the last inequality follows from Lemma~\ref{lemma:cont_replace}.

Hence, in this case, we have

$w(u,S_u) > \frac{\alpha \gamma - 1}{\gamma} \cdot \vartheta(\frac{t}{k}) \cdot \mathsf{m}
\geq \frac{\alpha \gamma - 1}{\gamma} \cdot \vartheta(0) \cdot \mathsf{m}
\geq \frac{\alpha}{\alpha - 1} \cdot \mathsf{m}$,
where the last inequality follows from Fact~\ref{fact:cont_technical}~(b).

\noindent \textbf{Case 2.} $t > \lambda k$.  After integrating (\ref{eq:widehatA_grow}) on $\tau \in (\tau_0 - \lambda k, \tau_0]$,
we have

$w(\widehat{A}(\tau_0)) \geq \exp\{(\alpha \gamma - 1) \cdot \lambda\} \cdot
w(\widehat{A}(\tau_0- \lambda k))$.

Note that $w(\widehat{A}(\tau_0- \lambda k))
\geq w(\widehat{S}(\tau_0- \lambda k)) \geq (k - \lambda k) \cdot \mathsf{m}$,
where the last inequality follows from Lemma~\ref{lemma:cont_replace}.

Hence, in this case,
we have

$w(u,S_u) > (\alpha \gamma - 1) \cdot \vartheta(\lambda) \cdot \mathsf{m}
=e^{\alpha \gamma - 2} \cdot \mathsf{m}
\geq \frac{\alpha}{\alpha - 1} \cdot \mathsf{m}$,
where the last inequality comes from Fact~\ref{fact:cont_technical}~(c).

\begin{fact}{\sc (Technical Inequalities)}
	\label{fact:cont_technical}
	The following inequalities can be verified easily, as the
	variables ($\alpha \approx 3.14619$ and $\gamma \approx 0.78188$) are absolute constants.
	\begin{compactitem}
		\item[(a)] $\alpha \gamma > 2$.
		\item[(b)] $\frac{\alpha \gamma - 1}{\gamma} 
		\geq \frac{\alpha}{\alpha - 1}$.
		\item[(c)] $e^{\alpha \gamma - 2}
		\geq \frac{\alpha}{\alpha-1}$.
	\end{compactitem}
\end{fact}

\subsection{Rounding Procedure: Proof of Lemma~\ref{lemma:rounding}}
\label{sec:rounding}

Recall that the goal is that given valid $S \in \measpace$,
we wish to show that $\widetilde{S} \subset \Omega$
produced by the rounding procedure in Definition~\ref{defn:rounding}
satisfies $\expct{f(\widetilde{S})} \geq \widef(S)$.

As we shall see later, the procedure to obtain $\widetilde{S}$
is related to sampling without replacement (in the limiting case) and
the definition of $\widef$ is related to independent sampling.

\noindent \textbf{Sampling Distributions.}
Given a finite ground set $\U$, we define the following random subsets.

\begin{compactitem}
\item[(a)] \textbf{Sampling without Replacement.}
For an integer $k > 0$, denote $\cho_k(\U)$ as the random subset
obtained by sampling a $k$-subset from $\U$ uniformly at random.
In other words, it is sampling $\U$ for $k$ times without replacement.

\item[(a)] \textbf{Independent Sampling.}
Given $p \in [0,1]$, denote $\ind_p(\U)$ as the random subset obtained
by including each element in $\U$ independently with probability $p$.
\end{compactitem}

\begin{lemma}{\sc (Sampling without Replacement vs
	Independent Sampling.)}
\label{lemma:k_cardi_vs_independent}
Suppose 
$g:2^\U \to \mathbb{R}^+$
is a submodular function.
Moreover, $|\U| = n$ and $k \in \Z_+$ such that $p = \frac{k}{n} \in [0,1]$.
Then, we have
\[
\expct{g(\cho_k(\U))} \geq \expct{g(\ind_p(\U))}.
\]
\end{lemma}

\begin{proof}
For $0 \leq i \leq n$,
define $g_i := \frac{1}{\binom{n}{i}} \sum_{T \in {\U \choose i}} g(T)$.

Observe that $g_i = \expct{g(\ind_p(\U)) \mid |\ind_p(\U)| = i}$,
since all subsets of size $i$ are equally likely in independent sampling.

\begin{lemma}
\label{lemma:prob_shift}
	$g_{i} + g_{i-2} \leq 2 g_{i-1}$, for all $ 2 \leq i \leq n$.
\end{lemma}

\begin{proof}
Define $\mathcal{N} := \{(P,Q) :P,Q \subseteq \U, |P|=|Q| = i-1, |P \cup Q| = i, |P \cap Q| = i-2\}$.

Observe that $N := \abs{\mathcal{N}} = \binom{n}{i-2}\cdot(n-i+2)\cdot (n-i+1)$.
By submodularity of $g$, we have the following

\[
\sum_{(P,Q)\in \mathcal{N}} (g(P) + g(Q)) \geq \sum_{(P,Q)\in \mathcal{N}} (g(P\cup Q) + g(P\cap Q)).
\]

Because of symmetry, subsets of $\mathcal{U}$ with the same cardinality
appear the same number of times.
Hence, the inequality above becomes

\[
2N g_{i-1} \geq N g_{i} + N g_{i-2} \iff 2 g_{i-1} \geq g_{i} + g_{i-2}.
\] 
\end{proof}

For $0 \leq i \leq n$,
define $a_i := \Pr[\abs{\cho_k(\U)} = i]$
and $b_i := \Pr[\abs{\ind_p(\U)} = i]$.
Observe that $a_k = 1$
and $b_i = {n \choose i} p^i (1-p)^{n-i}$.

Moreover, we have
$\sum_{i=0}^n a_i = 1 = \sum_{i=0}^n b_i$ and

$\sum_{i=0}^n i  \cdot a_i = \expct{\abs{\cho_k(\U)}} = k = np = 
\expct{\abs{\ind_p(\U)}} = \sum_{i=0}^n i \cdot b_i.$

\begin{fact}
For any $\{c_i\}_{i=0}^n$ such that $\sum_{i=0}^n c_i = 0$
and $\sum_{i=0}^n i \cdot c_i = 0$, and any $\{g_i\}_{i=0}^n$,
we have

$\sum_{i=0}^n c_i g_i
=
\sum_{i = 2}^{n} \sum_{j=i}^{n}(j-i+1)c_j \cdot (g_{i} + g_{i-2} - 2 g_{i-1}).$
\end{fact}

\begin{proof}
We write $g_{-1} = g_{-2}=0$
and use the backward difference operator
$\nabla g_i := g_i - g_{i-1}$.
%

Observing that $\sum_{j=0}^n c_j = \sum_{j=0} j c_j = 0$,
we can add two initial terms to the RHS
such that

\begin{align*}
\mathrm{RHS} & = \sum_{i=0}^n \sum_{j=i}^{n}(j-i+1) \cdot c_j \cdot (\nabla g_i - \nabla g_{i-1}) \\
& = \sum_{j=0}^n \sum_{i=0}^j (j-i+1) \cdot c_j \cdot (\nabla g_i - \nabla g_{i-1}),
\end{align*}

where the last inequality follows from changing the order of summation.
We next consider the coefficient of $c_j$ as follows:
\begin{align*}
\sum_{i=0}^j (j-i+1) (\nabla g_i - \nabla g_{i-1})
& = \sum_{i=0}^j \nabla g_i - (j+1) \nabla g_{-1} \\
& = g_j,
\end{align*}

where the last equality follows from a telescoping sum and $g_{-1} = \nabla g_{-1} = 0$.

Hence, we have
$\mathrm{RHS} = \sum_{j=0}^n c_j g_j = \mathrm{LHS}$, as required.
\end{proof}

Hence, using the above fact, we have
%
%
\begin{align*}
&\expct{g(\cho_k(\U))} - \expct{g(\ind_p(\U))}
 = \sum_{i=0}^{n}(a_i-b_i)g_i \\
&\quad = \sum_{i = 2}^{n} \sum_{j=i}^{n}(j-i+1)(a_j - b_j) \cdot (g_{i} + g_{i-2} - 2 g_{i-1}).
\end{align*}

Using Lemma~\ref{lemma:prob_shift}, it suffices to show that $e_i := \sum_{j=i}^{n}(j-i+1)(a_j - b_j) \leq 0$ holds for all $i \geq 2$.

Observe that
\[
e_i = \sum_{j=i}^{n} \Pr[|\cho_k(\U)| \geq j] - \sum_{j=i}^{n} \Pr[|\ind_p(\U)| \geq j].
\]

For $i > k$, the first term in the above expression is 0.  Hence, $e_i \leq 0$.

For $1 \leq i \leq k$,
consider $e_i - e_{i+1} = \Pr[|\cho_k(\U)| \geq i] - \Pr[|\ind_p(\U)| \geq i]
= 1 - \Pr[|\ind_p(\U)| \geq i] \geq 0$.

Hence, it follows that for $2 \leq i \leq k$,
$e_i \leq e_1 = \expct{|\cho_k(\U)|} -  \expct{|\ind_p(\U)|} = 0$, as required.
%
%
%
\end{proof}


\begin{lemma}{\sc (Restatement of Lemma~\ref{lemma:rounding})}
Given valid $S \in \measpace$, 
the rounding procedure in Definition~\ref{defn:rounding}
generates $\widetilde{S}$ such that
$\expct{f(\widetilde{S})} \geq \widehat{f}(S)$.
\end{lemma}

\begin{proof}
Recall that we use $\widetilde{\cdot}$ to represent the randomness
used in the rounding procedure in Definition~\ref{defn:rounding},
and we use $\rand(\cdot)$ to represent the randomness used
to define $\widef$.
Observe that for both $f(\widetilde{S})$ and $\widehat{f}(S)$,
the randomness involved for different $\Omega_l$'s are independent.
We shall use a hybrid argument.

Fix $l \in [L]$.
We condition on the randomness $\mathcal{R}_l := \bigcup_{l' \in \{1, \ldots, l-1\}} \rand(S|_{\Omega_{l'}})
\cup \bigcup_{l' \in \{l+1, \ldots, L\} } \widetilde{S|_{\Omega_{l'}}}$.
Define $f_l : 2^{\Omega_l} \to \R_+$ by
$f_l(X) := f(\mathcal{R}_l \cup X)$.
In order to apply the hybrid argument,
it suffices to prove that for each $l \in [L]$,

\begin{equation}
\label{eq:hybrid}
\expct{f_l(\widetilde{S|_{\Omega_l}})} \geq \expct{f_l(\rand(S|_{\Omega_l}))}.
\end{equation}

Observe that the expectations on both sides of the inequality (\ref{eq:hybrid})
are continuous in $S|_{\Omega_l}$.
Hence, without loss of generality, we assume that  for all $u \in \Omega_l$,
$S_u$ is rational.  This means that for arbitrarily large $n$,
we can form a partition $\U$ of $S|_{\Omega_l}$ into $n$ parts with equal measure
such that each $x \in \U$ is associated with only one item in $\Omega_l$.
We write $k := k_l$ and each $x \in \U$ has measure~$\frac{k}{n}$.

Define $g: 2^\U \to \R_+$ as
$g(P) := f_l(\{u \in \Omega_l: \exists x \in P: \textrm{$x$ is associated with $u$}\})$.
The submodularity of $g$ follows from the submodularity of $f_l$ (and $f$).

Define $\widehat{\cho}_k(\U)$ to be a random sampling of $\U$ for $k$ times independently with replacement.  Therefore, it follows that
$\expct{g(\widehat{\cho}_k(\U))} = \expct{f_l(\widetilde{S|_{\Omega_l}})}$.

Let $p := \frac{k}{n}$.  Fix some $u \in \Omega_l$ and let $r_u := \frac{S_u \cdot n}{k}$
be the number of elements in $\U$ that are associated with $u$.
It follows that the probability that at least one of these $r$ elements
appears in $\ind_{p}(\U)$ is 
\begin{align*}
1 - (1-p)^{r_u}
\geq 1 - e^{- p{r_u}}
= 1 - e^{- S_u} = \Pr[u \in \rand(S|_{\Omega_l})].
\end{align*}

By the monotonicity of $f$, it follows that
$\expct{g(\ind_{p}(\U))} \geq \expct{f_l(\rand(S|_{\Omega_l}))}$.

Let $\eta_n := \Pr[\abs{\widehat{\cho}_k(\U)} = k]$.
As $n$ tends to infinity, the probability of collision when sampling $k$ items
independently from a set of size $n$ tends to 0.  Hence,
as $n$ tends to infinity, $\eta_n$ tends to 1.

Since $g$ is non-negative, we have 
\begin{align*}
\expct{g(\widehat{\cho}_k(\U))} \geq \eta_n \cdot \expct{g({\cho}_k(\U))}
\geq \eta_n \cdot \expct{g({\ind}_p(\U))},
\end{align*}
where the last inequality follows from Lemma~\ref{lemma:k_cardi_vs_independent}.

Finally, we have
\begin{align*}
\expct{f_l(\widetilde{S|_{\Omega_l}})}
=
\expct{g(\widehat{\cho}_k(\U))} 
\geq \eta_n \cdot \expct{g({\ind}_p(\U))}
\geq \eta_n \cdot
\expct{f_l(\rand(S|_{\Omega_l}))}.
\end{align*}

Since this holds for arbitrarily large $n$, as $n$ tends to infinity, we have the required result.
\end{proof}

\section{Hardness for Uniform Matroids}
\label{sec:hardness_uni}

In this section, we give hardness results
for deterministic monotone algorithms (satisfying Definitions~\ref{defn:mono_alg})
on uniform matroid constraints.
Specifically, we show in the following
theorem that the best ratio is $\frac{1}{\alpha_\infty}$,
where $\alpha_\infty \in [3,4]$ is the root of
$\alpha = e^{\alpha-2}$.

\begin{theorem}
\label{theorem:submodular_hardness}
Suppose $\alpha \geq 1$ and $\alpha > e^{\alpha - 2}$ (i.e., $\alpha < \alpha_\infty$).
Then, there exists $k > 0$ such that with $k$-uniform matroid constraint,
no deterministic monotone algorithm can have competitive ratio~$\frac{1}{\alpha}$.
\end{theorem}

\noindent \textbf{Explanation.} Before going into the details, we give an intuition on where the $\alpha_\infty$ comes from. The key insight is that 
in our hard instance, it suffices 
to compare $f(\OPT)$ with $f(\OPT \cup A)$, if we consider strictly monotone algorithms.
We consider an instance in which each arriving item is a subset of some ``objects'', each of which has
some non-negative weight.  The objective function on a set of items is the weight
of the union of the corresponding subsets of objects.

In each phase $i$, $2k$ distinct singleton items come, each containing an object with weight $w_i = (1-\ve)^{-i}$. Note that the weight grows exponentially. If the deterministic algorithm accepts $x_i \cdot k$ of them, then the adversary gives a ``large'' item which is the union of the $x_i \cdot k $ items the algorithm chooses in this phase. However, due to the monotonicity of the algorithm, this large item cannot be included into the solution, while it may appear in the $\OPT$ and only occupy one of the $k$ quotas. Intuitively, a deterministic algorithm should exhibit convergent behavior after a large number of phases,
in the sense that $x_i$ converges to some $x$ as $i$ increases,
because the algorithm faces essentially the same scenario in every phase.

Hence, after the $n$-th round, 
\begin{align*}
f(S_n)
&= k w_n\cdot x (1 + (1-\ve) + \cdots + (1-\ve)^{\frac{1}{x}-1}) \\
&= k w_n \cdot x\frac{1-(1-\ve)^{\frac{1}{x}}}{\ve} \approx k w_n \cdot  \frac{x}{\ve}(1-e^{-\frac{\ve}{x}}).
\end{align*}

On the other hand, $f(\OPT_n)$ roughly equals $(k-n) w_n + k w_n \cdot x (1 + (1-\ve) + (1-\ve)^2 + \cdots)$, where the first term corresponds to $(k-n)$ singleton items in the last round and the second term corresponds to the ``large'' items. Note that the ``large'' items actually captures $A_n$ while each of them takes only one out of $k$ quotas. This is what we mean by comparing to $f(\OPT \cup A)$.
When $k$ is much larger than $n$, we have $f(\OPT_n) \approx k w_n (1+\frac{x}{\ve})$.

Thus, the competitive ratio is bounded by 
$$
\frac{f(S_n)}{f(\OPT_n)} = \frac{\frac{x}{\ve}(1-e^{-\frac{\ve}{x}})}{1+\frac{x}{\ve}} \leq \frac{1}{\alpha_\infty},
$$
where the inequality holds when $\frac{\ve}{x} = \alpha_\infty - 2$.

One issue in the above sketch proof is that we consider $k$ to be
much larger than $n$, which we also assume to be large.  To make the proof formal,
we choose the parameters carefully.  On a high level, assuming the existence
of a $\frac{1}{\alpha}$-competitive algorithm for all uniform matroids for some fixed $\alpha$,
the parameter $k$ is chosen to be sufficiently large, and
we only consider about $\delta k$ phases for some small enough $\delta > 0$.

In the formal proof below, we first introduce some notations and give the construction of our instance. We assume the existence of a $\frac{1}{\alpha}$-competitive strict monotone algorithm. This gives a family of constraints on the $x_i$ variables since the algorithm has to maintain $\alpha$ ratio after every round. However, we don't immediately have the property that the algorithm behaves the same in each round. Alternatively, we derive a lower bound for the $x_i$ variables (Lemma~\ref{lemma:beta_pos}) by induction and use it crucially to give a lower bound for $\alpha$.

\noindent \textbf{Parameters.} Suppose $\ve, \delta \in (0,1)$
are parameters that can vary.
For $i \geq 1$, define $w_i := (1 - \ve)^{-i}$.

\noindent \textbf{Ground Set $\Omega$ and Value Function $f$.}~
In our construction, each element in the ground set $\Omega$ is a union of a finite number of bounded intervals in $\R_+$.
We define the function $\varphi : \R_+ \rightarrow \R_+$
by $\varphi(x) := \sum_{i \geq 1}  w_i \cdot \chi_{[i-1, i)}(x)$, i.e,
if $i - 1 = \floor{x}$, then $\varphi(x) = w_i$.
Each element $A \in \Omega$ corresponds to a subset of $[0, +\infty)$.
For a finite $S \subset \Omega$,
the value function is $f(S) := 2 \int_{\cup_{A \in S} A} \varphi(x) dx$.
That is, $f(S)$ is a weighted coverage function and, thus, is submodular.


\noindent \textbf{Instance for $k$-Uniform Matroid.}~
For each $k \geq 1$, we assume that
there is an algorithm with competitive ratio $\frac{1}{\alpha}$.
The instance depends on $\delta$, $\epsilon$ and $k$.
The next arriving element can be chosen adversarially depending
on the algorithm's previous action.  Moreover,
the adversary can stop at any moment, and hence the algorithm
needs to maintain the ratio after every round.

\ignore{
\noindent \emph{Filler Elements.} As in Section~\ref{subsec:hardness_monotone_deterministic},
these elements are needed only for weakly monotone algorithms.
 The first $k$ elements have geometrically increasing weights, but
they are negligible compared to subsequent items.
Recall that we make the technical assumption that if
the algorithm wants to take an arriving item and it does not conflict
with the existing feasible set $S$, then no element is removed from $S$.
This is to force the algorithm to select all of the filler elements such that
later on when the ``real action'' happens, the feasible set is already of size $k$.
However, we use $w_0$ to denote the total weight of these filler elements.
To satisfy the monotonicity property,
the algorithm might be prevented from replacing
a filler element with an arriving item because the filler element 
has (negligible) positive weight.  Eventually, we would let $w_0$ tend to zero.
However, to avoid having too many parameters in our calculation, we just write $w_0 := 0$.
}

\noindent For $T := \floor{\delta k}$, the elements arrive in $T$ phases.
For $1 \leq i \leq T$, the following happens in phase~$i$.

\begin{compactitem}
\item[(a)] There are $2k$ elements $\{[i-1 + \frac{j-1}{2k},
i-1 + \frac{j}{2k}) : j \in [2k]\}$ arriving one by one.
Observe each is an interval in $[i-1, i)$ with measure $\frac{1}{2k}$.
Since in the construction these $2k$ elements in phase~$i$ are fixed,
we can assume that if the algorithm
selects an interval during phase $i$, then it will not discard it
before the next phase; otherwise, the algorithm needs not choose it in the first place.  Moreover, if the algorithm 
needs to remove an interval from its feasible set,
it will remove one from the earliest phase.
\item[(b)] Suppose $B_i$ is the collection of intervals selected
by the algorithm in step~(a).  If $B_i$ is non-empty,
the next arriving element is the union of the intervals in $B_i$. 
Since the algorithm is strictly monotone, it will discard this element.
\end{compactitem}

We write $x_0 := 0$. 
For $i \geq 1$,
we define $x_i := \frac{|B_i|}{k}$ to be twice the measure of the union of
intervals in $B_i$;
at the end of step~(a) of phase~$i$,
denote $S_i$ as the feasible set maintained by the algorithm,
$\OPT_i$ as the current optimal solution 
and $A_i := \cup_{j \leq i} B_i$ as the intervals that have
ever been picked by the algorithm so far.

\noindent \textbf{Defining the sequence $\{\beta_m\}$.}
We next define a sequence $\{\beta_m\}_{m \geq 0}$
by $\beta_1 := \frac{\alpha - 1}{1 - \delta}$
and $\beta_{m+1} := \frac{\alpha - 1 - \alpha(1-\varepsilon)^{1+\frac{\beta_m}{\varepsilon}}}{1 - \delta}$. 
Observe that the definition of the sequence depends only on $\alpha$, $\varepsilon$ and $\delta$,
and is independent of $k$ and the algorithm.

\begin{lemma}{\sc ($\{\beta_m\}$ is decreasing)}
\label{lemma:beta_dec}
For $m \geq 1$, $\beta_{m+1} < \beta_m$.
\end{lemma}

\begin{proof}
	We prove by induction on $m$.
	For $m=1$, 
	$\beta_2  = \frac{\alpha - 1 - \alpha(1-\varepsilon)^{1+\frac{\beta_1}{\varepsilon}}}{1 - \delta} < \frac{\alpha - 1}{1 - \delta} = \beta_1$.

	Suppose $\beta_{m} < \beta_{m-1}$ for some $m \geq 2$. We have 
	\begin{align*}
		\beta_{m+1} < \beta_m  &\Leftrightarrow \alpha - 1 - \alpha(1-\varepsilon)^{1+\frac{\beta_m}{\varepsilon}} < \alpha - 1 - \alpha(1-\varepsilon)^{1+\frac{\beta_{m-1}}{\varepsilon}} \\
		& \Leftrightarrow (1-\varepsilon)^{\frac{\beta_{m-1}}{\varepsilon}} < (1-\varepsilon)^{\frac{\beta_m}{\varepsilon}} \\
		& \Leftrightarrow \beta_m < \beta_{m-1},
	\end{align*}
	which is true by inductive hypothesis.  This completes the inductive proof.
\end{proof}

The following lemma is crucial to the hardness proof.
Even though the definition of the $\beta_m$'s
is independent of the algorithm,
we use the assumption on the algorithm's competitive ratio
to place constraints
on the $x_i$'s and infer that each $\beta_m$ is positive.

\begin{lemma}{\sc ($\{\beta_m\}$ is positive)}
	\label{lemma:beta_pos}
	For $m \geq 1$, $\beta_{m} > 0$.
\end{lemma}

\noindent \textbf{Constraints on $x_i$'s.}
Suppose $\mathbf{x} := \{x_i\}_{i=1}^T$,
and for notational convenience, we
write $x_0 = 1$ and $w_0 = 0$.
For $1 \leq n \leq T$,
define $i_n = i_n(\mathbf{x})$ to be the smallest index such that
$\sum_{i = i_n}^{n} x_i < 1$; if $x_n=1$, set $i_n = n + 1$,
and we interpret the summation $\sum_{i=n+1}^n$ as an empty sum equal to zero.
Then, the value of the feasible set at the end of phase $n$ is

$f(S_n(\mathbf{x})) = \sum_{i = i_n}^{n} x_i w_i + (1 - \sum_{i = i_n}^{n} x_i) w_{i_n-1}$.

On the other hand, another feasible solution
is to take the sets in all the step~(b)'s from phase $1$ to phase $n$,
together with $k-n$ sets in step~(a) of phase~$n$.
Hence, $f(\OPT_n) \geq \sum_{i=1}^n x_i w_i + (1 - \frac{n}{k}) \cdot w_n
\geq \sum_{i=1}^n x_i w_i + (1 - \delta) w_n$.
Since the algorithm has competitive ratio $\frac{1}{\alpha}$,
we have $\alpha \cdot f(S_n) \geq f(\OPT_n)$.  Hence, we have shown that
given any $k > 0$, for $T = \floor{\delta k}$, there exists
a sequence of numbers $\mathbf{x} = \{x_i\}_{i=1}^T$ in $[0,1]$ satisfying the following:

\begin{align}
	\label{eq:xn}
	\forall n \in [T], \alpha \cdot f(S_n(\mathbf{x}))
	= \alpha(\sum_{i = i_n}^{n} x_i w_i + (1 - \sum_{i = i_n}^{n} x_i) w_{i_n-1})
	\geq
	\sum_{i=1}^n x_i w_i + (1 - \delta) w_n.
\end{align}


The following lemma allows us to assume that
all equalities in (\ref{eq:xn}) hold. 


\begin{lemma}
	\label{lemma:achieve_eq}
	Suppose there exists a solution $\{x_i\}_{i=1}^{T}$ for (\ref{eq:xn}). Then, there exists a solution $\{x_i'\}_{i=1}^{T}$ such that all equalities hold. 
\end{lemma}

\begin{proof}
	Suppose $n$ is the smallest index
	such that the inequality in (\ref{eq:xn}) is strict.
	We will show that the $n$-th inequality can be made into equality
	by decreasing $x_n$ and perhaps increasing $x_{n+1}$.
	Since inequalities with indices smaller than $n$ do not
	involve $x_n$ or $x_{n+1}$, those equalities will be maintained.
	On the other hand, we will show that inequalities with indices larger than $n$
	will not be violated.  Hence, we can go through the inequalities
	from smaller to larger indices to transform all strict inequalities
	into equalities.
	
	Fixing the values of $x_1, x_2, \ldots, x_{n-1}$,
	we consider the difference of both sides of the $n$-th inequality as
	a function of $x_n$ given by:
	\begin{align*}
	h(x) := \alpha \cdot f(S_n(\mathbf{x}_{[n-1]}, x)) - \sum_{i=1}^{n-1} x_i - x - (1 - \delta) w_n,
	\end{align*}
	which is continuous.
	
	From our assumption, $h(x_n) > 0$;
	on the other hand, $h(0) = \alpha \cdot f(S_{n-1}(\mathbf{x}_{[n-1]}))
	- \sum_{i=1}^{n-1} x_i - (1 - \delta)w_n <
	\alpha \cdot f(S_{n-1}(\mathbf{x}_{[n-1]}))
	- \sum_{i=1}^{n-1} x_i - (1 - \delta)w_{n-1} = 0$,
	where the last equality holds from the choice of $n$.
	Therefore, $h(x) = 0$ for some $x \in (0, x_n)$;
	we let $\widehat{x}$ to be the largest number in $(0, x_n)$
	such that $h(\widehat{x}) = 0$.

	\noindent \textbf{Stage 1: $x_{n+1} < 1$.}
	We decrease $x_n$ and increase $x_{n+1}$ continuously such that
	$w_{n+1}x_{n+1} + w_n x_n$ remains constant.
	This stage ends when $x_{n+1}$ reaches 1
	or $x_n$ reaches $\widehat{x}$, whichever happens first.
	(If the latter happens first, then there is no need for Stage 2.)
	
	As remarked above, all inequalities with indices smaller than $n$
	are not affected and so they remain equalities.
	Consider the $m$-th inequality,
	where $m \geq n+1$.  Observe that
	the right hand side is $\sum_{i=1}^m x_i w_i + (1 - \delta) w_n$,
	which does not change.  Hence, it suffices
	to show that the left hand side does not decrease.
	
	Observe that $f(S_m(\mathbf{x})) = 
	\sum_{i = i_m}^{m} x_i w_i + (1 - \sum_{i = i_m}^{m} x_i) w_{i_m-1}$.
	
	We consider the following cases. We remark that
	$i_m$ could change during Stage 1.
	
	\begin{compactitem}
		\item[(a)] Case $i_m \geq n + 2$. In this case, $f(S_m(\mathbf{x}))$
		is independent of $x_n$ and $x_{n+1}$ and so does not change.
		
		\item[(b)] Case $i_m = n+1$.  In this case, $f(S_m(\mathbf{x}))$ depends only on $x_{n+1}$. As $x_{n+1}$ increases at rate 1,
		$f(S_m(\mathbf{x}))$ increases at rate $w_{n+1} - w_n$.
		However, observe that as $x_{n+1}$ increases and $x_n$ decreases,
		$i_m$ could change from $n+1$ to $n$.
		
		\item[(c)] Case $i_m \leq n$. In this case,
		the first term $\sum_{i = i_m}^{m} x_i w_i$ does not change.
		However, as $x_n$ decreases and $x_{n+1}$ increases to keep
		$w_{n+1}x_{n+1} + w_n x_n$ constant, it follows
		that $x_n + x_{n+1}$ decreases.
		Hence, the second term $(1 - \sum_{i = i_m}^{m} x_i) w_{i_m-1}$
		increases.  Observe that this could cause $i_m$ to further decrease,
		but $f(S_m(\mathbf{x}))$ never decreases.
	\end{compactitem}

	\noindent \textbf{Stage 2: $x_{n+1} = 1$.}  Suppose $x_{n+1}$ reaches 1 first
	before $x_n$ reaches $\widehat{x}$.  When this happens,
	we keep $x_{n+1}$ at 1 and only decreases $x_n$ (continuously) to $\widehat{x}$.
	Consider the $m$-th inequality where $m \geq n+1$.
	Observe that since $x_{n+1} = 1$, $i_m \geq n+2$, and hence the left hand side
	does not change.  On the other hand, as $x_n$ decreases, the right hand side
	decreases.  Therefore, the $m$-th inequality is not violated.
	
	This completes the proof of Lemma~\ref{lemma:achieve_eq}.
\end{proof}

\begin{proofof}{Lemma~\ref{lemma:beta_pos}}
	We prove the following stronger statement.
	Define 
	$N_m = 1 + \sum_{i=1}^{m-1} \ceil{\frac{\beta_i}{\varepsilon}}$.
	Suppose $k$ is sufficiently large such that $T = \floor{\delta k} > N_m$,
	and $\{x_i\}_{i=1}^T$ is a sequence in $[0,1]$ satisfying
	all equalities in (\ref{eq:xn}).
	Then, for all $N_m < n \leq T$, $\beta_m x_n \geq \ve$ (which implies
	that $\beta_m > 0$).
	We prove this by induction on $m$.

	For $m=1$ and $n > N_1 = 1$,
	from $f(S_{n}(\mathbf{x})) \leq f(S_{n-1}(\mathbf{x})) + x_{n} w_{n}$,
	we use equalities in (\ref{eq:xn}) to derive the following.
	\begin{align*}
		\sum_{i=1}^{n} x_i w_i + (1 - \delta) w_{n}
		 &= \alpha \cdot f(S_{n}(\mathbf{x}))
		 \leq \alpha \cdot f(S_{n-1}(\mathbf{x}))+ \alpha x_{n} w_{n} \\
		& = \sum_{i=1}^{n-1} x_i w_i + (1 - \delta) w_{n-1} + \alpha x_{n} w_{n},
	\end{align*}
	
	\noindent which is equivalent to
	$(\alpha - 1) x_{n} w_{n} \geq (1 - \delta) (w_{n} - w_{n-1}) = (1 - \delta) \cdot \ve w_{n}$.  Rearranging gives $\beta_1 x_{n} \geq \ve$.

	Now suppose that for some $m \geq 1$, for all $N_m < i \leq T$, $\beta_m x_i \geq \varepsilon$. 
	
	Consider  $T \geq n > N_m + \ceil{\frac{\beta_{m}}{\varepsilon}}$.
	Then, for $1 \leq j \leq \ceil{\frac{\beta_{m}}{\varepsilon}}$,
	$x_{n-j} \geq \frac{\ve}{\beta_m}$.
	Hence, it follows
	that $\sum_{j=1}^{\ceil{\frac{\beta_{m}}{\varepsilon}}}
	x_{n-j} \geq 1$.
	Therefore, $i_{n-1} \geq n - \ceil{\frac{\beta_{m}}{\varepsilon}} + 1$.
	
	Observe that in transforming
	the solution from $S_{n-1}$ to $S_n$,
	elements associated with $w_j$ is replaced by elements associated with $w_n$,
	where $j \geq n - \ceil{\frac{\beta_{m}}{\varepsilon}}$.
	Hence,
	we have $f(S_n(\mathbf{x})) \leq f(S_{n-1}(\mathbf{x}))
	+ x_n \cdot (w_n - w_{n-\ceil{\frac{\beta_{m}}{\varepsilon}}})$.
	Again, using equalities (\ref{eq:xn}),
	we have:
	\begin{align*}
		\sum_{i=1}^{n} x_i w_i + (1 - \delta) w_{n}
		 &= \alpha \cdot f(S_{n}(\mathbf{x})) 
		\leq \alpha \cdot f(S_{n-1}(\mathbf{x}))+ \alpha x_{n} \cdot (w_n - w_{n-\ceil{\frac{\beta_{m}}{\varepsilon}}}) \\
		& = \sum_{i=1}^{n-1} x_i w_i + (1 - \delta) w_{n-1} + \alpha x_{n} \cdot (w_n - w_{n-\ceil{\frac{\beta_{m}}{\varepsilon}}}).
	\end{align*}
	
	Rearranging gives
	$\beta_{m+1} x_n \geq 
	\frac{\alpha - 1 - \alpha(1-\ve)^{\ceil{\frac{\beta_{m}}{\varepsilon}}}}{1 - \delta} \cdot x_n \geq \ve$,
	completing the inductive proof.
\end{proofof}

\noindent \textbf{Completing the Proof of 
Theorem~\ref{theorem:submodular_hardness}.}
Recall that for some $\alpha \geq 1$, we assume that for any $k \geq 0$,
there is a deterministic monotone algorithm
for the $k$-uniform matroid with competitive ratio $\frac{1}{\alpha}$.  Then,
for any $\delta, \ve \in (0,1)$,
we define a sequence $\{\beta_m\}_{m \geq 1}$
(depending on only $\alpha$, $\delta$ and $\ve$).
In Lemma~\ref{lemma:beta_dec}, we
show that the sequence is decreasing.
In Lemma~\ref{lemma:beta_pos},
using the assumption on the competitive ratio of the algorithm, we show that each $\beta_m$ is positive.
Hence, by the monotone convergence theorem,
the sequence converges to some limit $\beta$,
which satisfies the following equation:
$\beta= \frac{\alpha - 1 - \alpha(1-\varepsilon)^{1+\frac{\beta}{\varepsilon}}}{1 - \delta}$.  After rearranging, we have $\alpha - 1 = g(\beta)$, where
\begin{align*}
g(t) := t(1-\delta) + \alpha(1-\ve)^{1+\frac{t}{\varepsilon}}.
\end{align*}

Then, $g'(t) = (1 - \delta) + 
\alpha \cdot \frac{\ln(1 - \ve)}{\ve} \cdot (1 - \ve)^{1+\frac{t}{\varepsilon}}$.

Writing $c(\ve) := -\frac{\ve}{\ln(1-\ve)}$,
$g$ attains its minimum when $g'(t) = 0$,
i.e., $(1 -\ve)^{1+ \frac{t}{\ve}} = \frac{c(\ve) \cdot (1 - \delta)}{\alpha}$.

Hence,
we have $\alpha - 1 = g(\beta)
\geq (1 - \delta) \cdot \big( c(\ve) \cdot \ln \frac{\alpha}{c(\ve) \cdot (1 - \delta)} + c(\ve) - \ve \big)$,
where the inequality holds for all $\ve, \delta \in (0,1)$.

Since the relevant quantities are all continuous in $\ve$ and $\delta$,
as $\ve$ and $\delta$ tend to zero,
$c(\ve)$ tends to 1, and the above inequality becomes
$\alpha - 1 \geq \ln \alpha + 1$,
which is equivalent to $e^{\alpha -2} \geq \alpha$,
as required.
\qed

\section{Hardness for Partition Matroids}
\label{sec:hardness_partition}

In this section, we give hardness results for deterministic algorithms
on partition matroids.
Specifically, the ground set $\Omega := \cup_{i \geq 0} \Omega_i$ is a union 
of disjoint sets $\Omega_i$'s such that
a (finite) set $S$ is independent \emph{iff} for all $i$, $|S \cap \Omega_i| \leq 1$.

We consider a universe $\U$ of items, each
of which has a weight given by $\nu : \U \rightarrow \R_+$.
A subset $X \subseteq \U$ has weight $\nu(X) := \sum_{x \in X} \nu(x)$.
Then, for $i \geq 0$, $\Omega_i := \{(u,i): u \in \U\}$.
For $S \subset \Omega$, we 
define $f(S) := \nu(\{x : (x,i) \in S\})$.

\subsection{Hardness for Monotone Algorithms}
\label{subsec:hardness_monotone_deterministic}

We show that for general matroids,
the competitive ratio $\frac{1}{4}$ is optimal
for monotone algorithms satisfying Definition~\ref{defn:mono_alg} (recalling that a monotone algorithm that achieves this ratio is given in Section~\ref{sec:matroid}). In particular, we show the following hardness result.

\begin{theorem}\label{thm:lowerbound_monotone}
For any $\alpha < 4$,
no monotone deterministic algorithm 
can have competitive ratio \textbf{strictly larger than} $\frac{1}{\alpha}$.
\end{theorem}

\noindent \textbf{Adversarial Model.} Given any $1 \leq \alpha < 4$,
we construct a finite sequence of elements.  For any algorithm,
an adversary can adaptively decide when to stop the arrival of items,
at which moment the competitive ratio will be at most $\frac{1}{\alpha}$.

\noindent \textbf{Instance Construction.} Given $\alpha < 4$,
we shall pick some large enough $n$ (to be decided later), and consider $\Omega := \cup_{i=0}^n \Omega_i$.

The sequence of elements come in $n$ phases.
For $1 \leq n$, in phase $i$, two elements arrive in the order: $(x_i,0)$, $(x_i, i)$, for some $x_i \in \U$.
We shall define the weights of the $a_i := \nu(x_i)$ carefully.
If the algorithm does not take the element $(x_i,0)$, then the adversary stops the sequence, and we shall see the competitive ratio will be at most $\frac{1}{\alpha}$.
However, if the algorithm takes $(x_{i},0)$, then
it cannot take the next element $(x_i,i)$ due to strict monotonicity.

\noindent \textbf{Defining $a_i = \nu(x_i)$ and the invariant.~}
We choose $a_1 = 1$.  Recall that $1 \leq \alpha < 4$. 
We show that if the algorithm has competitive ratio strictly greater than $\frac{1}{\alpha}$, then the following invariant holds:
after phase $i$,
the algorithm will have selected $(x_i,0)$.
As described above, this is true after phase 1.

After phase $i$, the value achieved by the algorithm is $a_i$ under $f$,
while the optimal solution is $\{(x_j,j): j \in [i]\}$ having
value $\sum_{j=1}^i a_j$.
We next define the weight $a_{i+1} = \nu(x_{i+1})$ such that the following holds:
\begin{align*}
\sum_{j=1}^{i+1} a_j = {\alpha} a_i .
\end{align*}

For instance, if ${\alpha}$ is just a little less than 4,
then $a_2$ is close to 3.

In phase $(i+1)$, the element $(x_{i+1}, 0)$ arrives first.
If the algorithm does not take it, then the adversary stops the sequence.
In this case, the algorithm has only selected $(x_i,0)$, whose value is $a_i = \frac{1}{{\alpha}}
\cdot \sum_{j=1}^{i+1} a_j $, which is at most $\frac{1}{\alpha}$ fraction
of the optimal value.  Hence, the algorithm must replace $(x_i,0)$
with $(x_{i+1},0)$.
As described above,
a monotone algorithm cannot take the next element $(x_{i+1}, i+1)$. 
Hence, we show that the invariant holds after phase $(i+1)$.

\noindent \textbf{Choosing $n$.}  We next show that there exists some $n$
such that after phase $n$, the competitive ratio is strictly less than $\frac{1}{\alpha}$.  Observe that
the weights $a_i$'s are determined totally by the recursion:
$a_1=1$ and $a_{i+1} = \alpha a_i - \sum_{j=1}^i a_j$.

By considering the difference
of the recursive definitions of $a_{i+2}$ and $a_{i+1}$,
we can obtain the following second order recursion:
$a_{i+2} - \alpha a_{i+1} + \alpha a_i = 0$.

Since $\Delta = \alpha^2 - 4 \alpha < 0$,
the characteristic equation has complex roots.
By Lemma~\ref{lemma:u_neg}, 
this sequence will eventually return a negative number.
We can pick $n$ to be the smallest integer such that $a_{n+1} < 0$.
Hence, after phase $n$,
the algorithm has value $a_n = \frac{1}{{\alpha}} \sum_{j=1}^{n+1} a_j < \frac{1}{{\alpha}} \sum_{j=1}^{n} a_j$,
which is $\frac{1}{\alpha}$ fraction of the
optimal value.  Hence,
to complete the hardness proof, it suffices to show the following lemma.

\begin{lemma}\label{lemma:u_neg}
Suppose $P, Q > 0$ such that $\Delta := P^2 - 4Q < 0$.
The sequence $\{a_n\}$
is defined by the recursion $a_{n+2} - P a_{n+1} + Q a_n = 0$
where both $a_1$ and $a_2$ are real and at least one is non-zero.
Then, there exists $n > 0$ such that $a_n < 0$.
\end{lemma}

\begin{proof}
Since $\Delta < 0$,
the characteristic equation $\rho^2 - P \rho + Q = 0$
has complex roots $\rho$ and $\overline{\rho}$.
Since $\Real(\rho) = \frac{P}{2} > 0$,
we can write $\rho = r e^{\im \phi}$,
where $r > 0$ and $0 < \phi < \frac{\pi}{2}$.

A standard technique to solve recurrence
relation gives that $a_n$ is a linear combination
of $\rho^n = r^n e^{\im n \phi}$ and $\overline{\rho}^n = r^n e^{-\im n \phi}$~\cite{graham1989patashnik}.
Since $a_n$ is real, it follows that there exist real numbers $A$ and $B$ such that
$a_n = r^n (A \cos n \phi + B \sin n \phi)$.
Since at least one of $a_0$ and $a_1$ is non-zero,
at least one of $A$ and $B$ is non-zero.

Finally, since $0 < \phi < \frac{\pi}{2}$,
as $n$ increases, $n \phi$ will eventually reach all
of the following 4 intervals: $(0, \frac{\pi}{2})$, $(\frac{\pi}{2}, \pi)$,
$(\pi, \frac{3\pi}{2})$, $(\frac{3\pi}{2}, 2 \pi)$.
Hence, there exists $n > 0$ such that $\cos n \phi$ has opposite sign as $A$ (if $A \neq 0$)
and $\sin n \phi$ has opposite sign as $B$ (if $B \neq 0$),
which implies that $a_n < 0$, as required.
\end{proof}

\subsection{Hardness for General Deterministic Algorithms}
\label{sec: hardness_general}

Similar to Section~\ref{subsec:hardness_monotone_deterministic}, we 
show that for algorithms that are not necessarily monotone,
the competitive ratio cannot be better than $\frac{3 - \sqrt{5}}{2} \approx 0.382$.
Specifically, we also use a partition matroid and show the following.

\begin{theorem} \label{thm:lowerbound_general}
For any $\alpha < \frac{3 + \sqrt{5}}{2}$,
no deterministic algorithm can have competitive
ratio \textbf{strictly larger than} $\frac{1}{\alpha}$.
\end{theorem}

\noindent \textbf{Adversarial Model.}  Fix $1 \leq \alpha < \frac{3 + \sqrt{5}}{2}$.
Unlike the case in Section~\ref{subsec:hardness_monotone_deterministic},
the sequence of arriving items will adapt according to the action of the algorithm.
The elements arrive in phases. For $i \geq 1$, in phase $i$, the following steps happen.

\begin{compactitem}
\item[(a)] First, $x_i^{(0)}$ and  $x_i^{(1)}$
are distinct items in $\U$ with the same
value $a_i = \nu(x_i^{(0)}) = \nu(x_i^{(1)})$ to be decided later.
The elements $(x_i^{(0)},0)$ and $(x_i^{(1)},0)$ in $\Omega_0$ arrive (one after another).
We shall show that the algorithm must select at least one of them,
say $(x_i^{(\chi_i)},0)$, for $\chi_i \in \{0,1\}$. Otherwise,
the adversary terminates the sequence.

\item[(b)] Next, there is some item $y_i \in \U$
with value $b_i = \nu(y_i)$ to be decided later.
Then, the elements $(y_i, 2i-1)$ and $(x_i^{(\chi_i)}, 2i-1)$ in $\Omega_{2i-1}$
arrive (one after another), where $(x_i^{(\chi_i)}, 0)$ is the element selected
by the algorithm in step~(a).

\item[(c)] If the algorithm selects an element $(z, 2i-1)$ in step~(b), choose $\widehat{z} := z$,
then the element $(z, 2i) \in \Omega_{2i}$ arrives; otherwise, choose $\widehat{z} := x_i^{(\chi_i)}$.
If $z$ is $x_i^{(\chi_i)}$,
then the adversary terminates the sequence.
\end{compactitem}

\noindent \textbf{Invariant.}  We show that if the
algorithm has competitive ratio \textbf{strictly larger than} $\frac{1}{\alpha}$,
then after each step in phase $i$, the following holds.

\begin{compactitem}
\item [(a)] The feasible set maintained by the algorithm contains
$\{(y_j, 2j-1): j \in [i-1]\}$ and some $(x_i^{(\chi_i)}, 0)$.
The $f$ value achieved by the algorithm is
$\sum_{j=0}^{i-1} b_j + a_i$,
while the optimal value is $\sum_{j=0}^{i-1} (a_j + b_j) + a_i$ attained
by the solution $\{ (x_j^{(\chi_j)}, 2j-1), (y_j, 2j): j \in [i-1]\} +
(x_i^{(1-\chi_i)}, 0)$.

\item[(b)] The algorithm selects $(y_i, 2i-1)$.

\item[(c)] The feasible set maintained by
the algorithm contains $\{(y_j, 2j-1): j \in [i]\}$,  $(x_i^{(\chi_i)}, 0)$
and $(y_i, 2i-1)$.
The optimal solution is 
$\{ (x_j^{(\chi_j)}, 2j-1), (y_j, 2j): j \in [i]\} +
(x_i^{(1-\chi_i)}, 0) + (y_i, 2i)$.

\end{compactitem}

\noindent \textbf{Defining $a_i$ and $b_i$ to maintain the invariant.}
We define $a_1 := 1$.  Observe that in step~(a) of phase~1,
the algorithm must pick at least 1 element.  Otherwise,
the algorithm has value 0, while the optimal value is 1.

\noindent \emph{Inductive argument.} We assume that for some 
$i \geq 1$, the invariant holds up to the moment after
step~(a) of phase~$i$, when $\{a_j\}_{j=1}^i$ and $\{b_j\}_{j=1}^{i-1}$
are already defined.  We shall show that
the invariant continues to hold after step~(a) of phase~$i+1$, and
define $b_{i}$ and $a_{i+1}$.  We define $b_i$ such that
the following holds:
\begin{align}\label{eqn:nodouble}
\alpha(a_i + \sum_{j=1}^{i-1}b_j) = \sum_{j=1}^{i} a_j + a_i + \sum_{j=1}^{i} b_j.
\end{align}

If $b_i \leq 0$, the adversary can terminate immediately,
because $\alpha(a_i + \sum_{j=1}^{i-1}b_j) \leq
\sum_{j=1}^{i} a_j + a_i + \sum_{j=1}^{i-1} b_j$,
which is the optimal value achieved at the end of step~(a) of phase $i$.
However, at this moment,
the algorithm has value
$a_i + \sum_{j=1}^{i-1}b_j$, and so the competitive ratio is at most $\frac{1}{\alpha}$.  Hence, we can assume $b_i > 0$.

Observe that the algorithm must take $(y_i, 2i-1)$ in step~(b).
Otherwise, the sequence terminates after step~(c), and
the right hand side of (\ref{eqn:nodouble}) is
the optimum value. The algorithm attains value $a_i + \sum_{j=1}^{i-1}b_j$,
and hence has competitive ratio $\frac{1}{\alpha}$.  However,
recall that we assume the algorithm has ratio strictly larger than $\frac{1}{\alpha}$.

Hence, the algorithm selects $(y_i, 2i-1)$ and it does not help
to select $(y_i, 2i)$ in step~(c).  However,
the optimal solution could include
$(x_i^{(1 - \chi_i)}, 0)$ in step~(a), $(x_i^{(\chi_i)}, 2i-1)$ in
step~(b) and $(y_i, 2i)$ in step~(c).

We next consider the beginning of step~(a) in phase $i+1$.
We define $a_{i+1}$ such that the following holds:
\begin{align}\label{eqn:must_take_new}
\alpha(a_i + \sum_{j=1}^{i} b_j) = \sum_{j=1}^{i+1} a_j + \sum_{j=1}^{i} b_j.
\end{align}

Observe that subtracting (\ref{eqn:nodouble}) from (\ref{eqn:must_take_new})
gives $a_{i+1} - a_i = \alpha b_i > 0$.
Hence, the optimal solution will replace the old element
in $\Omega_0$ arrived in phase $i$ with one of the new elements arrived in
step~(a) of phase $i+1$.

Moreover, the algorithm must select
some element $(x_{i+1}^{\chi_{i+1}}, 0)$.
Otherwise, the sequence terminates, and the
right hand side of (\ref{eqn:must_take_new})
is the optimal value,
while the algorithm achieves value
$a_i + \sum_{j=1}^{i} b_j$, which is exactly
$\frac{1}{\alpha}$ fraction of the optimal value.
Recall that the algorithm should achieve ratio strictly larger
than $\frac{1}{\alpha}$.

This completes the inductive argument and the recursive
definitions of $a_i$ and $b_i$.

\noindent \textbf{There exists negative $b_n$.}
As in Section~\ref{subsec:hardness_monotone_deterministic},
we show that there exists $n > 0$ such that $b_n < 0$.
Suppose $n$ is the smallest integer such that this happens.
Then, in the above inductive argument, it follows
that after step~(a) of phase $n$, the competitive ratio is 
strictly less than $\frac{1}{\alpha}$.

To apply Lemma~\ref{lemma:u_neg},
we shall form a second order recurrence relation for $b_n$.
If we consider (\ref{eqn:nodouble}) minus (\ref{eqn:must_take_new})($i \leftarrow i -1$),  we have:
$\alpha(a_i - a_{i-1}) = a_i + b_i$.

From the inductive argument,
we have $a_i - a_{i-1} = \alpha b_{i-1}$.  Hence,
the above equation becomes $\alpha^2 b_{i-1} = a_i + b_i$.
Replacing $i$ with $i+1$ gives $\alpha^2 b_i = a_{i+1} + b_{i+1}$.

Taking the difference between the last two equations gives:
\begin{align*}
\alpha^2 (b_i - b_{i-1})
&= (a_{i+1} - a_i) + b_{i+1} - b_i \\
&= \alpha b_i + b_{i+1} - b_i.
\end{align*}

Rearranging gives the required second order recurrence relation on $b_i$:
\begin{align*}
b_{i+1} - (\alpha^2 - \alpha + 1) b_i + \alpha^2 b_{i-1} = 0.
\end{align*}

To apply Lemma~\ref{lemma:u_neg},
observe that $P := \alpha^2 - \alpha + 1 > 0$,
$Q := \alpha^2$,
and $\Delta := (\alpha^2 - \alpha + 1)^2 - 4 \alpha^2
= (\alpha^2 + \alpha + 1) (\alpha^2 - 3 \alpha + 1)$,
which is negative for $1 \leq \alpha < \frac{3 + \sqrt{5}}{2}$,
as required.

\section{Monotone Algorithm with General Matroid Constraint}
\label{sec:matroid}

For general matroids, we give a (strictly) monotone algorithm
with competitive ratio $\frac{1}{4}$.
Observe that essentially the same algorithm
is given in~\cite{ChakrabartiK14,DBLP:conf/icalp/ChekuriGQ15}.
However, as we wish to emphasize monotonicity
and will also need to generalize the techniques
in Section~\ref{sec:multi-offline}, we give a proof here.

\begin{algorithm2e}
	\caption{Algorithm for General Matroids}
	\label{alg:submodular_matroid} 
	Initialize $S \leftarrow \varnothing$; $A \leftarrow \varnothing$; \\
	\For{each round when $u$ arrives}{
		\eIf {$u + S \in \mathcal{I}$ and $w(u) > 0$} {
		 $S \leftarrow S + u$; 
		 $A \leftarrow A + u$;
		}{
		$T  := \{v \in S: S - v + u \in \mathcal{I} \}$; \\
		$u' := \arg\min_{v \in T}w_S(v)$;
		 
		\If {$w(u) \geq 2 w_S(u')$} { 
		$S \leftarrow S - u' + u$; 
		$A \leftarrow A + u$;
		}
	}
	}
\end{algorithm2e}

\noindent \textbf{Explanation.}
Algorithm~\ref{alg:submodular_matroid} is a greedy algorithm.  In each round
when a new element $u$ arrives, $u$ is added to $S$ if
it does not violate the matroid constraint.  Otherwise,
the algorithm considers the set $T$ of elements currently in $S$ 
that could potentially be replaced by $u$.  We use some value functions carefully to decide if an element in $T$ should be replaced by $u$.

In addition to $S$, the algorithm maintains a set $A$ of elements
that have ever been added to $S$.
When an element $u$ arrives, we consider $w(u) := f(u | A(u))$.
Observe that even when $S + u \in \I$,
$u$ is added to $S$ only when $w(u) > 0$;
this ensures that at any moment, for any $v \in S$, $w_S(v) > 0$.
We keep the elements $S = \{v_1, v_2, \ldots \}$ in the order
they are added.  Then, for the element $v_i \in S$,
we consider $w_S(v_i) = f(v_i | \{v_1, v_2, \ldots, v_{i-1}\})$.
We replace the element $u'$ in $T$ of minimum value with $u$
if $w(u) \geq 2 w_S(u') > w_S(u')$,
where the last strict inequality holds because $w_S(u') > 0$.



\begin{lemma}[Values of $A$ and $S$]
\label{lemma:A_S}
Suppose $c > 1$, and
in each round the algorithm may only replace $u'$ in $S$ with the new $u$
such that $w(u) \geq c \cdot w_S(u')$. (Note that in Algorithm~\ref{alg:submodular_matroid},
$c$ is set to 2.) Then,
for each $n \geq 0$, $w(A_n) \leq \frac{c}{c-1} \cdot w(S_n)$.
\end{lemma}

\begin{proof}
Observe that $S_0 = A_0 = \varnothing$.
Hence, it suffices to show that for all $n \geq 0$,
\[
w(A_n) - w(A_{n-1}) \leq \tfrac{c}{c-1} \cdot (w(S_n) - w(S_{n-1})).
\]
For the case $S_n = S_{n-1} + u_n$, 
$w(A_n) - w(A_{n-1}) = w(u_n) = w(S_n) - w(S_{n-1})$,
and hence the inequality holds.

Otherwise, $S_n = S_{n-1} - u' + u_n$ for some $u' \in S_{n-1}$.
Since the algorithm replaces $u'$ with $u_n$,
it follows that $w(u_n) \geq c \cdot w_{n-1}(u')
= c \cdot f(u' | A(u') \cap S_{n-1})
\geq c \cdot f(u' | A(u')) = c \cdot w(u')$,
where the inequality follows from the submodularity of $f$.

Rearranging the inequality $w(u_n) \geq c \cdot w(u')$
gives $w(u_n) \leq \frac{c}{c-1} \cdot (w(u_n) - w(u'))$.
Finally, observing that $w(u_n) = w(A_n) - w(A_{n-1})$
and $w(S_n) - w(S_{n-1}) = w(u_n) - w(u')$ proves
the required inequality.
\end{proof}

\begin{lemma}[Circuit in greedy algorithm]
	\label{lemma:circuit}
	In Algorithm~\ref{alg:submodular_matroid},
	the set $T+u$ is dependent in every iteration.
\end{lemma}

\begin{proof}
	Suppose otherwise, and we apply the matroid augmentation property
	to add elements from $S$ to $T + u$ to form an independent set
	of the form $S - t' + u$ such that $t' \notin T$.  However, this
	contradicts the definition of $T$.
\end{proof}

\begin{lemma}[Matroid Properties]
\label{lemma:matroid_prop}
Suppose $(\Omega, \I)$ is a matroid.  Then, the following holds.
\begin{compactitem}
\item[(a)] Suppose $v \notin S$, and the sets $S+v$ and $T$ are in $\I$,
but $T+v \notin \I$.  Then, there exists $t \in T \setminus S$
such that $S + t \in \I$.

\item[(b)] Suppose $\mathcal{T} = \{T_1, T_2,\ldots, T_k \}$ is a family of independent sets, and $\{ v_1, v_2, \ldots, v_k \}$ is also independent.
	Suppose further that for all $i \in [k]$, $T_i + v_i \notin \mathcal{I}$.
	Then, there exist
		$k$ distinct elements forming an independent set $\{t_1,\ldots, t_k\}$
		such that for all $i \in [k]$, $t_i \in T_i$.		
\end{compactitem}
\end{lemma}

\begin{proof}
For statement (a),
we apply the matroid augmentation property 
to potentially add elements from $S+v$ to $T$
to form independent $T' \supseteq T$ such that $|T'| \geq |S + v|$.
Since $T+v \not\in \I$, it follows that $v \notin T'$.
We apply the augmentation property again to add an element from $T'$ to $S$,
and conclude that there exists $t \in T' \setminus S = T \setminus S$
such that $S + t \in \I$.

For statement (b), we apply a hybrid argument
to transform $R_0 = \{ v_1, v_2, \ldots, v_k \}$
into the desired set.
For $i \in [k]$, we shall construct
an independent set $R_i = \{t_1, \ldots, t_i, v_{i+1}, \ldots, v_k\}$
containing $k$ distinct elements such that
for all $j \in [i]$, $t_j \in T_j$.

Assuming that $R_{i-1}$ is already constructed, and we next construct $R_i$.
We apply statement~(a) to $S := R_{i-1} - v_i$, $T := T_i$ and $v := v_i$.
Then, there exists $t_i \in T_i \setminus S$
such that $R_i := S + t_i \in \I$, as required.

Hence, $R_k = \{t_1, t_2, \ldots, t_k\}$ has the required properties in statement~(b).
\end{proof}

\noindent \textbf{Auxiliary Value Function.}
For each $n \geq 1$, we define the
value function $\widehat{w}_n$ on elements in $A_n$
as follows.  If $u \in S_n$, then $\widehat{w}_n(u) = w_n(u)$;
otherwise, $u$ is removed from $S$ in some previous round,
and $\widehat{w}_n(u) = w_i(u)$, where $i$ is the largest index
such that $u \in S_i$.  From Lemma~\ref{lemma:monotone_ws},
we have $\widehat{w}_n \leq w_n$.

\begin{lemma}[Greedy is optimal with respect to $\widehat{w}$ and $A$]
\label{lemma:greed_opt}
Suppose in Algorithm~\ref{alg:submodular_matroid}, an element $u' \in S$ is replaced with a new $u$
only if $w(u) \geq w_S(u') = \min_{v \in T} w_S(v)$.  Then,
for $n \geq 1$, the set $S_n$ is an independent set in $A_n$
with maximum value under the (linear) function $\widehat{w}_n$.
In other words, $\widehat{w}_n (S_n) = \max_{I \subseteq A_n: I \in \I} \widehat{w}_n(I)$.
\end{lemma}

\begin{proof}
	The intuition is based on a greedy matroid algorithm (for non-negative linear objective function)
	that considers elements in an arbitrary order. At any moment,
	the algorithm maintains an optimal independent set $S$ among
	all processed elements so far.  When the next element $u$ is considered,
	if $S+u$ is independent, then $u$ is added to $S$.  Otherwise,
	an element with minimum value in the circuit (minimal dependent set) in $S+u$ is removed.

	We fix $n$ and write $\widehat{w} := \widehat{w}_n$,
	which is non-negative because $f$ is monotone.
	We prove a stronger statement that for all $i \in [n]$,
	$S_i$ is an optimal independent set in $A_i$ under $\widehat{w}$,
	by induction on $i$.

	For $i=1$, $S_1 = A_1 = \{u_1\}$ is optimal under value function $\widehat{w}$.
	For the induction hypothesis,
	we assume that the statement is true for some $i \geq 1$.
	Consider an independent set $I$ in $A_{i+1}$.  
	We can
	assume $u_{i+1} \in I$; otherwise,
	by the induction hypothesis,
	we immediately have $\widehat{w}(I) \leq \widehat{w}(S_i)
	\leq \widehat{w}(S_{i+1})$.  We have 2 cases to consider.

	\noindent (1) The simple case is when $S_{i+1} = S_i + u_{i+1} \in \I$.
	From the induction hypothesis, we have
	$\widehat{w}(I-u_{i+1}) \leq \widehat{w}(S_i)$, which
	implies that $\widehat{w}(I) \leq \widehat{w}(S_{i+1})$.
	
	\noindent (2) Consider the case when $S_{i+1} = S_i - u' + u_{i+1}$ for some $u' \in S_i$.
	Let $T := \{u \in S_i: S_i - u + u_{i+1} \in \I\}$.
	Since $u'$ is removed from $S_i$,
	it follows that $\widehat{w}(u') = w_i(u') \leq w_i(t)$
	for all $t \in T$.  Moreover, by Lemma~\ref{lemma:monotone_ws}, for all $t \in T$
	$w_i(t) \leq \widehat{w}(t)$.
	
	On the other hand,
	the greedy algorithm replaces $u'$ with $u_{i+1}$, and this means that
	we have $\widehat{w}(u') = w_i(u') \leq w(u_{i+1})$.
	Again, by Lemma~\ref{lemma:monotone_ws}, this implies that
	$\widehat{w}(u') \leq \widehat{w}(u_{i+1})$.
	
	Therefore, the algorithm is actually removing an element $u'$
	in $T+u_{i+1}$ with minimum value under $\widehat{w}$.  For completeness,
	we still finish the proof using a standard argument for the aforementioned
	matroid greedy algorithm.

	Recall that $u_{i+1} \in I$ and
	by Lemma~\ref{lemma:circuit},
	we have $T + u_{i+1} \notin \I$.
	Hence, 
	Lemma~\ref{lemma:matroid_prop}(a) implies that
	there exists $t \in T \setminus I$ such that $I - u_{i+1} + t$ is an independent set
	(in $A_i$).  
	Hence, by the induction hypothesis,
	it follows that $\widehat{w}(I - u_{i+1} + t) \leq \widehat{w}(S_i)$.
	Since $\widehat{w}(t) \geq \widehat{w}(u')$,
	this implies that $\widehat{w}(I) \leq \widehat{w}(S_{i+1})$,
	thereby finishing the proof.
\end{proof}

\begin{theorem}
\label{th:matroid_ratio}
Algorithm~\ref{alg:submodular_matroid} is $\frac{1}{4}$-competitive.
\end{theorem}

\begin{proof}
Suppose the algorithm has added $n$ elements to $A$ when it terminates.
Then, it returns the independent set $S_n$.  Suppose $\OPT$
is an optimal solution among the whole sequence of elements under $f$.

In Algorithm~\ref{alg:submodular_matroid},
we assume that only element $u$ with $w(u) \geq c \cdot w_S(u')$
can be picked.  We shall see that the competitive ratio is optimized
when $c = 2$.

Since $f$ is monotone and submodular,
we have 
\begin{align*}
f(\OPT) \leq f(A_n \cup \OPT) \leq f(A_n) + \sum_{u \in \OPT \setminus A_n} f(u | A_n).
\end{align*}
For the first term, we use Lemmas~\ref{lemma:w_and_f}
and~\ref{lemma:A_S} to 
conclude that
$f(A_n) = w(A_n) + f(\varnothing) \leq \frac{c}{c-1} \cdot w(S_n) + f(\varnothing) \leq \frac{c}{c-1} \cdot f(S_n) - \frac{1}{c-1} \cdot f(\varnothing) \leq \frac{c}{c-1} \cdot f(S_n)$.

We write $\widehat{\OPT} := \{u \in \OPT \setminus A_n: w(u) > 0\}$.
Observe that $f(u|A_n) \leq f(u|A(u)) = w(u)$,
and so $w(u) = 0$ implies that $f(u|A_n) \leq 0$.
For the second term, 
for $u \in \widehat{\OPT}$,
let $T(u)$ be the set $T$ defined in the algorithm
in the round that $u$ arrives (and is discarded). 
By Lemma~\ref{lemma:circuit}, we have $T(u) + u \notin \I$.
Since $\widehat{\OPT} \in \I$,
we apply
Lemma~\ref{lemma:matroid_prop}(b)
to show the existence of
independent $\{t_u : u \in \widehat{\OPT}\}$
such that $t_u \in T(u)$ for $u \in  \widehat{\OPT}$.

For $u \in \widehat{\OPT}$,
by the submodularity of $f$,
we have $f(u | A_n) \leq w(u)$,
which is at most $c \cdot w_{S(u)}(t_u)$,
because the algorithm discards $u$.
Moreover, by Lemma~\ref{lemma:monotone_ws},
we have $w_{S(u)}(t_u) \leq \widehat{w}_n(t_u)$.

Hence, it follows that
$\sum_{u \in \widehat{\OPT}} f(u | A_n) \leq 
 c \cdot \sum_{u \in \widehat{\OPT}} \widehat{w}_n(t_u)$,
which by Lemma~\ref{lemma:greed_opt},
is at most $c \cdot \widehat{w}_n(S_n) = c \cdot w_n(S_n)
\leq c \cdot f(S_n)$,
where the last inequality comes from
Lemma~\ref{lemma:w_and_f} and $f(\varnothing) \geq 0$.

Therefore, for $c=2$, we have
$f(\OPT) \leq (\frac{c}{c-1} + c) \cdot f(S_n) = 4 \cdot f(S_n)$,
as required.
\end{proof}

\section{Submodular Online Bipartite Matching with Matroid Constraints}
\label{sec:multi-offline}

The online problem we have considered so far
is actually a special case of the following
submodular online bipartite matching problem with only one offline node.

\noindent \textbf{Submodular Online Bipartite Matching Problem
with Matroid Constraints and Free Disposal.}
Let $\Lambda$ be the set of \emph{agents} (offline nodes),
and \emph{items} (online nodes) from $\Omega$
arrive one by one.  Each agent $\lambda \in \Lambda$
is equipped with a non-negative monotone
submodular function $f^{\lambda}: 2^\Omega \rightarrow \R^+$.
Moreover, each $\lambda$
is also associated with a matroid $(\Omega, \I^{\lambda})$,
and again without loss of generality, every singleton
in $\Omega$ is independent in $\I^{\lambda}$.

Each agent $\lambda \in \Lambda$ maintains
$S^{\lambda} \in \I^{\lambda}$,
which is initially empty.
When an online item $u \in \Omega$ arrives,
the algorithm can either discard it or assign it
to one of the agents $\lambda$,
in which case, $u$ is included into $S^{\lambda}$
and some element might have to be removed
from $S^{\lambda}$ to ensure $S^{\lambda} \in \I^{\lambda}$.
The goal is to maximize $\sum_{\lambda \in \Lambda} f^{\lambda}(S^{\lambda})$.

\noindent \textbf{Notation.}
We use the same notation as in Sections~\ref{sec:matroid} 
and~\ref{sec:cardinality}.  A superscript $\lambda$
is used to distinguish the objects associated with different agents.
For instance $w^\lambda(u) := f^\lambda(u | A^\lambda(u))$.
For each $\lambda \in \Lambda$,
we use $\mathcal{G}^\lambda$ to denote the online submodular maximization algorithm
used by agent~$\lambda$ with respect to its submodular function $f^\lambda$
and independent sets $\I^\lambda$.  Recall that each agent~$\lambda$
maintains its copy of $S^\lambda$ and $A^\lambda$.
In Algorithm~\ref{alg:multi-offline},
we use individual agents' algorithms as subroutines.

\begin{algorithm2e}
	\caption{Algorithm for Multiple Offline Nodes}
	\label{alg:multi-offline} 
	
	For each agent $\lambda \in \Lambda$, initialize algorithm
	$\mathcal{G}^\lambda$; \\ 
	
	\For{each round when $u$ arrives}{
		$L(u) \leftarrow \varnothing$; \\
		\For{each agent  $\lambda \in \Lambda$}{
		
		Pass $u$ to $\mathcal{G}^\lambda$; \\
		
		\If{$\mathcal{G}^\lambda$ is going to accept $u$ and remove $v^\lambda$
		from $S^\lambda$}{
			
			(If $\mathcal{G}^\lambda$ is not going to remove any element from $S^\lambda$,
			then $v^\lambda = \bot$ and $w^\lambda_{S^\lambda}(\bot) = 0$.) \\

		  $L(u) \leftarrow L(u) \cup \{(\lambda, v^\lambda)\}$;  \\
			
			(Do not let $\mathcal{G}^\lambda$ update its $S^\lambda$ and $A^\lambda$ yet,
			until it is confirmed that $u$ will be assigned to $\lambda$.) \\
		}
		%
		}
		\If{$L(u) \neq \varnothing$}{
			$(\lambda, v^\lambda) \leftarrow \arg\max_{(\lambda, v^\lambda) \in L(u)}
			\{w^\lambda(u) - w^\lambda_{S^\lambda}(v^\lambda)\}$; (ties resolved arbitrarily) \\
			Let $\mathcal{G}^\lambda$ update $S^\lambda \leftarrow S^\lambda + u - v^\lambda$; 
			$A^\lambda \leftarrow A^\lambda + u$;
		}
	}
\end{algorithm2e}

\noindent \textbf{Interpretation.}
We can view that each agent~$\lambda$ runs its own instance of
online algorithm $\mathcal{G}^\lambda$ in the background.
When an element $u$ arrives, it is passed to every agent~$\lambda$,
who might propose to accept $u$ and replace some element $v^\lambda$ currently
in $S^\lambda$.  Out of all the agents that
have proposed, the agent~$\lambda$ is selected such that
$w^\lambda(u) - w^\lambda_{S^\lambda}(v^\lambda)$ is maximized;
then, $u$ is assigned to agent~$\lambda$ and $\mathcal{G}^\lambda$ updates
its $S^\lambda$ and $A^\lambda$ accordingly.

\begin{theorem}[Online Bipartite Matching with Matroid Constraints]
\label{th:multi_matroid}
Suppose Algorithm~\ref{alg:multi-offline} is 
run such that each agent $\lambda \in \Lambda$
uses $\mathcal{G}^\lambda$ from Algorithm~\ref{alg:submodular_matroid}
for general matroids or Algorithm~\ref{alg:k_uni}
for $k$-uniform matroids.
Then, if each $\mathcal{G}^\lambda$ has competitive ratio at least
$\frac{1}{\alpha}$, then 
Algorithm~\ref{alg:multi-offline} has
competitive ratio is at least $\frac{1}{\alpha + 1}$.
\end{theorem}

\begin{proof}
We fix some offline optimal assignment in
which for each $\lambda \in \Lambda$,
the set of elements assigned to agent~$\lambda$ is
$\OPT^\lambda$.  At the end of Algorithm~\ref{alg:multi-offline},
we use $S^\lambda$ to denote the independent set maintained by agent~$\lambda$,
and $A^\lambda$ to denote the set of elements that have been accepted
at some point by agent~$\lambda$.

We fix some $\lambda \in \Lambda$. Define $\widehat{\OPT}^\lambda := \{u \in \OPT^\lambda \setminus A^\lambda: w^\lambda(u) > 0\}$.  We further partition $\widehat{\OPT}^\lambda := \widehat{\OPT}^\lambda_r \cup 
\widehat{\OPT}^\lambda_l$, where the elements in $\widehat{\OPT}^\lambda_r$
are rejected by $\mathcal{G}^\lambda$,
and the elements in $\widehat{\OPT}^\lambda_l$
are proposed by $\mathcal{G}^\lambda$, but are eventually assigned to another agent.
Given an element $u$ such that $L(u) \neq \varnothing$ in Algorithm~\ref{alg:multi-offline},
let $\lambda_u$ denote the agent that $u$ is assigned to in that round.

As in the proof of Theorem~\ref{th:matroid_ratio},
since $f^\lambda$ is monotone and submodular,
we have 

\begin{equation}
f^\lambda(\OPT^\lambda) \leq f^\lambda(\OPT^\lambda \cup A^\lambda)
\leq f^\lambda(A^\lambda) + \sum_{u \in \widehat{\OPT}^\lambda} w^\lambda(u).
\label{eq:opt_sum}
\end{equation}

We next separate the analysis into the two cases,
whether Algorithm~\ref{alg:submodular_matroid}
or Algorithm~\ref{alg:k_uni} is used for each $\mathcal{G}^\lambda$.

\noindent \textbf{Case (a): Algorithm~\ref{alg:submodular_matroid}
for general matroid.}  We have $\alpha = 4$.

Using Lemmas~\ref{lemma:w_and_f}
and~\ref{lemma:A_S},
we have $f^\lambda(A^\lambda) \leq 2 f^\lambda(S^\lambda) - f^\lambda(\varnothing)
\leq 2 f^\lambda(S^\lambda)$, as $f^\lambda$ is non-negative.

For each $u \in \widehat{\OPT}^\lambda$,
let $T(u) \subseteq S^\lambda(u)$ be the elements
that are in conflict with $u$ with respect to the matroid $(\Omega, \I^\lambda)$.
In other words, if $S^\lambda(u) + u \notin \I^\lambda$,
then $T(u) + u$ is the circuit in $S^\lambda(u) + u$;
if $S^\lambda(u) + u \in \I^\lambda$, then $T(u) = \varnothing$.

Applying Lemma~\ref{lemma:matroid_prop},
for each $u \in \widehat{\OPT}^\lambda$ such that $T(u) \neq \varnothing$,
there exists $t_u \in T(u)$ such that
$u \neq v$ implies that $t_u \neq t_v$, and
$\widehat{S}^\lambda := \{t_u : u \in \widehat{\OPT}^\lambda, T(u) \neq \varnothing\} \in \I^\lambda$.
For notational convenience,
if $T(u) = \varnothing$, we write $t_u = \bot$
and $w^\lambda(t_u) = 0$.

If $u \in \widehat{\OPT}^\lambda_r$,
then $u$ is rejected by $\mathcal{G}^\lambda$,
and hence, we have $w^\lambda(u) \leq 2 w^\lambda_{S^\lambda(u)}(t_u)$.

If $u \in \widehat{\OPT}^\lambda_l$,
then $u$ is assigned to another agent $\lambda_u$,
who might have replaced another element~$v_u$.
Hence, we have $w^\lambda(u) - w^\lambda_{S^\lambda(u)}(t_u)
\leq w^{\lambda_u}(u) - w^{\lambda_u}_{S^{\lambda_u}(u)}(v_u)
\leq w^{\lambda_u}(u) - w^{\lambda_u}(v_u)$,
where the last inequality comes from
Lemma~\ref{lemma:monotone_ws}.

Observing that $w^\lambda_{S^\lambda(u)} \leq \widehat{w}^\lambda(u)$
because of Lemma~\ref{lemma:monotone_ws},
we have
\[
\txts
\sum_{u \in \widehat{\OPT}^\lambda} w^\lambda(u)
\leq  2 \cdot \widehat{w}^\lambda(\widehat{S}^\lambda)
+ \sum_{u \in \widehat{\OPT}^\lambda_l} (w^{\lambda_u}(u) - w^{\lambda_u}(v_u)) \enspace.
\]

Since $\widehat{S}^\lambda$ is a subset of $A^\lambda$
that is independent in $\I^\lambda$,
by Lemma~\ref{lemma:greed_opt},
$\widehat{w}^\lambda(\widehat{S}^\lambda)
\leq \widehat{w}^\lambda({S}^\lambda) = f^\lambda(S^\lambda) - f(\varnothing)
\leq f^\lambda(S^\lambda)$, where the equality comes from
Lemma~\ref{lemma:w_and_f}.

Summing (\ref{eq:opt_sum}) over $\lambda \in \Lambda$,
we have

\begin{equation}
\sum_{\lambda \in \Lambda} f^\lambda(\OPT^\lambda)
\leq 4 \sum_{\lambda \in \Lambda} f^\lambda(S^\lambda)
+  \sum_{\lambda \in \Lambda} \sum_{u \in \widehat{\OPT}^\lambda_l} (w^{\lambda_u}(u) - w^{\lambda_u}(v_u)).
\label{eq:opt_sum_gen_mat}
\end{equation}

\noindent \textbf{Case (b): Algorithm~\ref{alg:k_uni} for
$k$-uniform matroids.}

If $u \in \widehat{\OPT}^\lambda_r$,
then $u$ is rejected by $\mathcal{G}^\lambda$,
and hence, we have: 

$w^\lambda(u) \leq \frac{1}{k} \cdot ( \alpha \cdot w^\lambda_{S^\lambda(u)}(S^\lambda(u)) - w^\lambda(A^\lambda(u)))
\leq \frac{1}{k} \cdot (\alpha \cdot \widehat{w}^\lambda(S^\lambda) - w^\lambda(A^\lambda))$,
where the last inequality comes from
Lemma~\ref{lemma:new_A_S}.

If $u \in \widehat{\OPT}^\lambda_l$,
then $u$ is assigned to another agent $\lambda_u$,
who replaces another element~$v_u$.
Hence, we have
\begin{align*}
w^\lambda(u) - w^\lambda_{S^\lambda(u)}(t_u)
\leq w^{\lambda_u}(u) - w^{\lambda_u}_{S^{\lambda_u}(u)}(v_u)
\leq w^{\lambda_u}(u) - w^{\lambda_u}(v_u),
\end{align*}
where
\mbox{$t_u := \arg \min_{v \in S^\lambda(u)} w^\lambda_{S^\lambda(u)}(v)$},
and the last inequality comes from
Lemma~\ref{lemma:monotone_ws}.
Observing that Lemma~\ref{lemma:large_replace}
implies that 
$w^\lambda_{S^\lambda(u)}(t_u) \leq \frac{1}{k} \cdot ( \alpha \cdot w^\lambda_{S^\lambda(u)}(S^\lambda(u)) - w^\lambda(A^\lambda(u)))$,
we have
\begin{align*}
w^\lambda(u) \leq \frac{1}{k} \cdot (\alpha \cdot \widehat{w}^\lambda(S^\lambda) - w^\lambda(A^\lambda))
+ w^{\lambda_u}(u) - w^{\lambda_u}(v_u).
\end{align*}

Observing that $|\widehat{\OPT}^\lambda| \leq k$,
(\ref{eq:opt_sum}) becomes:
\begin{align*}
f^\lambda(\OPT^\lambda)
&\leq f^\lambda(A^\lambda)
+ \alpha \cdot \widehat{w}^\lambda(S^\lambda) - w^\lambda(A^\lambda)
+ \sum_{u \in \widehat{\OPT}^\lambda_l} (w^{\lambda_u}(u) - w^{\lambda_u}(v_u)) \\
&\leq \alpha \cdot f^\lambda(S^\lambda) + 
\sum_{u \in \widehat{\OPT}^\lambda_l} (w^{\lambda_u}(u) - w^{\lambda_u}(v_u)),
\end{align*}
where the last inequality comes from Lemm~\ref{lemma:w_and_f}
and the fact that $f^\lambda$ is non-negative.

Summing over $\lambda \in \Lambda$,
we have

\begin{equation}
\sum_{\lambda \in \Lambda} f^\lambda(\OPT^\lambda)
\leq \alpha \sum_{\lambda \in \Lambda} f^\lambda(S^\lambda)
+  \sum_{\lambda \in \Lambda} \sum_{u \in \widehat{\OPT}^\lambda_l} (w^{\lambda_u}(u) - w^{\lambda_u}(v_u)).
\label{eq:opt_sum_unif_mat}
\end{equation}

\noindent \textbf{Combining the two cases.}
Hence, it remains to give an upper bound
on the second sum on the right hand sides of~
(\ref{eq:opt_sum_gen_mat}) and~(\ref{eq:opt_sum_unif_mat}).
Observing that $\OPT^\lambda$'s are disjoint,
we have
\begin{align*}
\sum_{\lambda \in \Lambda} \sum_{u \in \widehat{\OPT}^\lambda_l} (w^{\lambda_u}(u) - w^{\lambda_u}(v_u))
\leq 
\sum_{\lambda \in \Lambda} \sum_{u \in A^\lambda} (w^\lambda(u) - w^\lambda(v_u))
=\sum_{\lambda \in \Lambda} w^\lambda(S^\lambda)
\leq \sum_{\lambda \in \Lambda} f^\lambda(S^\lambda),
\end{align*}
where the last inequality comes from Lemma~\ref{lemma:w_and_f},
and the equality comes from a telescoping sum with each element $u$
replacing the one $v_u$ in $S^\lambda(u)$.

Therefore,
we have $\sum_{\lambda \in \Lambda} f^\lambda(\OPT^\lambda)
\leq (\alpha+1) \sum_{\lambda \in \Lambda} f^\lambda(S^\lambda)$, as required.
\end{proof}
\section{Algorithms for Non-Monotone Submodular Function}
\label{sec:non-mono}

In this section, we consider the case when the objective function $f$ is non-monotone but still submodular.
As in~\cite{BuchbinderFS15a}, we consider randomized algorithms that are not necessarily monotone.
For randomized algorithms, the competitive ratio
is the expected value of the algorithm's feasible
set divided by the optimal value.

The idea is to consider an auxiliary function
defined as follows.  For $0 \leq p \leq 1$ and set~$S$,
let~$\ind_p(S)$ be the random subset obtained by including
each element in~$S$ independently with probability~$p$.
Define $\widehat{f}_p(S) := \expct{f(\ind_p(S))}$.
Observe that evaluating~$\widehat{f}_p$ takes
exponential number of oracle accesses to~$f$,
but a sampling method is given in~\cite{BuchbinderFS15a}
to estimate~$\widehat{f}_p$.
However, for ease of exposition, in our presentation,
we assume that~$\widehat{f}_p$ is also returned by some oracle.
We also use the following result.

\begin{lemma}[Lemma 2.3 in~\cite{Feige07maximizingnon-monotone}]
\label{lemma:non_mono_exp}
Suppose $f$ is a submodular function.
Then, for any sets $A$ and~$B$ (not necessarily disjoint),
and $0 \leq p, q \leq 1$. We have
\begin{align*}
\expct{f(\ind_p(A) \cup \ind_q(B))} \geq (1-p)(1-q)f(\varnothing) + p(1-q)f(A) + q(1-p) f(B) + pq f(A\cup B).
\end{align*}
\end{lemma}

\subsection{Modification for General Matroids}
\label{sec:non-mono_matroid}

Algorithm~\ref{alg:submodular_matroid} is modified
in the following ways.
\begin{compactitem}
\item Instead of $f$,
we use $g := \widehat{f}_{\frac{1}{2}}$ as the objective function.

\item We only use a single value function $w(u) := g(u | A(u))$, i.e.,
we replace all occurrences of $w_S(\cdot)$ by $w(\cdot)$, since we
no longer need the algorithm to be monotone.  This will
actually simplify the proofs.

\item For an arriving element $u$,
if $w(u) = g(u|A(u))$ is negative,
then the element is definitely discarded.

\item The set $S$ is the same as before, but takes an auxiliary role.  The actual feasible set $\widehat{S}$,
which is a subset of $S$, is maintained by the algorithm as follows.
When the algorithm includes an element $u$ in $S$,
then with probability $\frac{1}{2}$ the element $u$
is included in $\widehat{S}$; when an element
$u'$ is removed from $S$, then the element $u'$ is
also removed from $\widehat{S}$ (if $\widehat{S}$ contains $u'$).
\end{compactitem}

\begin{theorem}
\label{th:non-mono_matroid_ratio}
The modified algorithm has competitive ratio~$\frac{1}{16}$.
\end{theorem}

\begin{proof}
Observe that in Section~\ref{sec:matroid},
the monotonicity of $f$ is necessary
to prove the competitive ratio only
when we need $f(A_n \cup \OPT) \geq f(\OPT)$.
(Monotonicity of $f$ is also used elsewhere to show
that Algorithm~\ref{alg:submodular_matroid} is monotone, but that
is not crucial to the competitive ratio.)

However, by Lemma~\ref{lemma:non_mono_exp},
we have $g(A_n \cup \OPT) \geq \frac{1}{4} f(\OPT)$.
Moreover, $\expct{f(\widehat{S})} = g(S)$.
Hence, it follows that the modification causes another factor of~$\frac{1}{4}$ to the previous competitive ratio~$\frac{1}{4}$.
\end{proof}

\subsection{Improvement for $k$-Uniform Matroids}

Observe that as long as $f$ is submodular and $f(\varnothing) \geq 0$,
keeping the best singleton still gives us a competitive ratio of $\frac{1}{k}$.
Hence, for small values of $k$, we can just return the best singleton;
for sufficiently large $k$, we can apply the randomized technique
described in Section~\ref{sec:non-mono_matroid} to Algorithm~\ref{alg:k_uni}
to achieve a competitive ratio of $\frac{1}{4 \alpha_k}$.  However,
we can exploit the special structure of uniform matroids as in~\cite{BuchbinderFS15a}
to improve the ratio.

We make the following modifications to Algorithm~\ref{alg:k_uni}.

\begin{compactitem}
\item Set $\rho := 3$. Instead of $f$,
we use $g := \widehat{f}_{\frac{1}{\rho}}$ as the objective function.


\item We only use a single value function $w(u) := g(u | A(u))$, i.e.,
we replace all occurrences of $w_S(\cdot)$ by $w(\cdot)$.

\item We allow $S$ to hold at most $\widehat{k} := \rho k$ elements.

\item We set $\alpha = \alpha_k$ to be the unique root of
the equation $(1+\frac{a - \rho  -1 }{\rho k+1}) ^ {\rho k + 1} = a$ that is
at least $\rho + 1$.


\item The condition for taking a new arriving item
$u$ becomes
$w(u) > \frac{1}{\rho k}\left( \alpha \cdot w(S) - \rho \cdot w(A) \right)$.

\item Observe that $S$ is no longer feasible.  We assume the slots in $S$
are indexed by $[\rho k]$.  We define a random subset $J \subset [\rho k]$
of size $k$ as follows.
For $i \in [k]$, pick $c_i \in [\rho]$ uniformly at random.
Then, $J := \{(i-1)\rho + c_i : i \in [k]\}$,
and the algorithm maintains the feasible $\widehat{S}$, which
contains elements of $S$ occupying slots indexed by $J$.  
By~\cite[Lemma~4.10]{BuchbinderFS15a},
$\expct{f(\widehat{S})} \geq g(S)$.
\end{compactitem}

Instead of Lemma~\ref{lemma:large_replace},
we have the following technical lemma, whose proof we defer to the end of the section.

\begin{lemma}
\label{lemma:new_large_replace}
If a new element $u$ is selected to replace some element in the current $S$,
then $w(u) > \frac{\alpha}{\alpha - \rho} \cdot \min_{v \in S} w(v)$.
\end{lemma}

\begin{lemma}
\label{lemma:nonmono_A_S}
The sequence $\{\alpha \cdot w(S_n) - \rho \cdot w(A_n)\}_n$
is monotonically increasing.
\end{lemma}

\begin{proof}
As in the proof of Lemma~\ref{lemma:new_A_S},
it suffices to show that for $n \geq 1$,
$\alpha \cdot( w(S_{n+1}) - w(S_n)) \geq \rho \cdot w(u_{n+1})$.

The case when $S_{n+1} = S_n + u_{n+1}$ is easy,
because the result follows from $\alpha \geq \rho$.

Suppose $S_{n+1} = S_n - u' + u_{n+1}$, because
$u_{n+1}$ replaces $u'$.  Then, the result follows
from Lemma~\ref{lemma:new_large_replace}.
\end{proof}

\begin{theorem}
\label{th:new_k_uni}
The modified Algorithm~\ref{alg:k_uni} has competitive ratio
$\frac{1}{\alpha}(1 - \frac{1}{\rho})$.
\end{theorem}

\begin{proof}
We follow the proof structure of Theorem~\ref{th:uni_ratio}.  Suppose $\OPT$
is an optimal solution (containing at most $k$ elements).
Recall that $\widehat{\OPT} := \OPT \setminus A_n$.

From Lemma~\ref{lemma:non_mono_exp},
$\frac{1}{\rho} (1 - \frac{1}{\rho}) \cdot
f(\OPT) \leq g(A_n \cup \OPT)$,
which is at most
$g(A_n) + \sum_{u \in \widehat{\OPT}} g(u|A_n)$,
because $g$ is submodular, and $A_n$ and $\widehat{\OPT}$
are disjoint.

Next, for $u \in \widehat{\OPT}$,
since $u \notin A(u) \subseteq A_n$,
\begin{align*}
g(u|A_n) \leq w(u) \leq 
\frac{1}{\rho k}\left( \alpha \cdot w(S(u)) - \rho \cdot w(A(u)) \right)
\leq \frac{1}{\rho k}\left( \alpha \cdot w(S_n) - \rho \cdot w(A_n) \right ),
\end{align*}
where the last inequality follows from Lemma~\ref{lemma:nonmono_A_S}.

Hence, we have
\begin{align*}
\frac{1}{\rho} (1 - \frac{1}{\rho}) \cdot
f(\OPT) \leq
g(A_n) + \frac{\alpha}{\rho} \cdot w(S_n) - w(A_n)
\leq 
\frac{\alpha}{\rho} \cdot g(S_n) - (\frac{\alpha}{\rho} - 1) g(\varnothing)
\leq \frac{\alpha}{\rho} \cdot g(S_n),
\end{align*}
where the second inequality follows from
Lemma~\ref{lemma:w_and_f},
and the last inequality follows because $\alpha \geq \rho$
and $g(\varnothing) = f(\varnothing) \geq 0$.

Finally,
as noted above~\cite[Lemma 4.10]{BuchbinderFS15a},
$\expct{f(\widehat{S}_n)}
\geq g(S_n) \geq \frac{1}{\alpha} (1 - \frac{1}{\rho}) f(\OPT)$, as required.
\end{proof}

\begin{corollary}
For non-monotone $f$ with uniform matroid,
there exists a randomized algorithm with
competitive ratio at least $\min_k \max \{\frac{1}{k},
\frac{1}{\alpha_k} (1 - \frac{1}{\rho}) \} = \max\{
\frac{1}{9}, \frac{1}{\alpha_{9}}(1 - \frac{1}{\rho})
\} > 0.1145 $.
\end{corollary}


\begin{proofof}{Lemma~\ref{lemma:new_large_replace}}
We follow the same proof structure as Lemma~\ref{lemma:large_replace}.
Recall that the first $\rho k$ elements are dummies.
Define $\beta := \alpha^{\frac{1}{\rho k+1}} = 1 + \frac{\alpha - \rho -1}{\rho k+1}$.
Let $n$ be the smallest integer at least $\rho k$
such that at least one of the following statements does
not hold.

\begin{compactitem}
\item[(A)] $w(u_{n+1}) > \frac{\alpha}{\alpha - \rho} \min_{u\in S_n} w(u)$.
	
\item[(B)] $w(A_{n+1}) \geq \beta \cdot w(A_n)$.
\end{compactitem}

Using a similar argument as before,
we have:
\begin{equation}
w(u_{n+1}) > \frac{1}{\rho k} \cdot \{\alpha (1 - \frac{1}{\beta^{\rho k}} - \rho \} \cdot w(A_n) = (\beta - 1) \cdot w(A_n).
\label{eq:new_wun}
\end{equation}

Hence, statement (B) holds for $n$, because
$w(A_{n+1}) = w(A_n) + w(u_{n+1}) \geq \beta \cdot w(A_n)$. We next proof an analog of Lemma~\ref{lemma:min_fraction}.

\begin{lemma}
\label{lemma:new_min_fraction}
For all $0\leq i\leq \rho k-1$,
$\min_{u\in S_n}{w(u)} \leq \frac{w(S_{n-i})}{\rho k-i}$.
\end{lemma}

\begin{proof}  The proof is actually simplified, because
we only have one value function $w(\cdot)$.
Observe that the monotonicity of $f$
is only used in Lemma~\ref{lemma:min_fraction} 
to prove $w_{j+1}(S_{j+1}) - w_j(S_j) \leq w_{j+1}(u_{j+1})$ for the case when
$S_{j+1} = S_j - u' + u_{j+1}$.

However, now this becomes
$w(S_{j+1}) - w(S_j) = w(u_{j+1}) - w(u') \leq w(u_{j+1})$,
which is true because the first $\rho k$ dummy elements
ensure that only elements with non-negative $w(\cdot)$ values will be selected.
\end{proof}

\ignore{

However, the analysis cannot go through since  Lemma~\ref{lemma:min_fraction} does not hold directly. The crucial part that we use $f$ is monotone in the lemma is in proving $w_{j+1}(S_{j+1}) - w_j(S_j) \leq w_{j+1}(u_{j+1})$.

	Instead, we use the following argument to replace Lemma~\ref{lemma:min_fraction}. By the selection of $i$, for $n-i<j\leq n$, we have $w(u_j) \geq \frac{1}{k}(\alpha w(S_{j-1}) - p w(A_{j-1})) > (\alpha \gamma - p)\frac{w(A_{j-1})}{k}$. This implies that at most $\frac{k}{\alpha\gamma - p}$ elements in set $S_{j-1}$ are larger than $u_j$, otherwise
	$w(S_{j-1}) \geq \sum_{u\in S_{j-1}: w(u_j)\leq w(u)}{w(u)} \geq \frac{k}{\alpha\gamma -p} w(u_j) > w(A_{j-1})\geq w(S_{j-1})$, where the last inequality is because for all $u\in A_{j-1}$, $w(u)>0$ by Statement 1.

}

We next prove statement (A).
Define $\gamma := \frac{(\alpha- \rho)(\alpha - \rho - 1)}{\alpha}\cdot \frac{\rho k}{\rho k+1}$,
and $\mathsf{m} := \min_{v \in S_n} w(v)$.


Again, the easy case is when $w(S_n) \leq \gamma \cdot w(A_n)$.
Then, from~(\ref{eq:new_wun}),
we have $w(u_{n+1}) > (\beta - 1) \cdot w(A_n) \geq \frac{\beta-1}{\gamma} \cdot w(S_n) = \frac{\alpha}{\alpha - \rho} \cdot \frac{w(S_n)}{\rho k}
\geq \frac{\alpha}{\alpha - \rho} \cdot \mathsf{m}$,
where the last inequality comes from Lemma~\ref{lemma:new_min_fraction}.
Hence, we can assume $w(S_n) > \gamma \cdot w(A_n)$ from now on.
Recall that since $u_{n+1}$ is selected by the algorithm,
we have $w(u_{n+1}) > \frac{1}{\rho k} \cdot (\alpha \cdot w(S_n) - \rho \cdot w(A_n))
\geq \frac{1}{\rho k} \cdot (\alpha \gamma - \rho) \cdot w(A_n)$.  Hence,
we next give a lower bound on $w(A_n)$.

Define $\delta := 1+\frac{1}{\rho k}(\alpha \gamma- \rho)$.
Suppose $0< i \leq n$ is the smallest integer such that $w_{n-i}(S_{n-i}) \leq \gamma \cdot w(A_{n-i})$.  
Using a similar argument as before,
we have for $0 \leq j < i$, 
$w(A_{n-j+1}) \geq \delta \cdot w(A_{n-j})$.

Define the function $\vartheta(x) := (1-x) \delta^{\rho kx}$ for $x \in [0,1]$,
and $\lambda := 1 - \frac{1}{\rho k \ln \delta}$.  
Recall that $\vartheta$ attains its maximum at $\lambda$.  We consider two cases.

\paragraph{Case 1.} $i \leq  \lambda \rho k$. 
In this case,
we have 
\begin{align*}
w(A_n) \geq \delta^{i-1} \beta \cdot w(A_{n-i}) \geq
\frac{\delta^{i-1} \beta}{\gamma} \cdot w_{n-i}(S_{n-i})
\geq \frac{\beta \rho k}{\delta \gamma} \cdot \delta^i (1 - \frac{i}{\rho k}) \cdot
\mathsf{m} =
\frac{\beta \rho k}{\delta \gamma} \cdot \vartheta(\frac{i}{\rho k}) \cdot
\mathsf{m},
\end{align*}
where the last inequality follows from
Lemma~\ref{lemma:new_min_fraction}.

To finish with this case,
we have
\begin{align*}
w(u_{n+1})
&> \frac{1}{\rho k} \cdot (\alpha \gamma - \rho) \cdot w(A_n) 
\geq \frac{1}{k} \cdot (\alpha \gamma - \rho) \cdot \frac{\beta k}{\delta \gamma} \cdot \vartheta(\frac{i}{k}) \cdot
\mathsf{m} \\
&\geq \frac{\beta}{\delta \gamma} 
\cdot (\alpha \gamma -\rho) \cdot \vartheta(0) \cdot \mathsf{m} \\
&\geq \frac{\alpha}{\alpha-\rho} \cdot \mathsf{m},
\end{align*}
where the last inequality follows from Lemma~\ref{lemma:new_technical}(b).

\paragraph{Case 2.} $i > \lambda \rho k$.
In this case, set $\ell := \floor{\lambda \rho k}$.  Then,
we have
\begin{align*}
w(A_n) \geq \delta^{\ell} \cdot w(A_{n-\ell}) \geq
\delta^{\ell} \cdot w(S_{n-\ell})
\geq \rho k \cdot \vartheta(\frac{1}{\rho k} \cdot \floor{\lambda \rho k}) \cdot 
\mathsf{m},
\end{align*}
where the last inequality follows from Lemma~\ref{lemma:min_fraction}.
Hence,
to finish with this case,
we have
\begin{align*}
w(u_{n+1}) > (\alpha \gamma - \rho ) \cdot \vartheta(\frac{1}{k} \cdot \floor{\lambda k}) \cdot \mathsf{m}
\geq \frac{\alpha}{\alpha-\rho} \cdot \mathsf{m},
\end{align*}
where the last inequality follows from
Lemma~\ref{lemma:new_technical}(c).

This finishes the proof of statement (A).
\end{proofof}

\begin{lemma}[$\alpha_k$ Is Decreasing]
\label{lemma:alpha_k_dec}
$\alpha_k$ is decreasing with respect to $k$, for any $\rho \geq 1$.
\end{lemma}

\begin{proof}
Recall that $\alpha_k$ is defined to be
the unique root of equation
$\alpha = (1+\frac{\alpha - \rho -1}{\rho k + 1})^{\rho k + 1}$ that is at least $\rho + 1$.

We are going to prove that for any $1\leq k \leq k'$, $\alpha_{k} \geq \alpha_{k'} $.
Since $\alpha_{k} = (1 + \frac{\alpha_{k}-\rho-1}{\rho k +1})^{\rho k + 1}
\leq (1 + \frac{\alpha_{k}-\rho-1}{\rho k' + 1})^{\rho k' +1}$,
we have
$(1+\frac{\alpha_{k} - \rho - 1}{\rho k' + 1})^{\rho k' + 1} - \alpha_{k} \geq 0
=(1+\frac{\alpha_{k'} - \rho - 1}{\rho k' + 1})^{\rho k' + 1} - \alpha_{k'}
$.

Observe that the function $(1+\frac{x - \rho - 1}{\rho k' + 1})^{\rho k' + 1} - x$ of $x$
is non-decreasing when $x\geq \rho +1$. Therefore, $\alpha_{k} \geq \alpha_{k'}$.
\end{proof}

\begin{lemma}[New Technical Lemmas]
\label{lemma:new_technical}
For $\rho=3$, we have the following.

\begin{compactitem}
\item[(a)] $\alpha \gamma \geq \rho$.

\item[(b)] $\frac{\beta}{\delta \gamma} 
\cdot (\alpha \gamma - \rho) \geq \frac{\alpha}{\alpha - \rho}$.

\item[(c)] $(\alpha \gamma - \rho ) \cdot \vartheta(\frac{1}{\rho k} \cdot \floor{\lambda \rho k}) 
\geq \frac{\alpha}{\alpha- \rho}$.
\end{compactitem}
\end{lemma}


\begin{proof}
By Lemma~\ref{lemma:alpha_k_dec}, we know $\alpha_k$ is decreasing with respect to $k$.

\noindent For (a),
we observe that $\alpha_{\infty} > 5.749$, which implies $\alpha > 5.749$.
Hence, $\alpha\gamma > (5.749-3)\cdot (5.749-4)\cdot \frac{3}{4} > 3.6 >3$.

\noindent For (b), we prove the equivalent inequality
$\alpha\gamma(1 - \frac{\delta}{(\alpha - 3)\beta})\geq 3$.

We first consider the case when $k\geq 100$. Then $\alpha \leq \alpha_{100} < 5.756$.
Observe that
$\alpha\gamma = (\alpha-3)(\alpha-4)\cdot \frac{3k}{3k+1} \leq 2.756\cdot 1.756 < 4.84$.
Hence,
$\alpha\gamma(1 - \frac{\delta}{(\alpha-3)\beta})
= \alpha\gamma(1 - \frac{3k+\alpha\gamma-3}{(\alpha-3)(3k+\alpha-3)}\cdot \frac{3k+1}{3k})
> \alpha\gamma(1 - \frac{3k + 1.84}{(\alpha-3)(3k+2.749)} \cdot \frac{3k+1}{3k})
> \alpha\gamma(1 - \frac{3k+1}{3(\alpha-3)k})
= (\alpha-3)(\alpha-4)\cdot \frac{3k}{3k+1}- (\alpha-4)
\geq (\alpha-3)(\alpha-4)\cdot \frac{300}{301} - (\alpha-4)
> 2.749\cdot 1.749 \cdot \frac{300}{301} - 1.756
> 3.03 > 3
$.

We verify the case when $k<100$ by plotting $I(k) := \alpha\gamma(1 - \frac{\delta}{(\alpha-3)\beta})$ in Figure~\ref{fig:new_tec_ine_b}.

\begin{figure}[H]

\centering
\begin{tikzpicture}
\begin{axis}[
    xlabel={$k$},
    ylabel={$I(k)$},
    xmin=1, xmax=100,
    ymin=3.0, ymax=4.0,
    legend pos=north west,
    ymajorgrids=true,
    grid style=dashed,
]
 
\addplot[
    color=blue,
    ]
    coordinates {
(1,3.718457)(2,3.474854)(3,3.361741)(4,3.296850)(5,3.254827)(6,3.225411)(7,3.203675)(8,3.186960)(9,3.173708)(10,3.162944)(11,3.154027)(12,3.146521)(13,3.140114)(14,3.134582)(15,3.129758)(16,3.125513)(17,3.121749)(18,3.118389)(19,3.115371)(20,3.112645)(21,3.110171)(22,3.107916)(23,3.105852)(24,3.103955)(25,3.102206)(26,3.100588)(27,3.099087)(28,3.097692)(29,3.096390)(30,3.095173)(31,3.094033)(32,3.092963)(33,3.091957)(34,3.091009)(35,3.090114)(36,3.089268)(37,3.088466)(38,3.087707)(39,3.086985)(40,3.086299)(41,3.085646)(42,3.085024)(43,3.084430)(44,3.083863)(45,3.083321)(46,3.082802)(47,3.082304)(48,3.081827)(49,3.081370)(50,3.080930)(51,3.080508)(52,3.080101)(53,3.079710)(54,3.079333)(55,3.078969)(56,3.078619)(57,3.078280)(58,3.077953)(59,3.077637)(60,3.077332)(61,3.077036)(62,3.076750)(63,3.076473)(64,3.076204)(65,3.075944)(66,3.075691)(67,3.075446)(68,3.075208)(69,3.074977)(70,3.074752)(71,3.074534)(72,3.074322)(73,3.074115)(74,3.073914)(75,3.073718)(76,3.073528)(77,3.073342)(78,3.073161)(79,3.072985)(80,3.072813)(81,3.072645)(82,3.072481)(83,3.072321)(84,3.072165)(85,3.072013)(86,3.071864)(87,3.071718)(88,3.071576)(89,3.071437)(90,3.071301)(91,3.071168)(92,3.071038)(93,3.070910)(94,3.070786)(95,3.070664)(96,3.070544)(97,3.070427)(98,3.070312)(99,3.070200)(100,3.070090)

    };
 
\end{axis}

\end{tikzpicture}
\caption{Plot of $I(k)$}
\label{fig:new_tec_ine_b}
\end{figure}

\noindent For (c), we prove $\vartheta(\frac{1}{3k}\cdot \floor{3\lambda k})
\geq \frac{\alpha}{(\alpha-3)(\alpha\gamma-3)}$.

When $k<1000$, we prove the inequality by plotting $H(k):= \vartheta(\frac{1}{3k}\cdot \floor{3\lambda k})
- \frac{\alpha}{(\alpha-3)(\alpha\gamma-3)}$ in Figure~\ref{fig:new_tec_ine_c}.

\begin{figure}[H]

\centering
\begin{tikzpicture}
\begin{axis}[
    xlabel={$k$},
    ylabel={$H(k)$},
    xmin=1, xmax=1000,
    ymin=0, ymax=0.8,
    legend pos=north west,
    ymajorgrids=true,
    grid style=dashed,
]
 
\addplot[
    color=blue,
    ]
    coordinates {

(1,0.658647)(2,0.524392)(3,0.407088)(4,0.334590)(5,0.299263)(6,0.265935)(7,0.240421)(8,0.225701)(9,0.210394)(10,0.197539)(11,0.189579)(12,0.180825)(13,0.173100)(14,0.168142)(15,0.162480)(16,0.157327)(17,0.153953)(18,0.149990)(19,0.146969)(20,0.143869)(21,0.140939)(22,0.138693)(23,0.136335)(24,0.134080)(25,0.132347)(26,0.130492)(27,0.128702)(28,0.127327)(29,0.125829)(30,0.124373)(31,0.123256)(32,0.122020)(33,0.120812)(34,0.119888)(35,0.118852)(36,0.117833)(37,0.117057)(38,0.116174)(39,0.115303)(40,0.114642)(41,0.113881)(42,0.113128)(43,0.112559)(44,0.111896)(45,0.111238)(46,0.110744)(47,0.110161)(48,0.109580)(49,0.109148)(50,0.108631)(51,0.108115)(52,0.107733)(53,0.107272)(54,0.106810)(55,0.106471)(56,0.106056)(57,0.105641)(58,0.105338)(59,0.104963)(60,0.104652)(61,0.104315)(62,0.103974)(63,0.103694)(64,0.103386)(65,0.103076)(66,0.102822)(67,0.102541)(68,0.102256)(69,0.102025)(70,0.101766)(71,0.101505)(72,0.101294)(73,0.101055)(74,0.100814)(75,0.100621)(76,0.100400)(77,0.100176)(78,0.099998)(79,0.099794)(80,0.099586)(81,0.099422)(82,0.099231)(83,0.099038)(84,0.098887)(85,0.098709)(86,0.098528)(87,0.098388)(88,0.098221)(89,0.098052)(90,0.097922)(91,0.097766)(92,0.097607)(93,0.097486)(94,0.097339)(95,0.097189)(96,0.097077)(97,0.096938)(98,0.096798)(99,0.096692)(100,0.096562)(101,0.096429)(102,0.096330)(103,0.096207)(104,0.096103)(105,0.095989)(106,0.095872)(107,0.095774)(108,0.095666)(109,0.095555)(110,0.095464)(111,0.095361)(112,0.095256)(113,0.095169)(114,0.095072)(115,0.094972)(116,0.094890)(117,0.094797)(118,0.094702)(119,0.094625)(120,0.094536)(121,0.094445)(122,0.094372)(123,0.094288)(124,0.094201)(125,0.094132)(126,0.094051)(127,0.093968)(128,0.093903)(129,0.093826)(130,0.093746)(131,0.093684)(132,0.093610)(133,0.093534)(134,0.093475)(135,0.093404)(136,0.093332)(137,0.093276)(138,0.093207)(139,0.093138)(140,0.093084)(141,0.093019)(142,0.092952)(143,0.092901)(144,0.092838)(145,0.092773)(146,0.092725)(147,0.092665)(148,0.092613)(149,0.092556)(150,0.092498)(151,0.092449)(152,0.092394)(153,0.092338)(154,0.092291)(155,0.092238)(156,0.092184)(157,0.092139)(158,0.092088)(159,0.092036)(160,0.091993)(161,0.091944)(162,0.091893)(163,0.091852)(164,0.091804)(165,0.091756)(166,0.091716)(167,0.091670)(168,0.091623)(169,0.091585)(170,0.091541)(171,0.091495)(172,0.091459)(173,0.091416)(174,0.091371)(175,0.091337)(176,0.091295)(177,0.091252)(178,0.091219)(179,0.091178)(180,0.091137)(181,0.091104)(182,0.091065)(183,0.091025)(184,0.090994)(185,0.090956)(186,0.090917)(187,0.090887)(188,0.090850)(189,0.090812)(190,0.090783)(191,0.090747)(192,0.090717)(193,0.090683)(194,0.090648)(195,0.090619)(196,0.090586)(197,0.090552)(198,0.090523)(199,0.090491)(200,0.090458)(201,0.090431)(202,0.090400)(203,0.090368)(204,0.090341)(205,0.090311)(206,0.090280)(207,0.090254)(208,0.090224)(209,0.090194)(210,0.090169)(211,0.090141)(212,0.090111)(213,0.090087)(214,0.090059)(215,0.090030)(216,0.090007)(217,0.089980)(218,0.089951)(219,0.089929)(220,0.089902)(221,0.089875)(222,0.089853)(223,0.089827)(224,0.089800)(225,0.089780)(226,0.089754)(227,0.089728)(228,0.089708)(229,0.089683)(230,0.089657)(231,0.089638)(232,0.089614)(233,0.089589)(234,0.089570)(235,0.089546)(236,0.089526)(237,0.089503)(238,0.089480)(239,0.089460)(240,0.089438)(241,0.089416)(242,0.089397)(243,0.089375)(244,0.089353)(245,0.089335)(246,0.089314)(247,0.089292)(248,0.089274)(249,0.089253)(250,0.089232)(251,0.089215)(252,0.089195)(253,0.089174)(254,0.089157)(255,0.089137)(256,0.089117)(257,0.089100)(258,0.089081)(259,0.089061)(260,0.089045)(261,0.089026)(262,0.089007)(263,0.088991)(264,0.088973)(265,0.088954)(266,0.088939)(267,0.088920)(268,0.088902)(269,0.088887)(270,0.088869)(271,0.088851)(272,0.088837)(273,0.088819)(274,0.088801)(275,0.088787)(276,0.088770)(277,0.088753)(278,0.088739)(279,0.088722)(280,0.088708)(281,0.088692)(282,0.088675)(283,0.088661)(284,0.088646)(285,0.088630)(286,0.088616)(287,0.088600)(288,0.088585)(289,0.088571)(290,0.088556)(291,0.088540)(292,0.088528)(293,0.088513)(294,0.088497)(295,0.088485)(296,0.088470)(297,0.088455)(298,0.088443)(299,0.088428)(300,0.088414)(301,0.088402)(302,0.088388)(303,0.088373)(304,0.088361)(305,0.088347)(306,0.088333)(307,0.088322)(308,0.088308)(309,0.088294)(310,0.088283)(311,0.088270)(312,0.088256)(313,0.088245)(314,0.088232)(315,0.088218)(316,0.088208)(317,0.088195)(318,0.088181)(319,0.088171)(320,0.088158)(321,0.088145)(322,0.088135)(323,0.088123)(324,0.088112)(325,0.088100)(326,0.088087)(327,0.088077)(328,0.088065)(329,0.088053)(330,0.088043)(331,0.088031)(332,0.088019)(333,0.088009)(334,0.087998)(335,0.087986)(336,0.087976)(337,0.087965)(338,0.087953)(339,0.087944)(340,0.087933)(341,0.087921)(342,0.087912)(343,0.087901)(344,0.087890)(345,0.087881)(346,0.087870)(347,0.087859)(348,0.087850)(349,0.087839)(350,0.087828)(351,0.087820)(352,0.087809)(353,0.087799)(354,0.087790)(355,0.087780)(356,0.087769)(357,0.087761)(358,0.087751)(359,0.087740)(360,0.087732)(361,0.087722)(362,0.087712)(363,0.087704)(364,0.087694)(365,0.087684)(366,0.087676)(367,0.087666)(368,0.087658)(369,0.087649)(370,0.087639)(371,0.087631)(372,0.087622)(373,0.087612)(374,0.087604)(375,0.087595)(376,0.087586)(377,0.087578)(378,0.087569)(379,0.087560)(380,0.087552)(381,0.087544)(382,0.087534)(383,0.087527)(384,0.087518)(385,0.087509)(386,0.087502)(387,0.087493)(388,0.087485)(389,0.087477)(390,0.087469)(391,0.087460)(392,0.087453)(393,0.087445)(394,0.087436)(395,0.087429)(396,0.087421)(397,0.087413)(398,0.087406)(399,0.087398)(400,0.087389)(401,0.087383)(402,0.087375)(403,0.087366)(404,0.087360)(405,0.087352)(406,0.087344)(407,0.087337)(408,0.087330)(409,0.087321)(410,0.087315)(411,0.087308)(412,0.087301)(413,0.087293)(414,0.087286)(415,0.087279)(416,0.087272)(417,0.087264)(418,0.087258)(419,0.087251)(420,0.087243)(421,0.087237)(422,0.087230)(423,0.087222)(424,0.087216)(425,0.087209)(426,0.087202)(427,0.087196)(428,0.087189)(429,0.087182)(430,0.087176)(431,0.087169)(432,0.087162)(433,0.087156)(434,0.087149)(435,0.087142)(436,0.087136)(437,0.087130)(438,0.087123)(439,0.087117)(440,0.087110)(441,0.087104)(442,0.087098)(443,0.087091)(444,0.087085)(445,0.087079)(446,0.087073)(447,0.087066)(448,0.087061)(449,0.087054)(450,0.087048)(451,0.087042)(452,0.087036)(453,0.087029)(454,0.087024)(455,0.087018)(456,0.087011)(457,0.087007)(458,0.087000)(459,0.086995)(460,0.086989)(461,0.086983)(462,0.086978)(463,0.086972)(464,0.086966)(465,0.086960)(466,0.086954)(467,0.086948)(468,0.086943)(469,0.086938)(470,0.086932)(471,0.086927)(472,0.086921)(473,0.086915)(474,0.086910)(475,0.086904)(476,0.086898)(477,0.086894)(478,0.086888)(479,0.086882)(480,0.086878)(481,0.086872)(482,0.086866)(483,0.086862)(484,0.086856)(485,0.086850)(486,0.086846)(487,0.086840)(488,0.086835)(489,0.086830)(490,0.086825)(491,0.086819)(492,0.086815)(493,0.086810)(494,0.086804)(495,0.086800)(496,0.086794)(497,0.086789)(498,0.086785)(499,0.086780)(500,0.086774)(501,0.086770)(502,0.086765)(503,0.086760)(504,0.086755)(505,0.086750)(506,0.086746)(507,0.086741)(508,0.086736)(509,0.086731)(510,0.086727)(511,0.086721)(512,0.086717)(513,0.086712)(514,0.086707)(515,0.086703)(516,0.086698)(517,0.086693)(518,0.086689)(519,0.086685)(520,0.086680)(521,0.086676)(522,0.086671)(523,0.086666)(524,0.086662)(525,0.086657)(526,0.086653)(527,0.086649)(528,0.086644)(529,0.086639)(530,0.086636)(531,0.086631)(532,0.086626)(533,0.086622)(534,0.086618)(535,0.086613)(536,0.086610)(537,0.086605)(538,0.086600)(539,0.086597)(540,0.086592)(541,0.086588)(542,0.086584)(543,0.086580)(544,0.086575)(545,0.086572)(546,0.086567)(547,0.086563)(548,0.086559)(549,0.086555)(550,0.086551)(551,0.086547)(552,0.086543)(553,0.086539)(554,0.086535)(555,0.086530)(556,0.086527)(557,0.086523)(558,0.086519)(559,0.086515)(560,0.086511)(561,0.086507)(562,0.086503)(563,0.086499)(564,0.086495)(565,0.086492)(566,0.086488)(567,0.086483)(568,0.086480)(569,0.086476)(570,0.086472)(571,0.086469)(572,0.086465)(573,0.086461)(574,0.086457)(575,0.086453)(576,0.086449)(577,0.086446)(578,0.086442)(579,0.086438)(580,0.086435)(581,0.086431)(582,0.086427)(583,0.086424)(584,0.086420)(585,0.086416)(586,0.086413)(587,0.086410)(588,0.086406)(589,0.086403)(590,0.086399)(591,0.086396)(592,0.086392)(593,0.086388)(594,0.086385)(595,0.086382)(596,0.086378)(597,0.086375)(598,0.086371)(599,0.086367)(600,0.086364)(601,0.086361)(602,0.086357)(603,0.086354)(604,0.086351)(605,0.086347)(606,0.086344)(607,0.086341)(608,0.086337)(609,0.086334)(610,0.086331)(611,0.086327)(612,0.086324)(613,0.086321)(614,0.086317)(615,0.086314)(616,0.086311)(617,0.086307)(618,0.086305)(619,0.086301)(620,0.086298)(621,0.086295)(622,0.086292)(623,0.086288)(624,0.086285)(625,0.086282)(626,0.086279)(627,0.086276)(628,0.086273)(629,0.086269)(630,0.086267)(631,0.086263)(632,0.086260)(633,0.086257)(634,0.086254)(635,0.086251)(636,0.086248)(637,0.086245)(638,0.086242)(639,0.086239)(640,0.086236)(641,0.086233)(642,0.086230)(643,0.086227)(644,0.086224)(645,0.086221)(646,0.086218)(647,0.086215)(648,0.086212)(649,0.086209)(650,0.086206)(651,0.086203)(652,0.086200)(653,0.086198)(654,0.086195)(655,0.086192)(656,0.086189)(657,0.086186)(658,0.086183)(659,0.086181)(660,0.086178)(661,0.086175)(662,0.086172)(663,0.086169)(664,0.086166)(665,0.086164)(666,0.086161)(667,0.086158)(668,0.086155)(669,0.086152)(670,0.086149)(671,0.086147)(672,0.086144)(673,0.086141)(674,0.086139)(675,0.086136)(676,0.086133)(677,0.086131)(678,0.086128)(679,0.086125)(680,0.086123)(681,0.086120)(682,0.086118)(683,0.086115)(684,0.086112)(685,0.086110)(686,0.086107)(687,0.086104)(688,0.086102)(689,0.086099)(690,0.086096)(691,0.086094)(692,0.086091)(693,0.086089)(694,0.086086)(695,0.086084)(696,0.086081)(697,0.086079)(698,0.086076)(699,0.086073)(700,0.086071)(701,0.086068)(702,0.086066)(703,0.086063)(704,0.086061)(705,0.086058)(706,0.086056)(707,0.086053)(708,0.086051)(709,0.086049)(710,0.086046)(711,0.086043)(712,0.086041)(713,0.086039)(714,0.086036)(715,0.086034)(716,0.086032)(717,0.086029)(718,0.086027)(719,0.086024)(720,0.086022)(721,0.086020)(722,0.086017)(723,0.086015)(724,0.086013)(725,0.086010)(726,0.086008)(727,0.086006)(728,0.086003)(729,0.086001)(730,0.085999)(731,0.085996)(732,0.085994)(733,0.085992)(734,0.085989)(735,0.085987)(736,0.085985)(737,0.085982)(738,0.085980)(739,0.085978)(740,0.085976)(741,0.085974)(742,0.085971)(743,0.085969)(744,0.085967)(745,0.085965)(746,0.085962)(747,0.085960)(748,0.085958)(749,0.085956)(750,0.085954)(751,0.085951)(752,0.085949)(753,0.085947)(754,0.085945)(755,0.085942)(756,0.085941)(757,0.085938)(758,0.085936)(759,0.085934)(760,0.085932)(761,0.085930)(762,0.085928)(763,0.085926)(764,0.085923)(765,0.085921)(766,0.085919)(767,0.085917)(768,0.085915)(769,0.085913)(770,0.085911)(771,0.085909)(772,0.085907)(773,0.085905)(774,0.085903)(775,0.085900)(776,0.085899)(777,0.085897)(778,0.085894)(779,0.085892)(780,0.085890)(781,0.085888)(782,0.085886)(783,0.085884)(784,0.085882)(785,0.085880)(786,0.085878)(787,0.085876)(788,0.085874)(789,0.085872)(790,0.085870)(791,0.085868)(792,0.085866)(793,0.085864)(794,0.085863)(795,0.085861)(796,0.085858)(797,0.085857)(798,0.085855)(799,0.085853)(800,0.085851)(801,0.085849)(802,0.085847)(803,0.085845)(804,0.085843)(805,0.085841)(806,0.085839)(807,0.085837)(808,0.085835)(809,0.085834)(810,0.085832)(811,0.085830)(812,0.085828)(813,0.085826)(814,0.085824)(815,0.085823)(816,0.085821)(817,0.085819)(818,0.085817)(819,0.085815)(820,0.085813)(821,0.085811)(822,0.085810)(823,0.085808)(824,0.085806)(825,0.085804)(826,0.085802)(827,0.085801)(828,0.085799)(829,0.085797)(830,0.085795)(831,0.085793)(832,0.085792)(833,0.085790)(834,0.085788)(835,0.085786)(836,0.085785)(837,0.085783)(838,0.085781)(839,0.085779)(840,0.085777)(841,0.085776)(842,0.085774)(843,0.085772)(844,0.085771)(845,0.085769)(846,0.085767)(847,0.085765)(848,0.085764)(849,0.085762)(850,0.085760)(851,0.085758)(852,0.085757)(853,0.085755)(854,0.085753)(855,0.085752)(856,0.085750)(857,0.085748)(858,0.085747)(859,0.085745)(860,0.085743)(861,0.085742)(862,0.085740)(863,0.085738)(864,0.085737)(865,0.085735)(866,0.085733)(867,0.085732)(868,0.085730)(869,0.085728)(870,0.085727)(871,0.085725)(872,0.085724)(873,0.085722)(874,0.085720)(875,0.085719)(876,0.085717)(877,0.085716)(878,0.085714)(879,0.085712)(880,0.085711)(881,0.085709)(882,0.085708)(883,0.085706)(884,0.085704)(885,0.085703)(886,0.085701)(887,0.085700)(888,0.085698)(889,0.085697)(890,0.085695)(891,0.085694)(892,0.085692)(893,0.085690)(894,0.085689)(895,0.085687)(896,0.085686)(897,0.085684)(898,0.085683)(899,0.085681)(900,0.085680)(901,0.085678)(902,0.085677)(903,0.085675)(904,0.085674)(905,0.085672)(906,0.085671)(907,0.085669)(908,0.085668)(909,0.085666)(910,0.085665)(911,0.085663)(912,0.085662)(913,0.085660)(914,0.085659)(915,0.085657)(916,0.085656)(917,0.085654)(918,0.085653)(919,0.085651)(920,0.085650)(921,0.085648)(922,0.085647)(923,0.085646)(924,0.085644)(925,0.085643)(926,0.085641)(927,0.085640)(928,0.085638)(929,0.085637)(930,0.085635)(931,0.085634)(932,0.085633)(933,0.085631)(934,0.085630)(935,0.085628)(936,0.085627)(937,0.085625)(938,0.085624)(939,0.085623)(940,0.085621)(941,0.085620)(942,0.085619)(943,0.085617)(944,0.085616)(945,0.085614)(946,0.085613)(947,0.085612)(948,0.085610)(949,0.085609)(950,0.085608)(951,0.085606)(952,0.085605)(953,0.085603)(954,0.085602)(955,0.085601)(956,0.085599)(957,0.085598)(958,0.085597)(959,0.085595)(960,0.085594)(961,0.085593)(962,0.085591)(963,0.085590)(964,0.085589)(965,0.085587)(966,0.085586)(967,0.085585)(968,0.085583)(969,0.085582)(970,0.085581)(971,0.085579)(972,0.085578)(973,0.085577)(974,0.085576)(975,0.085574)(976,0.085573)(977,0.085572)(978,0.085570)(979,0.085569)(980,0.085568)(981,0.085566)(982,0.085565)(983,0.085564)(984,0.085563)(985,0.085561)(986,0.085560)(987,0.085559)(988,0.085558)(989,0.085556)(990,0.085555)(991,0.085554)(992,0.085553)(993,0.085551)(994,0.085550)(995,0.085549)(996,0.085548)(997,0.085546)(998,0.085545)(999,0.085544)(1000,0.085543)

    };
 
\end{axis}

\end{tikzpicture}
\caption{Plot of $H(k)$}
\label{fig:new_tec_ine_c}
\end{figure}

Now we assume $k\geq 1000$. We observe that $5.749 < \alpha < 5.7497$.
Hence,
$\alpha\gamma > 2.749\cdot 1.749 \cdot \frac{3000}{3001} > 4.806$, and
$\alpha\gamma < 2.7497\cdot 1.7497 < 4.812$.
Then,
$\delta > 1 + \frac{1.806}{3k}$, and $\delta < 1 + \frac{4.812}{3k}$.
Furthermore,
$\lambda = 1 - \frac{1}{3k\ln{\delta}}
= 1 - \frac{1}{3k\ln{1+\frac{1}{3k}(\alpha\gamma-3)}}
> 1 - \frac{1}{3k\ln{1+\frac{1.806}{3k}}}
\geq 1 - \frac{1}{3000\ln{1+\frac{1.806}{3000}}}
> 0.446
$, and
$\lambda < 1 - \frac{1}{3k\ln{1+\frac{1.812}{3k}}}
< 1 - \frac{1}{1.812}
< 0.449
$.

Therefore, $\vartheta(\frac{1}{3k}\cdot \floor{3\lambda k})
\geq \vartheta(\lambda - \frac{1}{3k})
> \frac{1-\lambda}{\delta}\cdot \delta^{3k\lambda}
> \frac{1-\lambda}{\delta}\cdot (1+\frac{1.806}{3000})^{3000\cdot 0.446}
> \frac{1-0.449}{1 + \frac{4.812}{3000}} \cdot 2.237
> 1.230
$. On the other hand,
$\frac{\alpha}{(\alpha-3)(\alpha\gamma-3)}
< \frac{5.749}{2.749\cdot 1.806}
< 1.158
$.

This finishes the proof.

\end{proof}

\newpage

\bibliography{main,matching,buyback}
\bibliographystyle{alpha}


\end{document}